\documentclass[11pt]{article}
\usepackage{amssymb,dsfont,amsmath,amsfonts,hyperref,mdframed,mathrsfs,stmaryrd}
\usepackage[active]{srcltx}
\usepackage{slashed}
\usepackage{color}
\usepackage{amsthm}
\usepackage{scrtime}
\usepackage{cite}
\usepackage[normalem]{ulem}
\usepackage{framed}
\usepackage{cancel}
\usepackage{comment}
\usepackage{empheq}
\usepackage{rotating}
\usepackage{pdflscape}
\usepackage{afterpage}
\usepackage{csquotes}
\usepackage{enumitem}

\advance \textheight by \topskip
\numberwithin{equation}{section}

\let\oldsqrt\sqrt

\def\sqrt{\mathpalette\DHLhksqrt}
\def\DHLhksqrt#1#2{%
\setbox0=\hbox{$#1\oldsqrt{#2\,}$}\dimen0=\ht0
\advance\dimen0-0.2\ht0
\setbox2=\hbox{\vrule height\ht0 depth -\dimen0}%
{\box0\lower0.4pt\box2}}

\def\be{\begin{equation}}
\def\ee{\end{equation}}
\def\bea{\begin{eqnarray}}
\def\eea{\end{eqnarray}}

\def\a{\alpha}
\def\b{\beta}
\def\c{\gamma}
\def\d{\delta}
\def\e{\epsilon}
\def\k{\kappa}

\def\m{\mu}
\def\n{\nu}
\def\s{\sigma}

\def\vark{\varkappa}
\def\t{\tau}

\def\Comp{\mathbb{C}}
\def\Real{\mathbb{R}}

\def\ad{{\dot{\alpha}}}
\def\bd{{\dot{\beta}}}
\def\cd{{\dot{\gamma}}}
\def\dd{{\dot{\delta}}}

\def\au{{\underline{\alpha}}}
\def\bu{{\underline{\beta}}}
\def\cu{{\underline{\gamma}}}
\def\du{{\underline{\delta}}}

\def\ab{{\bar{a}}}
\def\bb{{\bar{b}}}
\def\jb{{\bar{\jmath}}}

\def\ub{{\bar{u}}}
\def\vb{{\bar{v}}}

\def\xb{{\bar{x}}}
\def\yb{{\bar{y}}}
\def\zb{{\bar{z}}}

\def\kb{\bar\kappa}
\def\varkb{\bar\varkappa}
\def\sb{{\bar{\sigma}}}

\def\ft#1#2{{\textstyle{{\scriptstyle #1}\over {\scriptstyle #2}}}} 

\newcommand{\dbtilde}[1]{\overset{\approx}{#1}}
\newcommand{\ytp}{{\widetilde{y}}}
\newcommand{\ytb}{{\dbtilde{y}}}

\def\mso{\mathfrak{so}}

\def\mhs{\mathfrak{hs}}
\def\mD{\mathfrak{D}}

\def\acd{D^{(0)}_{\rm ad}}
\def\tcd{D^{(0)}_{\rm tw}}

\input{glyphtounicode}
\newcommand{\eq}[1]{(\ref{#1})}

\newcommand{\nn}{\nonumber}

\newcommand{\bse}{\begin{subequations}}
\newcommand{\ese}{\end{subequations}}

\newcommand{\diff}{d}
\newcommand{\diffhat}{\hat{d}}
\newcommand{\dx}{dx}
\newcommand{\dZ}{dZ}
\newcommand{\dz}{dz}
\newcommand{\dzb}{d\bar{z}}
\newcommand{\starcomm}[2]{\left[#1\,,#2\right]_\star}
\newcommand{\staracomm}[2]{\left\{#1\,,#2\right\}_\star}
\newcommand{\picomm}[2]{\left[#1\,,#2\right]_{\pi}}

\DeclareMathOperator{\hc}{h.c.}
\DeclareMathOperator{\Tr}{Tr}
\DeclareMathOperator{\STr}{STr}

\begin{document}

\begin{center}


{\Large\sc Fronsdal fields from gauge functions} \\[10pt]
{\Large\sc in Vasiliev's higher spin gravity}

\vskip 1.0cm


{\sc David De Filippi$^1$, Carlo Iazeolla$^2$ and Per Sundell$^3$}

\vskip 1.0cm


{\small{
{\em $^1$ \hskip -.1truecm
\textit{Service de Physique de l'Univers, Champs et Gravitation, UMONS,
\\Place du Parc 20, 7000 Mons, Belgium}
\vskip 1pt }

{email: \href{david.defilippi@umons.ac.be}{\tt david.defilippi@umons.ac.be}}
}}

\vskip .5cm


{\small{
{\em $^2$ \hskip -.1truecm
\textit{NSR Physics Department, G. Marconi University \\
	via Plinio 44, Rome, Italy \quad \& \\
	INFN, Sezione di Napoli\\ Complesso Universitario di Monte S. Angelo, Napoli, Italy}
\vskip 1pt }

{email: \href{c.iazeolla@gmail.com}{\tt c.iazeolla@gmail.com}}
}}

\vskip .5cm


{\small{
{\em $^3$ \hskip -.1truecm
\textit{Departamento de Ciencias Fisicas, Universidad Andres Bello, 
Republica 220, Santiago de Chile}
\vskip 1pt }

{email: \href{per.anders.sundell@gmail.com}{\tt per.anders.sundell@gmail.com}}
}}

\vskip 1.0 cm

\end{center}


\paragraph{Abstract.} 
In this paper, we revisit a number of issues in Vasiliev's theory related to gauge functions, ordering schemes, and the embedding of Fronsdal fields into master fields.
First, we parametrize a broad equivalence class of linearized solutions using gauge functions and integration constants, and show explicitly how Fronsdal fields and their Weyl tensors arise from these data in accordance with Vasiliev's central on mass shell theorem.
We then gauge transform the linearized piece of exact solutions, obtained in a convenient gauge in Weyl order, to the aforementioned class, where we land in normal order.
We spell out this map for massless particle and higher spin black hole modes.
Our results show that Vasiliev's equations admit the correct free-field limit for master field configurations that relax the original regularity and gauge conditions in twistor space.
Moreover, they support the off-shell Frobenius--Chern--Simons formulation of higher spin gravity for which Weyl order plays a crucial role.
Finally, we propose a Fefferman-Graham-like scheme for computing asymptotically anti-de Sitter master field configurations, based on the assumption that gauge function and integration constant can be adjusted perturbatively so that the full master fields approach free master fields asymptotically.

\newpage
\tableofcontents

\section{Introduction}
\label{sec:intro}

\subsection{Motivations and summary of our main results} 

An outstanding problem in higher spin gravity (HSG) is the quest for a geometric formulation suitable for computing physical observables. 
This is a highly non-trivial and impactful task as HSG requires a ``stringy'' generalization of geometry, in which gauge transformations mix fields of different spins and different numbers of derivatives, in such a way that the standard concepts of Riemannian geometry tied to the notion of a privileged spin-2 field lose meaning.

Vasiliev's equations \cite{Vasiliev:1990en,Vasiliev:1990vu,Vasiliev:1992av,Vasiliev:1999ba,Vasiliev:2003ev,Bekaert:2005vh,Didenko:2014dwa} provide a fully non-linear description of higher spin geometries using master fields given by horizontal differential forms on non-commutative fibered manifolds. 
The master fields are functions of fiber coordinates $Y$, and the base manifold is an extension of spacetime with additional coordinates $Z$, with $Y$ and $Z$ each making up a non-commutative twistor space.
Originally, this construction was proposed as a method for encoding an infinite tower of Fronsdal fields together with highly non-local interactions into a remarkably simple set of constraints, but at the price of introducing ambiguities, entering via the resolution scheme for the $Z$-dependence, that affect the reading of the dynamics of physical fields from spacetime vertices and Witten diagrams. 
It remains an interesting open problem whether this approach to higher spin gravity will eventually bear fruit, e.g. by finding a suitable geometric principle for selecting (a class of) $Z$-space resolutions, and progress is currently being made \cite{Vasiliev:2017cae,Didenko:2018fgx,Didenko:2019xzz}.

An alternative strategy for obtaining  physical information from Vasiliev's equations is to gain guiding principles for higher spin geometry by studying their exact solutions.
While extracting spacetime vertices is quite a delicate operation in the sense mentioned above, the form of the equations offers powerful ways of constructing solution spaces. 
Moreover, examining these solutions using classical observables given by geometrically constructed higher spin invariants opens up for an approach to higher spin gravity that sidesteps the issues of spacetime interactions vertices, which is one of the basic motivations behind this work.

Over the years, families of classical solutions to Vasiliev's equations \cite{Prokushkin:1998bq,Sezgin:2005pv,Iazeolla:2007wt,Didenko:2009td,Iazeolla:2011cb,Iazeolla:2012nf,Gubser:2014ysa,Iazeolla:2015tca,Sundell:2016mxc,Iazeolla:2017vng,Iazeolla:2017dxc,Aros:2017ror,Aros:2019pgj} have been constructed, including candidate asymptotically anti-de Sitter (AAdS) higher spin black holes in four dimensions\footnote{%
The solutions that we refer to as higher spin black hole states are so called essentially because they possess a tower of Weyl tensors of all integer spins that include and generalize the spin-2 Weyl tensor of an AdS Schwarzschild black hole. However, at present there is no known higher-spin invariant quantity ensuring that the singularity of the individual Weyl tensors is physical, and whether there exists any invariant notion of an event horizon --- as well as an entropy attached to it --- remains an open problem. On the other hand, the fact that each such solution has identical black-hole asymptotics but is possibly non-singular and horizon-free may suggest an interpretation in terms of black-hole microstates, similar to fuzzballs \cite{Lunin:2001jy,Lunin:2002qf,Mathur:2005zp,Skenderis:2008qn}. In that sense, the name black-hole states may turn out to be appropriate in an even deeper sense. See \cite{Iazeolla:2017vng} for more details on this proposal and on our usage of the terminology.} and topologically non-trivial asymptotically locally anti-de Sitter (ALAdS) solutions.
Large solution spaces, that are of relevance to this paper, exhibit the following characteristics:
\begin{enumerate}[label=\alph*),ref=\alph*]
\item\label{it:vacuum gF} choices of gauge functions that trivialize all fluctuations in the spacetime gauge fields;
\item\label{it:int cst} zero-form integration constants (sometimes referred to as initial data) that contain local degrees of freedom in AAdS and ALAdS geometries;
\item usage of special non-polynomial classes of symbols of horizontal forms presented in terms of auxiliary parametric integrals, including singular potentials for delta function sources, induced from solving the equations in Weyl order  \cite{Iazeolla:2011cb,Iazeolla:2012nf,Iazeolla:2017vng,Aros:2017ror,Aros:2019pgj};
\item\label{it:holom gauge} 
a universal $Z$-space resolution that dresses any first-order initial datum into an exact solution in a convenient holomorphic gauge.
\end{enumerate}
While very effective and natural for the basic structure of the Vasiliev equations, these ingredients blur the identification of Fronsdal fields within Vasiliev's master fields in accordance with the Central On Mass Shell Theorem (COMST) \cite{Vasiliev:1990vu,Vasiliev:1992av,Vasiliev:1999ba}, which makes use of normal order \cite{Vasiliev:1990vu,Bekaert:2005vh,Vasiliev:2015wma} and a gauge in twistor space that differs from that it in \eqref{it:holom gauge}. 
More precisely, as noted in (a), most of the known exact solutions have been constructed in gauges in which the spacetime connection encodes only the AdS background.
Hence it remains uncorrected\footnote{In fact, the spacetime connection remains uncorrected modulo a specific field redefinition, required for the perturbatively-defined master fields to transform properly under the local Lorentz symmetry generators \cite{Vasiliev:1990vu,Vasiliev:1999ba,Sezgin:2005pv}. %
At linear order however, this additional effect is concentrated outside the $Z=0$ slice, i.e. irrelevant for the spacetime approach.
}, i.e., it is not glued to the propagating degrees of freedom carried by the Weyl zero-form.
The natural question is then whether such solutions admit an interpretation in terms of Fronsdal fields upon linearization. 

In this paper, we construct a family of linearized gauge functions and integration constants that encode propagating Fronsdal fields with non-trivial Weyl tensors.
The gauge functions are obtained by solving a gauge condition that relaxes the one used in the standard linearization of Vasiliev's equations.
In other words, the (relaxed) gauge functions provide the Chevalley-Eilenberg cocycle required by the COMST, that glues the on-shell curvatures of the Fronsdal fields, i.e. their Weyl tensors, to the corresponding components of the Weyl zero-form. 
In particular, our results spell out the relation at linear order between the gauge function solution method \cite{Vasiliev:1990bu,Sezgin:2005pv,Iazeolla:2007wt,Iazeolla:2011cb,Iazeolla:2012nf,Gubser:2014ysa,Sundell:2016mxc,Iazeolla:2017vng,Iazeolla:2017dxc,Aros:2017ror,Aros:2019pgj} and the ordinary perturbative analysis of the Vasiliev equations \cite{Vasiliev:1992av,Vasiliev:1999ba,Sezgin:2002ru,Didenko:2015cwv}. 
Equivalently, the relevant gauge functions appear as the gauge parameters that send field configurations obtained within a class of $Z$-space resolution schemes, satisfying certain assumptions, into field configurations obtained within the aforementioned relaxed version of Vasiliev's original resolution scheme, which we shall refer to as the standard scheme.
Hence, our construction encapsulates the propagating Fronsdal fields within (gauge equivalence classes of) gauge functions on twistor space with interesting properties that we will exhibit in the bulk of the paper.

Thus, in order to properly interpret families of exact solutions, it is in general important to connect various perturbative schemes by means of re-orderings and gauge transformations.
One may then examine to what extent these transformations are large by evaluating classical observables given by functionals that are higher-spin gauge invariant up to total derivatives on the full non-commutative manifold.

Being able to relate the perturbative schemes in different orderings is also of importance from the point of view of the off-shell extension of the Vasiliev theory put forward in \cite{Boulanger:2015kfa,Bonezzi:2016ttk,Arias:2017bvi}. More precisely, in those works it has been proposed to 
\begin{itemize}
\item[i)] embed Vasiliev's higher spin geometries into the moduli space of a flat Quillen superconnection \cite{Quillen:1985vya} valued in the direct product of the differential graded algebra of horizontal forms and an internal graded associative algebra;
\item[ii)] treating the flat superconnections as semi-classical boundary states of a Frobenius--Chern--Simons (FCS) gauge theory with a star-product local Batalin--Vilkovisky master action of Alexandrov--Kontsevich--Schwarz--Zaboronsky (AKSZ) type \cite{Alexandrov:1995kv,Boulanger:2012bj};
\item[iii)] assign each boundary state an on-shell action in the form of a topological vertex operator \cite{Sezgin:2011hq}\footnote{%
In the FCS model, the on-shell actions are derived from Chern classes, whereas more generally, within the context of the AKSZ formalism, they need only be topological vertex operators, i.e. functionals whose total variation vanishes on-shell, that is, they are partial actions in the sense of \cite{Kazinski:2005eb,Lyakhovich:2006sc}.};
\item[iv)] identify the on-shell action evaluated on ALAdS higher spin geometries with the generating functional of a dual holographic conformal field theory (CFT) \cite{Sundborg:2000wp,Sezgin:2002rt,Klebanov:2002ja}.
\end{itemize}
In the FCS model, the superconnection is represented by symbols defined using Weyl ordering.
On the other hand, as mentioned above, in the standard scheme the horizontal forms are represented using normal ordering, as it allows to formulate the field equations in terms of regular symbols only.
Hence, the identification of a Vasiliev branch within the FCS model relies on the possibility of going between Weyl and normal ordering in classical moduli spaces, which involves going beyond the class of analytic functions on twistor space\footnote{%
Indeed, the Weyl-ordered symbol of one of the fundamental ingredients of the Vasiliev formulation, the Klein operator \eqref{eq:k^2=1}, is a delta function \cite{Didenko:2009td}.}.

In this paper, we resolve the above problem at the linearized level by showing the gauge equivalence between $Z$-space resolutions schemes that arise naturally in Weyl and normal order\footnote{The act of exchanging any two ordering schemes is a similarity transformation on symbols generated by a symmetric polyvector field that preserves the trace operation, known as a Kontsevich gauge transformation, as it is local in twistor space (but not in spacetime).
The higher spin transformations arise as inner transformations generated by star product conjugations by functions on twistor space.
The Kontsevich gauge transformations can be made local also in spacetime by introducing additional spacetime gauge fields.
An interesting open problem is whether it is possible to source these additional gauge fields by means of any generalization of the cocycle appearing in Vasiliev's equations.}.
Indeed, the standard scheme can be connected by our procedure to what we refer to as the Weyl-ordered $Z$-space resolution scheme, as it originates from solving the Vasiliev equations using the standard resolution operator but applied in Weyl order (or, equivalently, keeping the $Y$ and $Z$ dependence factorized). 
This method has been used in a number of circumstances to obtain solution spaces in which the symbols have then been mapped from Weyl to normal order; for example, see \cite{Iazeolla:2017vng}. 
As we show in Section \ref{ssec:resolution}, the re-ordering step maps the Weyl-ordered $Z$-space resolution operator to an equivalent operator in normal order, different from the standard one; see also \cite{Didenko:2019xzz}, where a one-parameter family of $Z$-space resolution schemes in normal ordering, including the Weyl-ordered scheme, is mapped to Vasiliev's original resolution scheme via parameter-dependent star-product re-orderings. 
In other words, the aforementioned solution spaces can equivalently be obtained by working in normal order and using the new resolution operator, as we prove in Appendix \ref{App:fact}. 
For this reason, we refer to the latter resolution as Weyl-ordered, even if it operates on normal-ordered symbols. 
Once in normal order, our general results, presented in Section \ref{sec:COMST}, enable us to move to the relaxed Vasiliev gauge by means of a gauge transformation. 
The required gauge parameter removes the non-analyticities from the solutions obtained via the Weyl-ordered scheme (which are present in normal as well as Weyl-order), leading to a real-analytic generating function for Fronsdal gauge fields; see Eq. \eq{3.70}.
Essentially, this is possible because the singularities arising within the Weyl-ordered scheme are cohomologically trivial. 

In other words, motivated by the study of exact solutions, we prove that Vasiliev's equations admit a sensible free-field limit for master fields belonging to a class of symbols which is larger than formal polynomials, thereby allowing us to relax one the assumptions of Vasiliev's original analysis.  
We exemplify the procedure explicitly for initial data from specific higher-spin representations corresponding to linearized mode functions of massless particle  and  higher spin black hole states, and we explicitly construct the gauge function from which one can extract the linearized gauge fields. 

In order to deal with the Weyl-order-induced singularities of symbols and the aforementioned initial data that are non-polynomial functions and distributions in the non-commutative twistor variables, it is important to specify their functional presentation.
To this end, a \emph{regular computational scheme}  \cite{Iazeolla:2011cb,Iazeolla:2012nf,Iazeolla:2017vng,Iazeolla:2017dxc,Aros:2019pgj} based on auxiliary integral presentations of symbols has been set up in order to construct classical solutions from the classical moduli in \eqref{it:vacuum gF} and \eqref{it:int cst} using various gauges and orderings.
This scheme, to be spelled out in yet more detail in the body of the paper, consists of the following three prescriptions:
\begin{enumerate}[label=\roman*),ref=\roman*]
\item\label{it:regpres1 pres} the symbols of perturbatively defined master fields are given \emph{regular presentations} as auxiliary parametric integrals with kernels given by Gaussian operators;
\item\label{it:regpres1 star}
star products and traces are performed \emph{prior to} auxiliary integrals at each order of classical perturbation theory; and 
\item\label{it:regpres1 ambigu}
the auxiliary integrals must provide \emph{ambiguity-free}\footnote{%
One potential source of ambiguity is the rise of singularities in complex auxiliary integration planes upon performing star products, that may interfere with an otherwise ambiguity-free choice of closed integration contours.} regular presentations at every order of classical perturbation theory.
\end{enumerate}
Successfully implemented, the scheme can be used to
ensure that the on-shell master fields form free differential graded associative algebras \cite{Iazeolla:2017vng} with well-defined invariants.

In this paper, we shall refine and formalize the regular scheme as originally proposed and use it to map master field configurations built in Weyl order to corresponding unfolded Fronsdal fields on AdS spacetime glued to their Weyl curvatures in accordance with the COMST.    
We thereby close the apparent gap between the Weyl- and normal-ordered formalisms that has existed in the literature already at the linearized level.

Moreover, we propose a Fefferman--Graham-like scheme for the perturbative construction of AAdS solutions to Vasiliev's equations whose on-shell action yields physically meaningful holographic two-point functions in the leading order of classical perturbation theory. 
This scheme involves fixing boundary conditions in both spacetime and twistor space; in particular, it is natural to expect that such a procedure will help in resolving the ambiguities that arise in integrating the $Z$-dependence perturbatively, and that it is instrumental to properly singling out the superselection sector that may be captured by the dual CFT. 
It is then conceivable that implementing this scheme at higher orders may provide a rationale for fixing the resolution of the $Z$-dependence and for selecting a class of twistor space functions (which is crucial in a non-commutative field theory). 

Thus, the analysis carried out in this paper addresses a number of open issues in the literature on Vasiliev's theory concerning exact solutions, perturbative schemes, admissible classes of symbols, and choices of gauge and ordering prescriptions.
It also provides the starting block for an iterative procedure for constructing ALAdS higher spin geometries with non-trivial topology both in spacetime and twistor space and related on-shell actions.
Non-trivial twistor space topology can be created within the regular scheme by inclusion of non-polynomial star product algebra extensions of the Weyl algebra that reach beyond naive real-analytic non-polynomial completions (which in general do not form star product algebras).
Another advantage of the regular scheme is that it facilitates the evaluation of classical observables, allowing to directly address the hypothetical duality between the bulk HSG and the holographic CFT \cite{Colombo:2010fu,Colombo:2012jx,Didenko:2012tv,Didenko:2013bj,Bonezzi:2017vha}, thereby sidelining the passage via the non-local deformed Fronsdal theory on AdS   \cite{Boulanger:2015ova,Bekaert:2015tva,Vasiliev:2016xui,Sleight:2017pcz,Vasiliev:2017cae,Didenko:2018fgx}.
More precisely, the evaluation of zero-form charges, that is, observables built from higher spin Weyl tensors on-shell, can be performed directly on the Weyl-ordered configuration.
Indeed, the gauge function drops out from the zero-form charge, while the AAdS boundary conditions may require a modification of the Weyl-ordered solution itself that yields a non-trivial deformations of the zero-form invariants.
In this work, we show that they are protected for Weyl-ordered solutions based on uncorrected initial data, in the sense that, although there are sub-leading perturbative corrections to the observables, they are all proportional to the leading contribution.

\subsection{ALAdS geometries and AKSZ quantization}

Our working hypothesis is that the regular scheme gives rise to a global formulation of HSG based on the FCS model with
\begin{enumerate}[label=\roman*),ref=\roman*]
\item \label{it:alads} a moduli space of ALAdS higher spin geometries;
\item higher spin invariant functionals serving as observables;
\item an on-shell action giving rise to a partition function of the FCS model.
\end{enumerate}
We expect \eqref{it:alads} to include asymptotically massless particle and higher spin black hole states \cite{Didenko:2009td,Iazeolla:2011cb,Iazeolla:2012nf,Sundell:2016mxc,Iazeolla:2017vng,Iazeolla:2017dxc}.
At the linearized level, these states are carried by Weyl zero-forms that are localizable inside the bulk of the ALAdS spacetime \cite{Colombo:2010fu,Iazeolla:2011cb}, that is, that remain well-defined upon replacing conformal infinities by compact marked submanifolds.
Thus, their on-shell action should depend on the Weyl zero-form but not the vacuum gauge function, that is, it should be an on-shell closed and gauge invariant spacetime zero-form, also known as a zero-form charge \cite{Sezgin:2005pv}.
A natural candidate is the second Chern class on Vasiliev's internal twistor space, which can indeed be added to the FCS bulk action without changing the classical equations of motion while giving rise to a non-trivial on-shell action \cite{Boulanger:2015kfa}. 

In classical perturbation theory, the aforementioned on-shell action is given in the leading order by a bilinear function of the Weyl zero-form that defines a (positive or negative) definite bilinear form \cite{Iazeolla:2011cb} on the higher spin representations containing massless particle and higher spin black hole states.
Moreover, at the first sub-leading order, the back-reaction to the master fields from linearized particle states contains higher spin black hole states \cite{Iazeolla:2017vng}.

This suggests that the dual holographic CFT contains operators corresponding to higher spin black hole states, which form real higher spin multiplets, as well as particle states, which belong to ordinary unitarizable complex higher spin multiplets.
The holographic CFT would thus be non-unitary but nonetheless equipped with a well-defined partition function in Lorentzian signature.
As for its microscopic field content\footnote{%
A natural mechanism for the required quantization of Newton's constant in the bulk is an embedding of the second Chern class defining the on-shell action into a Chern--Simons form for the full Quillen superconnection master field of the FCS model.}, one may think of $N$ conformal matter fields coupled to three-dimensional conformal higher spin gauge fields \cite{Nilsson:2013tva} induced by large gauge functions in the bulk, captured by an on-shell action given by the fourth Chern class of the generalized higher spin Lorentz connection \cite{Sezgin:2011hq}, which encodes a mixture of localizable states and boundary states.
In the case of the minimal bosonic HSG model, the resulting anomalous dimensions could blow up as $N\to 1$ (unlike in the case of the Ising model-like $O(N)$-model), as suggested by the intriguing quantum shift in the inverse Newton's constant from $1/N$ to $1/(N-1)$ computed using the deformed Fronsdal theory on AdS$_4$ \cite{Giombi:2013fka}; for a recent treatment, see \cite{Ponomarev:2019ltz}.

Thus, assuming that massless higher spin particles (but not the graviton) acquire large masses by spontaneous symmetry breaking mechanism (triggered by composite Goldstone modes), an ALAdS higher spin geometry can thus be trusted in its strongly coupled core region (but not its weakly coupled ALAdS region) where it provides a semi-classical realization of a microstate of a quantum theory of gravity with asymptotical observers in a broken phase described by ordinary gravity.
In other words, we think of the core region as an unbroken bubble produced within a broken phase at finite temperature, i. e. created at the expense of switching on microstate on-shell actions while minimizing the free energy as the core region contains a large number of semi-classical microstates (labelled by topologies \cite{Aros:2019pgj} and higher spin charges).

We expect that the relation between the deformed Fronsdal and FCS formulations of HSG provides a prototype for a broader duality between relativistic quantum field theories (QFT) on metric backgrounds (including gravity and string theory) and topological field theories of AKSZ-type with infinitely many fields capable of describing local degrees of freedom, referred to as quasi-topological field theories (QTFT).
The key feature of a QTFT vis-\`a-vis an ordinary topological QFT is that its AKSZ gluing operation \cite{Alexandrov:1995kv} leaves a gluing mark in the interior of the bulk manifold (where the AKSZ momenta vanish identically).
Thus, letting $Z$ denote the corresponding hypothetical functor\footnote{%
To our best understanding, the AKSZ extension of Atiyah's geometric category of (unmarked) bordisms \cite{Atiyah:1989vu} remains to be defined.}, the QTFT partition function 
\be Z(S^1\times \Sigma)={\rm Tr}_{Z(\Sigma)} e^{iK|_{\{0\}\times \Sigma}}\,,\label{QTFTZ}\ee
where $Z(\Sigma)$ is the space of boundary states of the QTFT and $K$ is an on-shell action given by a positive definite functional on $Z(\Sigma)$ evaluated at the marked $\{0\}\times \Sigma\subset S^1\times \Sigma$. 
Thus, if the QFT/QTFT correspondence extends to macroscopic length scales, then it could provide new insights into holography, black hole physics and cosmology.
In particular, a QTFT providing scattering amplitudes and other local QFT observables could bypass the problematic identification of the normalization of the partition function\footnote{%
The normalization of the QTFT partition seems to be related to balancing the (infinite) numbers of even and odd forms by means of topological supersymmetry \cite{Arias:2015wha,Arias:2016agc}.} with the cosmological constant, as the latter quantity enters the QTFT action as a cubic vertex.

\subsection{Outline of the paper.} 
The paper is organized as follows: 

In Sec. \ref{sec:VE}, we formulate Vasiliev's HSG geometrically on-shell starting from first principles, including the COMST, the gauge function method, and the regular perturbative scheme.
Sec. \ref{sec:COMST} treats the linearization of the theory around its AdS vacuum.
In particular, we show how to use gauge functions to map zero-form integration constants to properly unfolded Fronsdal fields on-shell, stressing the fact that the COMST only fixes the linearized gauge function up to $O(Z^2)$ in normal order.
We also highlight occurrences of twistor space singularities in the linearized gauge function even though the physical master fields are real-analytic.
In Sec. \ref{sec:examples}, we spell out the aforementioned map on mass-shells consisting of massless particle and higher-spin black hole states, establishing that the linearized gauge function is real-analytic on twistor space for black hole states but non-real-analytic on twistor space for particle states.
In Sec. \ref{sec:NL}, we use the results of Sec. \ref{sec:COMST} to propose a Fefferman--Graham-like scheme for imposing AAdS boundary conditions on Vasiliev's master fields order by order in classical perturbation theory.
In Sec. \ref{sec:obs}, we review the construction of zero-form charges in Vasiliev's higher-spin gravity, and their perturbative expansion on the classical moduli spaces constructed in the previous sections.
In particular, we shall verify that the virtual twistor space spin-frame used in the holomorphic gauge indeed decouples from these observables up to the first sub-leading order in classical perturbation theory, in accordance with the map established in Sec. \ref{sec:COMST}. 
The paper is concluded in Sec. \ref{sec:ccl}. 
In Appendix \ref{App:conv} we spell out our conventions; in Appendices \ref{App:Tr}, \ref{App:param} and \ref{App:BH} we collect some relevant algebraic properties of those solutions and some useful identities; in Appendix \ref{App:fact} we show that the solutions of \cite{Iazeolla:2017vng} can be obtained by solving recursively to all orders the perturbative expansion based on the standard homotopy operator in Weyl ordering.

\section{Vasiliev's equations}
\label{sec:VE}

We review Vasiliev's on-shell formulation of HSG, starting with the formal definition in terms of locally defined horizontal forms on non-commutative fibered spaces and higher spin geometries supporting globally defined classical observables, after which we spell out the regular scheme for perturbative expansions that we shall implement in the following sections.

\subsection{Local formulation}

\paragraph{Correspondence space.} Vasiliev's master fields are differential forms defined locally on a direct product manifold ${\cal X}_4\times {\cal Z}_4$, where ${\cal X}_4$ is a commutative four-manifold with a real differential structure and ${\cal Z}_4$ is a non-commutative four-manifold with complex (almost) symplectic structure (that blows up at points at infinity).
The master fields are valued in an associative algebra given by distributions (including real-analytic functions as well as Dirac delta functions) on the non-commutative ${\cal Y}_4\cong \Comp^2$ with complex symplectic structure.
Thus, we may think of the master fields as horizontal forms on a total bundle space ${\cal C}$ with fiber ${\cal Y}_4$ and base ${\cal X}_4\times {\cal Z}_4$, referred to as the correspondence space, as reductions from the total space to either ${\cal X}_4\times {\cal Y}_4$ or ${\cal Z}_4\times {\cal Y}_4$ yield dual formulations of the full dynamics. 
To the latter end, we shall take ${\cal X}_4$ and ${\cal Z}_4$ to be closed manifolds with marked points where master field configurations are allowed to blow up, corresponding to boundaries and other defects.
\paragraph{Non-commutative differential Poisson structure.} We assume that ${\cal C}$ has a differential Poisson structure with trivial pre-connection; see \cite{Arias:2015wha} and references therein.
Thus, the quantized versions of the wedge product and the de Rham differential on ${\cal C}$, denoted by $\star$ and $d_{\cal C}$, respectively, obey the standard homotopy relations 
\be d_{\cal C}^2=0\,,\qquad d_{\cal C}(f\star g)=(d_{\cal C} f)\star g+(-1)^{{\rm deg} f} f\star (d_{\cal C} g)\,,\ee
\be f\star (g\star h)=(f\star g)\star h\,,\ee
where $f,g,h\in\Omega({\cal C}|{\cal R})$, a space of symbols on ${\cal C}$ forming the representation ${\cal R}$ of a star product algebra\footnote{%
The regular representation of a star product algebra is the algebra itself.}.
The triviality of the pre-connection implies that there exist local coordinates $(x^\mu;Y^{\au};Z^{\au})$ in charts ${\cal U}\subseteq {\cal X}_4\times {\cal Z}_4\times {\cal Y}_4$, and corresponding operator ordering schemes, such that
\be \Omega({\cal U}|{\cal R})\cong \Omega_{[0]}({\cal U}|{\cal R})\otimes \Comp[dx^\mu,dY^{\au},dZ^{\au}]\,,\ee
where $\Comp[dx^\mu,dY^{\au},dZ^{\au}]$ is the algebra generated by anti-commuting line-elements, and 
\be d_{\cal C}|_{\cal U}=dx^\mu\partial_\mu+dY^{\au} \partial^{(Y)}_\au+dZ^{\au}\partial^{(Z)}_\au\,,\ee
i. e.  the de Rham differential on ${\cal U}$ acting on the symbols in ${\cal R}$.

\paragraph{Horizontal forms.} Locally, the horizontal subalgebra is given by\footnote{%
The definition of a horizontal subalgebra of $\Omega({\cal C})$ requires the existence of a closed and central volume form on the fiber \cite{Arias:2016agc}. }
\be \Omega_{\rm hor}({\cal U}|{\cal R})\cong \Omega_{[0]}({\cal U}|{\cal R})\otimes \Comp[dx^\mu,dZ^{\au}]\,,\ee
that is, a horizontal form $f$ on ${\cal C}$ is represented locally by a symbol $f(x,Y,Z;dx,dZ)$ with $Y$ and $Z$ dependence given by a symbol in ${\cal R}$.
We denote the horizontal projection of $d_{\cal C}$ by $\diffhat$, which we decompose as 
\begin{align}
\diffhat&=d+q
\,,&
d:&=\dx^\mu\partial^x_\mu
\,,& 
q:&=\dZ^\au\partial^Z_\au 
\,,
\end{align}
where thus $d$ and $q$ are the de Rham differentials on ${\cal X}_4$ and ${\cal Z}_4$, respectively.

\paragraph{Holomorphic symplectic structure and chiral twisted convolutions.} Letting
\be Y^\au=(y^\a,\yb^\ad)\,\qquad Z^\au=(z^\a,-\zb^\ad)\,,\ee
be canonical, that is, 
\begin{align}
\starcomm{Y_\au}{Y_\bu}&=2iC_{\au\bu} 
\,,&
\starcomm{Z_\au}{Z_\bu}&=-2iC_{\au\bu} 
\,,&
\starcomm{Y_\au}{Z_\bu}&=0
\,,
\end{align}
where $C_{\au\bu}$ is the $Sp(4)$-invariant tensor; for spinor conventions, see Appendix \ref{App:conv}\footnote{%
Pairs of spinor indices are contracted from north-west to south-east as in Eq.\eqref{eq:NWSE}.}
The canonical commutation rules equip the space of arbitrary polynomials on ${\cal Y}_4\times{\cal Z}_4$ with an associative product, which defines the Weyl algebra on ${\cal Y}_4\times{\cal Z}_4$.
Higher spin master fields obeying non-trivial boundary conditions belong, however, to more general classes of symbols, which in general contain Dirac delta functions and their derivatives as well as regular functions. 
In these classes of symbols, the star product can be realized as a twisted convolution formula, 
\begin{equation}
\label{eq:NOprod}
(f\star g)(Y,Z)
=
\int_{\mathbb{R}^8}\frac{\diff^4U\diff^4V}{(2\pi)^4}
\,e^{iVU}\,f(Y+U,Z+U)\,g(Y+V,Z-V)
\,,
\end{equation}
where $f,g\in \Omega_{[0]}({\cal Y}_4\times {\cal Z}_4|{\cal R})$,  and the integration measure is \emph{chiral}, that is, $U_\au=(u_\a,\ub_\ad)$ and $V_\au=(v_\a,\vb_\ad)$ and $u_\a$ and $\bar u_\ad$ are integrated \emph{separately} over one copy of $\mathbb{R}^2$.
As we shall see, the chiral measure facilitates 
\begin{enumerate}[label=\roman*),ref=\roman*]
\item the definition of \emph{holomorphic} delta functions; and
\item Fourier transformation of phase space realizations of operators in \emph{Fock and anti-Fock spaces};
\end{enumerate}
which are crucial objects in the construction of ALAdS solutions with particle and black hole states.

\paragraph{Separation of variables versus Weyl order.} The star product \eqref{eq:NOprod} obeys
\begin{equation}
f(Y-Z)\star g(Y+Z)
=
f(Y-Z)\,g(Y+Z)
\,,
\end{equation}
that is, it provides the representation of an operator algebra in terms of symbols given in the normal order in which $Y-Z$ and $Y+Z$ are treated as creation and annihilation operators, respectively; for this reason, the above composition rule is referred to as the normal-ordered star product.
From $Y\star Z=Z\star Y$, it follows that if $f=f_1(Y)\star f_2(Z)$ and $g=g_1(Y)\star g_2(Z)$, then $f\star g= f_1\star g_1 \star f_2\star g_2$ where $f_1\star g_1$ and $f_2\star g_2$ are \emph{Groenewold--Moyal star products}.
Thus, the normal ordered star product is equivalent to the Groenewold--Moyal star product provided that the dependence of the master fields on the $Y$ and $Z$ variables can be \emph{separated}.

\paragraph{(Anti-)automorphisms.} The star product algebra admits the linear outer automorphisms\footnote{%
These automorphisms can be made inner by extending the star product algebra by outer Klein operators.}
\begin{align}
\pi\left(f(x,z,\zb;y,\yb;dx,dz,d\zb)\right)
&=
f(x,-z,\zb;-y,\yb;dx,-dz,d\zb)
\,,\\
\bar\pi\left(f(x,z,\zb;y,\yb;dx,dz,d\zb)\right)
&=
f(x,z,-\zb;y,-\yb;dx,dz,-d\zb)
\,,\end{align}
and the graded anti-automorphisms
\begin{align}
\tau\left(f(x,z,\zb;y,\yb;dx,dz,d\zb)\right)
&=
f(x,-iz,-i\zb;iy,i\yb;dx,-idz,-id\zb)
\,,\\
\left(f(x,z,\zb;y,\yb;dx,dz,d\zb)\right)^\dagger
&=
\left(f(x,\zb,z;\yb,y;dx,d\zb,dz)\right)^\ast
\,,
\end{align}
of which $\tau$ is linear and $\dagger$ is anti-linear.
Thus,
\begin{align}
\pi(f\star g)&=\pi(f)\star\pi(g)
\,,&
\bar\pi(f\star g)&=\bar\pi(f)\star\bar\pi(g)
\,,\\
\tau(f\star g)&=(-)^{\deg f\deg g}\tau(g)\star\tau(f)
\,,& 
(f\star g)^\dagger&=(-)^{\deg f\deg g}g^\dagger\star  f^\dagger
\,.
\end{align}
The associativity of the star product ensures that the graded bracket and twisted graded bracket defined by 
\begin{align}
\starcomm{f}{g}
:&= 
f\star g - (-)^{\deg f\deg g}g\star f
\,,\\
\picomm{f}{g}
:&=f\star g - (-)^{\deg f\deg g}g\star \pi(f)
\,,
\end{align}
obey graded Jacobi identities.

\paragraph{Twisted central closed two-form.} The star product algebra is assumed to contain the inner Klein operators
\begin{align}
\label{eq:k^2=1}
\k:&=e^{iyz}
\,,&
\kb:&=\k^\dagger=e^{-i\yb\zb}
\,,&
\k\star\k&=\kb\star\kb=1
\,,
\end{align}
such that 
\begin{align}
\label{eq:pifromk}
\pi(f_{[p;q,\bar q]})
&=
(-1)^{q}\k\star f_{[p;q,\bar q]}\star\k 
\,,&
\bar\pi(f_{[p;q,\bar q]})
&=
(-1)^{\bar q}\kb\star f_{[p;q,\bar q]}\star\kb
\,,
\end{align}
for horizontal forms $f_{[p;q,\bar q]}$ of degree $p$ on ${\cal X}_4$ and mixed holomorphic and anti-holomorphic degree $q$ and $\bar q$ on ${\cal Z}_4$, respectively.
It follows that the two-form
\be
\label{eq:firstdef J}
J:= -\frac{ib}4 dz^\a\wedge dz_\a \k- \frac{i\bb}4 d\zb^\ad\wedge d\zb_\ad \kb\,,\qquad 
b=e^{i\theta}\,,\qquad \bb=e^{-i\theta}
\,,
\ee
is de Rham closed, anti-hermitian and twisted central in the sense that for any horizontal form one has
\begin{equation}
\label{eq:pifromJ}
f\star J=J\star\pi(f)
\,.
\end{equation}
The inner Klein operators enjoy the following factorization property
\begin{align}
\label{eq:k fact}
\k&=\k_y\star\k_z
\,,&
\k_y:&=2\pi\delta^2(y)
\,,&
\k_z:&=2\pi\delta^2(z)
\,,&
\k_y\star\k_y&=
\k_z\star\k_z=1
\,,\\
\kb&=\kb_y\star\kb_z
\,,&
\kb_y:&=2\pi\delta^2(\yb)
\,,&
\kb_z:&=2\pi\delta^2(\zb)
\,,&
\kb_y\star\kb_y&=
\kb_z\star\kb_z=1
\,,
\end{align}
where the chiral delta functions are assumed to be real-analytic\footnote{%
In the sense that they preserve real-analyticity of the test function.}, i. e.  $\delta^2(My)=(\det M)^{-1} \delta^2(y)$ where $(My)^\alpha\equiv M^{\alpha\beta} y_\beta$, idem $\delta^2(z)$.
As a consequence, 
\be
\label{eq:def J}
J=\k_y\star j_z+\kb_y\star\jb_z
\,,\qquad
j_z:= -\frac{ib}4 dz^\a\wedge dz_\a \k_z
\,,\qquad
\jb_z:= -\frac{i\bb}4 d\zb^\ad\wedge d\zb_\ad \kb_z
\,
\ee
One also has
\begin{equation}
\label{eq:J^2}
J\star J=-\frac{1}{8}\k_y\star\kb_y
\star\left(\k_z\dz^\a\dz_\a\right)
\star\left(\kb_z\dzb^\ad\dzb_\ad\right)
\,.
\end{equation}

\paragraph{Master field equations.} The dynamical fields of the (duality unextended \footnote{%
In the FCS off-shell formulation, the field content is unified into a Quillen superconnection whose linearized field content consists of an equal number of even and odd forms (in isomorphic representations) \cite{Boulanger:2015kfa,Bonezzi:2016ttk}; this topological supersymmetry implies that the AKSZ partition function is finite at one-loop on manifolds with boundary.
}) Vasiliev system are two horizontal forms $\Phi$ and $A$ of degrees zero and one, respectively, obeying 
\begin{align}
\label{eq:VE}
\diffhat{A}+{A}\star{A}+{\Phi}\star{J}&=0
\,,&
\diffhat{\Phi}+\picomm{{A}}{{\Phi}}&=0
\,,
\end{align}
which are compatible with homotopy relations of $\diffhat$ and $\star$, thereby defining a Cartan integrable system. 
As a consequence, Cartan curvatures transform covariantly under 
\begin{align}
\label{eq:GIi}
\delta {A}
&=
\diffhat\epsilon + \starcomm{{A}}{{\epsilon}}
\,,&
\delta {\Phi} 
&= -\picomm{\epsilon}{{\Phi}}\,,
\end{align}
where $\epsilon$ are infinitesimal parameter defined locally on coordinate charts (hence not subject to any boundary conditions).

\paragraph{Bosonic models.} The equations of motion are compatible with the reality conditions
\begin{align}
\label{eq:RC}
\Phi^\dagger &= \pi(\Phi)
\,,&
A^\dagger &= -A
\,,
\end{align}
and linear projection conditions
\begin{align}
\label{eq:BP}
\pi\bar\pi(\Phi)\equiv\tau^2(\Phi)&=\Phi
\,,&
\pi\bar\pi(A)\equiv\tau^2(A)&=A
\,,
\end{align}
which define the bosonic model.
This model can be projected further by imposing
\begin{align}
\label{eq:MBP}
\tau(\Phi)&=\pi(\Phi)
\,,&
\tau(A)&=-A
\,,
\end{align}
which defines the minimal bosonic model.

\paragraph{Type A and B model.}
The parity operation $P$ is an automorphism of $\Omega_{\rm hor}({\cal U}|{\cal R})$ that acts non-trivially on its coefficient fields as well as the basis of ${\cal R}$.
The action on the latter is induced from
\be P(y^\alpha,\bar y^{\dot\alpha};z^\alpha,\bar z^{\dot\alpha})=(\bar y^{\dot\alpha},y^\alpha;-\zb^{\dot\alpha},-z^\alpha)\,;\ee
the action on the coefficient fields is then induced by constraining $P(\Phi)$ and $P(A)$.
There are two possibilities corresponding to taking the Lorentz singlet component of $\Phi$ to be either even or odd under $P$, viz.
\begin{align}
\mbox{Type A model (parity even scalar)}:&& P(\Phi,A,J)&=(\Phi,A,J)
\,,&
b&=1
\,,\\
\mbox{Type B model (parity odd scalar)}:&& P(\Phi,A,J)&=(-\Phi,A,-J)
\,,&
b&=i
\,.
\end{align}

\subsection{Unfolded Fronsdal fields and COMST} 

Upon decomposing 
\be A=A_{[1;0,0]}+A_{[0;1,0]}+A_{[0;0,1]}\,,\ee
and defining 
\begin{equation}
\label{eq:A=U+V}
U:=A_{[1;0,0]}= \dx^\mu U_\mu\,,\qquad V=A_{[0;1,0]}+A_{[0;0,1]}= \dZ^\au V_\au\,,
\end{equation}
the equations of motion split into 
\begin{align}
\label{eq:VE Phi(x,z)}
q\Phi+\picomm{V}{\Phi}&=0
\,,&
d\Phi+\picomm{U}{\Phi}&=0
\,,&&\\
\label{eq:VE UV}
qV+V\star V+\Phi\star J&=0
\,,&
qU+dV+\starcomm{U}{V}&=0
\,,&
dU+U\star U&=0
\,.
\end{align}
In the context of a perturbative expansion around AdS${}_4$, subject to suitable boundary and gauge conditions in twistor space that we shall exhibit in detail in the next section, the linearized initial data
\be 
\label{eq:phys proj sec 2}
W^{(1)}:=U^{(1)}|_{Z=0}\,,\qquad C^{(1)}:=\Phi^{(1)}|_{Z=0}\,,
\ee
turn out to obey 
\begin{align}
\label{eq:COMST}
\acd W^{(1)}&=
-\left.
\frac{ib}{4}e^{\a\ad}e_{\a}^{\phantom\a\ad}\partial^\yb_\ad \partial^\yb_\ad C^{(1)}\right\vert_{y=0}
-\left.
\frac{i\bb}{4}e^{\a\ad}e^{\a}_{\phantom\a\ad}\partial^y_\a \partial^y_\a C^{(1)}\right\vert_{\yb=0}
\,,\\
\label{eq:COMST C}
\tcd C^{(1)}&=0\,,
\end{align}
which decompose under $Sp(4)$ into unfolded equations of motion for a set of Fronsdal fields of spins $s\in \{0,1,2,3,\dots\}$ under the bosonic projection \eqref{eq:BP} and $s\in \{0,2,4,\dots\}$ under the minimal bosonic projection \eqref{eq:MBP}.
This result is known as the Central On Mass Shell Theorem (COMST).
In other words, the Vasiliev system can be subjected to boundary and gauge conditions in twistor space such that its linearization around anti-de Sitter spacetime describes the gluing of a Weyl zero-form $\Phi^{(1)}$ to an adjoint one-form $W^{(1)}$ via the two-form cocycle appearing in the constraint on the linearized curvature two-form $\acd W^{(1)}$ in \eqref{eq:COMST}.
A key feature of the linearization procedure is that the aforementioned cocycle assumes the canonical form as stated by the Central On Mass Shell Theorem (COMST) in a basis where the spin-$s$ Fronsdal field is identified as
\be
\label{eq:comp Fronsdal}
\phi_{\mu(s)}
:=\left.
e_{(\mu_1}{}^{\a_1\ad_1}\cdots
e_{\mu_{s-1}}{}^{\a_{s-1}\ad_{s-1}}
(\partial^y_\a)^{s-1}(\partial^\yb_\ad)^{s-1}W_{\mu_s)}
\right|_{Y=0}\,,\ee
where $e_\mu{}^{\a\ad}$ is the AdS${}_4$ vierbein, and the generalized spin-$s$ Weyl tensor as
\be 
\label{eq:comp Weyl}
C_{\a(2s)}:= \left.(\partial^y_\a)^{2s} C\right|_{Y=0}
\,,\qquad
C_{\ad(2s)}:= \left.(\partial^\yb_\ad)^{2s} C\right|_{Y=0}
\,.\ee

\subsection{Locally defined solution spaces }

\paragraph{Gauge functions and virtual configurations.} Vasiliev system can be integrated on a coordinate chart ${\cal U}_4\subset {\cal X}_4$ by applying a locally defined \emph{gauge function} 
\be 
\label{eq:def gF}
M:{\cal U}_4\times {\cal Z}_4\to {\cal G}({\cal Y}_4)\,,\ee
valued in a Cartan gauge group ${\cal G}({\cal Y}_4)$ to a \emph{particular solution} $(\Phi',V')$ to 
\be 
\label{eq:VE'}
q\Phi'+\picomm{V'}{\Phi'}=0
\,,\qquad 
qV'+V'\star V'+\Phi'\star J=0
\,,\qquad 
d\Phi'=0
\,,\qquad
dV'=0
\,,\ee
on ${\cal Z}_4$, referred to as a \emph{virtual twistor space configuration}, viz.
\begin{align}
\label{eq:gF}
U&=M^{-1}\star dM
\,,&
V&=M^{-1}\star qM+M^{-1}\star V'\star M
\,,&
\Phi&=M^{-1}\star\Phi'\star\pi(M)
\,.
\end{align}

\paragraph{Integration constants.}

The virtual configuration encodes
\begin{enumerate}[label=\roman*),ref=\roman*]
\item a \emph{zero-form integration constant}
\begin{equation}
C':={\cal P}\Phi'
\,,
\end{equation}
where ${\cal P}$ projects onto the $q$-cohomology in form degree zero.
This data can be given equivalently in terms of
\begin{align}
\Psi^\prime:&=C'\star\kappa_y
\,,&
qC^\prime=q\Psi^\prime=0
\,;
\label{eq:Psi' vs Phi'}
\end{align} 
\item a \emph{flat twistor space connection}\footnote{%
Examples of non-trivial flat twistor space connections are given in \cite{Iazeolla:2007wt}.}
\be \theta':= V'|_{\Psi'=0}\,,\ee
obeying
\be 
\label{eq:non trivial flat connection}
q\theta'+\theta'\star \theta'=0\,.\ee
\end{enumerate}
We refer to $(\Psi';\theta')$ as the \emph{integration constants} for the Vasiliev system on ${\cal U}_4\times {\cal Z}_4$.

\paragraph{Classical perturbative expansion.} We shall consider perturbative expansions
\be \Psi'=\sum_{n=1}^\infty \Psi^{\prime(n)}\,,\ee
inducing expansions
\be
\label{eq:perturbation'}
\Phi'=\sum_{n=1}^\infty\Phi{}'^{(n)}\,,\qquad V'=\sum_{n=0}^\infty V{}'^{(n)}\,,\qquad M=M^{(0)}\star \left(1+\sum_{n\geqslant 1}H^{(n)}\right)\,,
\ee
around 
\be \Phi{}'^{(0)}=0\,,\qquad V'{}^{(0)}=\theta'\,,\ee
where $(\Phi{}'^{(n)},V{}'^{(n)},H^{(n)})$ are $n$th order in $\Psi^{\prime(1)}$, and $M^{(0)}$ is a $\Psi'$-independent gauge function.
As we propose in Section \ref{sec:NL}, the perturbative expansions of $M$ and $\Psi'$ may be induced by boundary conditions on ${\cal C}$.

\paragraph{Holomorphic gauge and spin-frame on ${\cal Z}_4$.}
In ALAdS geometries, it is convenient to give the virtual twistor space configuration in the \emph{holomorphic gauge} \cite{Iazeolla:2017vng} 
\be
\label{eq:sol holom}
\Phi'=\Phi'(Y)=\Psi'\star \kappa_y\,,\qquad 
V'_\alpha=V'_\a(Y;z)=\sum_{n=1}^\infty \Psi'^{\star n}\star
v'_{\alpha,n}(z)\,,\ee
where we have assumed that $\theta'=0$, and $v':=dz^\alpha \sum_{n=1}^\infty \nu^n v'_{\alpha,n}$ is a particular solution to 
\be qv'+v'\star v'+\nu j_z=0\,,\label{eq:vprime}\ee
built from distributions on ${\cal Z}_4$ using a \emph{spin-frame} $u^\pm_\alpha$ (see Appendix \ref{App:conv}), that is, a holomorphic metric
\be ds^2_z:= {\cal D}_{\alpha\beta}dz^\alpha dz^\beta\,,\qquad {\cal D}_{\alpha\beta}:=2u^+_{(\alpha} u^-_{\beta)}\,,\qquad u^+ u^-=1\,.\ee

\subsection{Global formulation}\label{ssec:global}

\paragraph{Moduli spaces and Cartan gauge orbits.}

A classical solution space $\check{\cal M}$ consists of globally defined master field configurations on ${\cal C}$ obtained by gluing together chartwise defined master fields $(A,\Phi)$ using transition functions $T$ from a principal bundle $\check{\cal P}$ with structure group $\check{\cal G}\subseteq {\cal G}$. 
The space $\check{\cal M}$ is coordinatized by classical observables given by globally defined functionals
\be {\cal O}: (T;A,\Phi)\in \check{\cal M}\mapsto {\cal O}[T;A,\Phi]\in \Real\,.\ee
The configurations are constructed from classical moduli parameters given by gauge functions $M\in \Gamma({\cal M};{\cal X}_4\times {\cal Z}_4)$, where ${\cal M}:=[{\cal P}\times_{{\cal G}}{\cal G}]$ is associated\footnote{%
We recall that a principal $G$-bundle is a space $P$ on which $G$ acts freely and transitively from the right; its fibers are $G$-torsors and its projection map $\pi:P\to P/G$. 
Its local trivializations are equivariant isomorphisms $\phi_\xi:U_\xi\times G\to \pi^{-1}(U_\xi)$, viz. $\phi_\xi(p,g)=\phi_\xi(p,e)g$.
The gauge functions $M_\xi:U_\xi\to G$ are locally defined sections of the associated bundle $[P \times_G G ]$, where $[u,g]\sim [ug,e]$, viz. $M_\xi(p)=[\phi_\xi(p,M_\xi(p)),e]=[\phi_\xi(p,e),M_\xi(p)]$.
}
to a principal ${\cal G}$-bundle ${\cal P}$, and by virtual configurations $(T';A',\Phi')$ associated to $\check{\cal P}$.
Thus, $\check{\cal M}$ can be sliced into Cartan gauge orbits  
\begin{align}
(T^{(M)})_\xi^\eta&= (M_\xi)^{-1}\star (T')_\xi^{\eta}\star M_\eta\,,\\ 
(A^{(M)})_\xi&=(M_\xi)^{-1}\star(\diffhat  +A'_\xi)\star M_\xi
\,,\\  
(\Phi^{(M)})_\xi&=(M_\xi)^{-1}\star \Phi'_\xi\star\pi(M_\xi)
\,,
\end{align}
where $\xi$ and $\eta$ are chart indices, and $(T^{(M)};A^{(M)},\Phi^{(M)})$ obey boundary conditions, namely, $T^{(M)}$ must belong to $\check{\cal G}$ and the sections must fall off correctly in asymptotic regions of ${\cal C}$, which thus constrain the virtual data $(T';A',\Phi')$ and the gauge function $M$.
The Cartan gauge orbits decompose into equivalence classes
\begin{equation}
\label{eq:small gt}
[M_1]\sim[M_2]
\quad\Leftrightarrow\quad 
\forall{\cal O}\,,
\quad 
{\cal O}[T^{(M_1)};A^{(M_1)},\Phi^{(M_1)}]={\cal O}[T^{(M_2)};A^{(M_2)},\Phi^{(M_2)}]
\,.
\end{equation}
As usual, the gauge parameter $S:=M_2^{-1}\star M_1$ and the corresponding gauge transformation are said to be \emph{small}, or \emph{proper}, if $[M_1]\sim[M_2]$ in the above sense
and \emph{large}, or \emph{improper}, otherwise.
We shall refer to both types simply as gauge transformations, keeping in mind that the large ones comprise classical moduli.
Thus, the classical moduli space $\check{\cal M}$ has the structure of a locally fibered space, viz.
\be 
\label{eq:bundle Mcheck}
[\Gamma({\cal M};{\cal X}_4\times {\cal Z}_4)] \hookrightarrow \check{\cal M}\stackrel{\rm Proj}{\longrightarrow} {\cal I}\,,\ee 
where ${\cal I}$ consists of ${\cal G}$-equivalence classes of zero-form integration constants $(\Psi',v')$, that is, two pairs of integration constants are considered equivalent of they reside on the same Cartan gauge orbit.

The construction of classical field configurations thus requires the specification of spaces of virtual data and physical boundary conditions, to which we turn next.

\paragraph{Perturbative global formulation.}
In the context of a perturbative expansion \eqref{eq:perturbation'}, we take the master fields to belong to 
\be
\Gamma({\cal\check M};{\cal X}_4\times {\cal Z}_4)\equiv
{\cal E}:=\bigoplus_{n\geqslant 0}{\cal E}^{(n)}\,,\ee
which is a perturbatively defined differential graded associative algebra with a graded trace operation ${\rm STr}_{\cal E}$. 
Likewise, we assume that gauge functions and transition functions, respectively, belong to formally defined groups
\begin{align}
{\cal G}
:&=
\bigoplus_{n\geqslant 0}{\cal G}^{(n)}
\,,&
\check{\cal G}
:&=
\bigoplus_{n\geqslant 0}\check{\cal G}^{(n)}
\,,&
\check{\cal G}^{(n)}\subseteq{\cal G}^{(n)}
\,,
\end{align}
to which we associate corresponding connections that act on the sections in adjoint representations, that is, we assume that 
\be 
{\rm Ad}^\star_{{\cal G}}{\cal E}= {\cal E}
\,.\ee

\paragraph{Separation of twistor space variables and chiral traces.} Over a chart ${\cal U}_4\subset{\cal X}_4$, we shall assume that the dependence on ${\cal Y}_4$ and ${\cal Z}_4$ of master fields and gauge functions can be separated, viz.
\be 
f\mid_{{\cal U}_4\times {\cal Z}_4}=\sum_{\lambda,\lambda'} f_{\lambda,\lambda'}(x,dx)
\Delta_x^{\lambda}(Z,dZ)
\star\Theta_x^{\lambda'}(Y)
\,,
\label{eq:fact sum}
\ee
where, at a given point $x\in{\cal U}_4$, $\Theta_x^{\lambda'}$ span an associative algebra ${\cal A}({\cal Y}_4)$ of symbols on ${\cal Y}_4$ and $\Delta_x^{\lambda}(Z,dZ)$ span a differential graded associative algebra of forms on ${\cal Z}_4$; and where $f_{\lambda,\lambda'}$ are locally defined forms on ${\cal U}_4$.
Notice that the index can be discrete or continuous. In the latter case, Eq.\eqref{eq:fact sum} takes the form
\be 
\label{eq:fact int}
f\mid_{{\cal U}_4\times {\cal Z}_4}=\int d\lambda d\lambda'\, f_{\lambda,\lambda'}(x,dx)
\Delta_x^{\lambda}(Z,dZ)
\star\Theta_x^{\lambda'}(Y)
\,.
\ee
In particular, for bosonic symbols $f$ that are sections, the statement is that they belong to
\be {\cal E}({\cal U}_4):= \Omega({\cal U}_4)\otimes\frac12(1+\pi\bar\pi)\Big(\Omega({\cal Z}_4|J)\otimes {\cal A}({\cal Y}_4)\Big)\,,\label{calEU4}\ee
where
\begin{enumerate}[label=\roman*),ref=\roman*]
\item ${\cal A}({\cal Y}_4)$ is a star-product algebra of functions of $Y$ that is equipped with a (cyclic) trace operation
$\Tr_{{\cal A}({\cal Y}_4)}$ used to define classical observables\footnote{%
Asking for a finite trace is appropriate to construct solutions from initial data given in compact basis, such as the ones considered in Section \ref{sec:examples}. 
This condition is relaxed in the context of amplitude computations \cite{Colombo:2012jx,Bonezzi:2017vha}, where the initial datum for bulk-to-boundary propagators is given in non-compact basis, and where the resulting observable are expected to diverge at colliding points on the boundary.
}${}^{,}$\footnote{%
In Appendix \ref{App:Tr}, we give possible definitions of trace operations, including one that is relevant for field configurations obtained from massless particle and black hole states.}.
The Klein operators $\k_y$ and $\bar\k_{\yb}$ are also assumed to belong to ${\cal A}({\cal Y}_4)$.
\item $\Omega({\cal Z}_4|J)$ is the space of forms on ${\cal Z}_4$ given by
\be \Omega({\cal Z}_4|J)=\Omega(S^2|j_z)\otimes \Omega(\overline{S}^2|\bar j_{\bar z})\,,\qquad \Omega(\overline{S}^2|\bar j_{\bar z})=(\Omega(S^2|j_z))^\dagger\,,\ee
and\footnote{\label{ftn:chiral forms}The space $\Omega({S}^2)$, which consists of chiral forms that are bounded and integrable, forms a star product subalgebra of $\Omega({S}^2|j_z)$, and $\Omega_{[2]}({S}^2)\cap\left(\Omega_{[0]}({S}^2)\star j_z\right)\neq \emptyset$.
The element $\kappa_z$ is excluded from $\Omega_{[0]}({S}^2|j_z)$, as $\kappa_z\star j_z$ does not have a finite chiral trace; likewise $dz^\alpha \kappa_z$ is excised from $\Omega_{[1]}({S}^2|j_z)$.}
\be \Omega({S}^2|j_z):= \Omega({S}^2)\cup \Omega_{[0]}({S}^2)\star j_z\,,\qquad \Omega({S}^2)=L^1(S^2)\cap L^\infty(S^2)\,,\ee
equipped with the \emph{chiral} graded cyclic trace operation
\be \oint_{{\cal Z}_4}f(z,dz) \star \bar f(\bar z,d\bar z):= \oint_{S^2} f(z,dz)  \oint_{ \overline{S}^2} \bar f(\bar z,d\bar z)\,,\ee
i.e.,  we assume that ${\cal Z}_4\stackrel{\rm top}{\cong} S^2\times \overline{S}^2$ where $S^2$ and $\overline{S}^2$ are treated as two separate \emph{real} non-commutative two-spheres each given by the non-commutative $\Real^2$ with a (commuting) point added at infinity.
The relation to the usual trace $\int d^4Z$ associated with the Groenewold-Moyal product induced on $\Omega({\cal Z}_4)$ from \eqref{eq:NOprod} 
is given by the definition of integrals of forms.
In terms of the spin frame \eqref{eq:spin frame},
the real two-dimensional measure is
\begin{align}
\label{eq:d2z}
d^2z
&=
dz^+dz^-=\frac{1}{2}dz^\a dz_\a
\,,&
\int_{S^2}dz^\a dz_\a\k_z = 4\pi
\,.
\end{align}
\end{enumerate}
The full supertrace operation on ${\cal E}$ is then defined via the factorisation property \eqref{eq:fact sum} as
\begin{equation}
\STr_{\cal E} := \oint_{{\cal Z}_4}\Tr_{{\cal A}({\cal Y}_4)} \,.
\end{equation}

\paragraph{Fronsdal branch.} ${\cal E}$ is assumed to contain a branch ${\cal E}_{\rm Fr}$ whose sections have restrictions to submanifolds ${\cal X}_{4,\,{\rm Fr}}$ of ${\cal X}_4$ that admit interpretations in terms of Fronsdal fields defined as in (\ref{eq:comp Fronsdal},\,\ref{eq:comp Weyl})\footnote{%
In the original interpretation \cite{Vasiliev:1990vu,Vasiliev:1992av,Vasiliev:1999ba} of the theory as a deformation of Fronsdal theory, ${\cal X}_{4,\,{\rm Fr}}$ was taken to coincide with ${\cal X}_{4}$.
In Section \ref{ssec:AAdS}, however, we shall assume ${\cal X}_{4,\,{\rm Fr}}$ to be a neighborhood of the conformal boundary of AdS${}_4$.}, that is, the sections in ${\cal E}_{\rm Fr}$ are valued in a subspace ${\cal A}_{\rm Fr}({\cal Y}_4)$ of ${\cal A}({\cal Y}_4)$ that consists of non-polynomial functions on ${\cal Y}_4$ that are analytic at $Y=0$ and that need not form a star-product algebra\footnote{%
The space of real-analytic functions at $Y=0$ does not form a star product algebra.
For example, given a star-invertible Fourier transformable function $f$ (e.g. a generic gaussian in $Y$), one has  $f\star{\cal F}(f^{-1})=f\star(f^{-1}\star(\k_y\star\kb_\yb))=\kappa_y\kb_\yb$.}.
In the linearized analysis to be performed in Section \ref{sec:examples}, we shall work, however, with a subspace of ${\cal E}_{\rm Fr}$ whose elements restricted to ${\cal X}_{4,\,{\rm Fr}}$ take value in a star-product algebra, that is given by an operator algebra (of endomorphisms of an extended Fock space).


\subsection{Regular computational scheme.} 
\label{ssec:reg}


As shown in \cite{Sezgin:2005pv,Iazeolla:2007wt,Iazeolla:2008ix,Iazeolla:2011cb,Sundell:2016mxc,Iazeolla:2017vng,Aros:2017ror,Aros:2019pgj}, the virtual configurations corresponding to physically interesting solution spaces are non-polynomial in twistor space. In order to facilitate perturbative computations on-shell using virtual classical moduli, it is convenient to adopt the following rules\footnote{%
We would like to stress that (i) and (ii) do not imply that (iii) is automatically satisfied.  }:

\begin{enumerate}[label=\roman*),ref=\roman*]
\item \emph{Regular presentations:} When master fields and gauge functions are presented in factorized form with respect to $Y$ and $Z$ as in \eq{eq:fact sum}, the oscillator dependence of each factor is expanded over a set of elements closed under star product (see for example \cite{Iazeolla:2017vng}). For example, for the master fields of the solution spaces that have been studied in \cite{Iazeolla:2011cb,Iazeolla:2012nf,Sundell:2016mxc,Iazeolla:2017vng,Aros:2019pgj}, this step amounts to:
\begin{itemize}
    \item[$-$]  expanding their $Y$-dependence by means of a contour-integral presentation of elements $P_{m_1,m_2|n_1,n_2}$, $m_i,n_i\in \mathbb{Z}-1/2$, that correspond to endomorphisms $|m_1,m_2\rangle \langle n_1,n_2|$ of two-dimensional (anti-)Fock spaces. For such cases, ${\cal A}({\cal Y}_4)=End({\cal F}^{(+)}\oplus{\cal F}^{(-)})$. In section \ref{sec:examples} we shall review a subset of these solutions representing massless particle and higher-spin black-hole states: the regular presentation of their virtual configuration $\Phi'^{(1)}$ is given by \eq{eq:regHS} for particles, and by the same element star-multiplied by $\k_y$ for black holes;
    
    \item[$-$] expanding the $Z$-dependence of the $Z$-space connection in terms of open parametric integrals realizing deformed oscillators in $Z$-space, with deformation term proportional to $\k_z$  (as we review in Appendix \ref{App:fact param}). In particular, such integrals realize the deformed oscillators in terms of gaussian elements in $z$ (and $\zb$ for their complex conjugates) as in (\ref{eq:V_n sol},\,\ref{eq:f_n sol}) , and at first order in perturbation theory such integral realize a potential for a delta-function source, see Appendix \ref{App:altv1}.
    
\end{itemize}

The above two examples can be unified into a more general notation for regularly-presented master fields and gauge functions
\be 
\label{eq:gauss expansion}
{\cal T}[f](Y,Z,dZ):=\int_{S\in {\rm sym}_8(\Comp),\, T\in \Comp^8} d^{36}S d^8 T\,f(S,T;dZ)\,E_{S;T}\,,\ee
using chiral integration measures, where 
\be 
\label{eq:gauss basis}
E_{S;T}(\Xi):=\exp\left(\tfrac{i}2 \Xi^t S \Xi+ iT^t \Xi\right)\,, \qquad \Xi=\begin{pmatrix}Y\\Z\end{pmatrix}\,,
\ee
and $f(S,T;dZ)$ may contain additional parametric integrals. However, whether one can actually choose the kernel $f(S,T;dZ)$ 
so as to represent symbols taking value outside the algebra generated by the two above examples
is left for future studies. 
\item\label{it:regpres2 nested integrals}
\emph{Nested integration order:} The following operations involving the twistor variables $Y_\au$ and $Z_\au$ :
\begin{itemize}
\item[$-$] Derivatives
\begin{equation}
g(\partial_Y,\partial_Z){\cal T}[f](Y;Z,dZ):=\int d^{36}S d^8 T\,f(S,T;dZ)
g(\partial_Y,\partial_Z)E_{S;T}
\,;
\end{equation}
\item[$-$] Traces\footnote{%
The regularised trace discussed in App. \ref{App:Tr}
should be generalised in a way that is compatible with this prescription.}
\begin{equation}
\Tr_{{\cal A}({\cal Y}_4)} {\cal T}[f](Y;Z,dZ):=\int d^{36}S d^8 T\,\int_{{\cal Z}_4}f(S,T;dZ)
\Tr_{{\cal A}({\cal Y}_4)} E_{S;T}
\,,
\end{equation}
where the trace operation requires a factorization of $E_{S;T}$;
\item[$-$] Star products
\be{\cal T}[f]\star{\cal T}[g]:= \int  d^{36}S d^8 T \int d^{36}S' d^8 T'\,f(S,T;dZ) g(S',T';dZ) \,\left(E_{S;T}\star E_{S';T'}\right)\,,\ee	
where $E_{S;T}\star E_{S';T'}$ is computed using \eqref{eq:NOprod};
\item[$-$] Homotopy integrals (cf \eqref{eq:rhoV})
\begin{equation}
q^{(E+V)\ast}{\cal T}[f]
= 
\int_{S\in {\rm sym}_8(\Comp),\, T\in \Comp^8} d^{36}S d^8 T
\int_0^1 \frac{\diff t}{t}
\imath_{\vec E+\vec V} 
f(S,T;tdZ)
\left.E_{S;T}\right\vert_{Z\to tZ+(t-1)V}
\,,
\end{equation}
\end{itemize}
are performed prior to the auxiliary integrals.
\item\label{it:regpres2 ambigu}
\emph{Ambiguity-free nesting:} At each order of classical perturbation theory, the on-shell master fields must have unambiguous regular presentations, such as to generate a perturbatively defined differential graded associative algebra.  
\end{enumerate}
In the following, we shall employ the above calculational scheme in moving from the simple factorized regular presentations such as \eq{eq:regHS} and \eq{eq:V_n sol}, cast the resulting $Z$-space connection in normal order and then integrating the equations for the gauge fields subject to a gauge choice and to specific boundary conditions, thereby obtaining ``induced'' regular presentations for the spacetime one-form and for the gauge functions in normal ordering. 

We would like to make the following remarks:
\begin{enumerate}[label=\alph*),ref=\alph*]
\item Although rule \eqref{it:regpres2 nested integrals} requires to perform \emph{all} parametric integrals as the very last step, the result should stay unchanged if one first performs sub-integrals that do not interplay with the $Y_\au$ and $Z_\au$ variables.
This relaxed prescription allows one to define the $\circ$-product \eqref{eq:def circ} as in \cite{Prokushkin:1998bq} and to establish the algebra \eqref{eq:projalg} as in \cite{Iazeolla:2011cb,Iazeolla:2017vng}.
\item Rule \eqref{it:regpres2 ambigu} provides a non-trivial condition which in some cases may actually resolve apparent ambiguities in the choice of regular presentation; for an example, see App. E of \cite{Aros:2019pgj}.
\item The expansion is not unique, and in fact can always be rewritten in the form $f(S,T)=\delta^{36}(S)f(T)$, as the Gaussians can be Fourier transformed\footnote{%
$f(T)$ itself can contain distributions, which happens for example in the case where the support of $f(S,T)$ contains points $(S,T)$ where $S$ is not invertible. 
}.
This basis is also interesting because of the property
\begin{equation}
E_{0,T}
=
e^{i\varphi(T)}
E_{0,T_Y}
\star
E_{0,T_Z}
\,,\qquad 
\partial^Z_\au E_{0,T_Y}
=
\partial^Y_\au E_{0,T_Z}
=0
\,,
\end{equation}
where  $\varphi(T)$ can be computed using Eq.\eqref{eq:NOprod} which makes Eq.\eqref{eq:fact int} manifest.
\item A generalization of this kind may actually help avoiding the ambiguity in the expansion \eq{Wexp}, but a regularized trace adapted to elements as general as \eq{eq:gauss expansion} is yet to be defined.
\item The generic mode functions $E_{S,T}$ of the expansion \eqref{eq:gauss expansion} belong to the group algebra $\Comp SpH(8)$ where $SpH(8)$ is the semi-direct product of $Sp(8)$ with the Heisenberg group\footnote{%
$SpH(8)$ is itself a twist of $ISp(8;\Real)$ by the same cocyle as the one defining the Heisenberg algebra.}.
Indeed, one can show\footnote{%
The case of A squaring to a number was proven in \cite{Sundell:2016mxc} (see also \cite{Iazeolla:2008ix}). The general proof works in the same way, except that one writes a first order equation for $S(A)$ rather than postulating that it is proportional to $A$.}  that its generic element can be written
\begin{align}
\label{eq:gauss vs stargauss}
&
\exp_\star\left(\tfrac{i}{2}\Xi^t A\Xi+i\Lambda\Xi\right)
\\&=\nonumber
\frac{1}{\sqrt{\det \cosh(A)}}\exp\left(
\tfrac{i}{2}\Xi^tS(A)\Xi+iT(A,\Lambda)\Xi
-i\Lambda T(A,\Lambda)
-\tfrac{i}{2}T(A,\Lambda)AT(A,\Lambda)
\right)
\,;
\end{align}
in particular, if
\begin{equation}
\label{eq:group gauss(Y)}
A=\begin{pmatrix}A&0\\0&0\end{pmatrix}
\,,\qquad
\Lambda=\begin{pmatrix}0\\0\end{pmatrix}
\,,
\end{equation}
then $S(A)$ and $T(A,\Lambda)$ are given by
\begin{equation}
\label{eq:gaussian tanh}
S = 
\begin{pmatrix}
\tanh{A}&0\\0&0
\end{pmatrix}
= 
\begin{pmatrix}
\frac{1-e^{-2A}}{1+e^{-2A}}&0\\0&0
\end{pmatrix}
\,,\qquad
T=\begin{pmatrix}0\\0\end{pmatrix}
\,,
\end{equation}
which is the Cayley transform of the group element obtain from the generator $2iA$, thereby giving the identification \cite{Didenko:2003aa} between Weyl ordered gaussians and group elements.
\item The group algebra contains elements whose symbols in normal order are not Gaussians but that are nonetheless included in the integral representation \eqref{eq:gauss expansion} as boundary limits. An example of such elements are those of the form \eqref{eq:group gauss(Y)} for which $(1+e^{-2A})$ is not invertible. 
\item There is a set of matrices $S$ for which $E_{S,T}$ \eqref{eq:gauss basis} does not belong to the group algebra. 
For example, from \eqref{eq:gaussian tanh} it follows that this is the case of $S=\begin{pmatrix}B&0\\0&0\end{pmatrix}$ with $B$ outside the image of the $\tanh$ map.
It similarly happens for the $Z$-space counterpart of this example, as in the case of the elements \eqref{eq:param expand};
$\frac{i}{2}\frac{1-s}{1+s}{\cal D}$ is indeed outside the image of $\tanh$  for the limiting points $s=\pm1$ of the integration domain as well as for the central point $s=0$.
This parametric integral is thus expanded in the closure of the group algebra $\Comp Sp(8)$ rather than in the algebra itself.
\item The composition rule for $E_{S;T}\star E_{S';T'}$ obtained using \eqref{eq:NOprod} is the analytic continuation of the product in $SpH(8,\Comp)$ \cite{Didenko:2012tv}.
\item The higher-spin initial datum \eqref{eq:Phi' pt} is expanded as
\begin{align}
\Phi^{\prime(1)}_{\rm pt}
&=
{\cal T}\left[
\delta^{36}(S-S_\eta)
\delta^{8}(T+iX)
\right]
\,,&
\Xi S_\eta \Xi
&=
4i\eta E
\,,&
X\Xi
&= \chi y+\bar\chi\yb
\,.
\end{align}
This Weyl tensor are expanded within $SpH(8)$, 
as $S_\eta$ is in the image of $\tanh$ for $\eta$ encircling $\e$.
Notice however that the $\eta$ contour in Eq.\eqref{eq:regHS} is not the image of a closed contour in the group algebra,
as it unavoidably crosses a branch cut of the inverse $\tanh$ map.
Similarly, one possibility for the black hole modes \eqref{eq:Phi' bh} is to expand them over the Heisenberg group generators as
\begin{align}
\Phi^{\prime(1)}_{\rm bh}
=
{\cal T}\left[\frac{1}{2\pi}\delta^{36}(S)\delta^6\left(T(1-\Pi_\eta)-X(1-\Pi_h)\right)
\exp(-T\Pi_hX)
\right]
\,,
\\\nonumber
\Xi_1\Pi_\eta\Xi_2 = y_1(y_2-i\eta\sigma_0\yb_2)
\,,\qquad
\Xi_1\Pi_h\Xi_2 = y_1y_2
\,.
\end{align}
In both cases ($\Phi^{\prime(1)}_{\rm pt}$ and $\Phi^{\prime(1)}_{\rm bh}$), the expansion not only facilitates computations but also \emph{regularizes} the formally divergent star product ${\cal P}_1\star{\cal P}_{-1}$ \cite{Iazeolla:2011cb,Iazeolla:2017vng}.
\end{enumerate}

\section{Linearized solution spaces and unfolded Fronsdal fields}
\label{sec:COMST}

In this section, we linearize Vasiliev's equations around  anti-de Sitter spacetime and describe a linearized solution space that contains properly unfolded Fronsdal fields, as stated by the COMST.
We would like to stress that 
\begin{itemize}[label=---]
\item the linearized fields as well as the vacuum are configurations on ${\cal C}$ given by regular presentations that make sense in various ordering schemes;
\item the unfolded Fronsdal fields arise in an adjoint one-form $W^{(1)}$ and twisted adjoint zero-form $C^{(1)}$ that can be obtained from the master fields on ${\cal C}$ by localizing the latter to $Z=0$ in normal order; 
\item the COMST requires the twistor space connection on ${\cal C}$ to obey a \emph{relaxed} twistor space gauge condition that only determines the linearized gauge function up to $O(Z^2)$ in normal order.
\end{itemize}
In the next section, we shall map the linearized fields in the holomorphic gauge to the relaxed gauge in the cases of black hole and particle states, thereby corroborating the compatibility between the gauge function method and the COMST in these cases.
In Section \ref{sec:NL}, we shall extend these results to a proposal for a Fefferman--Graham-like perturbative construction of ALAdS solution spaces to Vasiliev's equations.

\subsection{Linearization around anti-de Sitter background}
\label{ssec:AdS}

\paragraph{Vacuum.}
The (proper) anti-de Sitter vacuum AdS${}_4$ of Vasiliev's equations is obtained by taking 
\be {\cal X}_4\stackrel{\rm top}{\cong} S^1\times S^3\,,\qquad {\cal X}'_4\stackrel{\rm top}{\cong} S^1\times (S^3\setminus\{N\})\stackrel{\rm top}{\cong}S^1\times\Real^3\,,\label{eq:topcalx4}\ee
where $N$ is a point on $S^3$, and choosing a vacuum gauge function
\be
\label{eq:AdS L}
L:{\cal X}'_4\times {\cal Z}_4\to SO(2,3)
\,,\qquad
qL=0
\,,
\ee
that is homotopic to\footnote{%
The section condition need only hold in a tubular neighbourhood of the boundary $S^1\times \{N\}$.} a section of $SO(1,3)\hookrightarrow SO(2,3)\to SO(2,3)/SO(1,3)\equiv AdS_4$.
The vacuum field configuration is given by
\be
\label{eq:AdS}
{U}^{(0)}=\Omega^{(0)}:=L^{-1}\star d L\,,\qquad
{\Phi}^{(0)}= {V}^{(0)}=0\,,
\ee
that is, in accordance with Eq. \eqref{eq:gF}, we have
\be
\label{eq:AdS'}
M^{(0)}=L
\,,\qquad{\Phi}^{\prime}= {V}^{\prime}=0\,.\ee
If $L$ is a section, then $\Omega^{(0)}=L^\ast \Theta$, the pull-back of the Maurer--Cartan form $\Theta$ on $SO(2,3)$ to ${\cal X}_4^\prime$, that is 
\begin{equation}
\label{eq:AdS U}
\Omega^{(0)}=\frac{1}{4i}\Omega^{(0)\underline{\alpha\beta}}Y_{\underline\alpha}Y_{\underline\beta}\equiv 
\frac{1}{4i}\left(
\omega^{(0)\a\b}y_\a y_\b
+\bar\omega^{(0)\ad\bd}\yb_\ad\yb_\bd 
+2e^{(0)\a\bd}y_\a\yb_\bd
\right)
\,,
\end{equation}
where $e^{(0)\a\bd}$ is a vierbein on AdS${}_4$ with compatible spin connection $(\omega^{(0)\a\b},\bar\omega^{(0)\ad\bd})$.

In what follows, we shall work with a vacuum gauge function corresponding to the stereographic coordinate for AdS${}_4$ \cite{Bolotin:1999fa}; for details, see Appendix \ref{App:conv}.

\paragraph{Linearized equations of motion}

In a perturbative expansion around the AdS${}_4$ vacuum, the linearized Vasiliev system reads
\begin{align}
\label{eq:qPhi1}
q {\Phi}^{(1)}
&= 0\,,\\
\label{eq:DPhi1}
\tcd {\Phi}^{(1)}
&= 0\,,\\
\label{eq:qV1}
q {V}^{(1)}+
 {\Phi}^{(1)}\star {J}
&=0\,,\\
\label{eq:qU1+DV1}
q {U}^{(1)}
+\acd {V}^{(1)}
&=0\,,\\
\label{eq:DU1}
\acd {U}^{(1)}
&=0\,,
\end{align}
which is a Cartan integrable set of curvature constraints.

The adjoint and twisted-adjoint background covariant derivatives of a master field $f$, i. e.  a horizontal differential form $f=f(x,Z,dx,dZ;Y)$ on ${\cal Y}_4\hookrightarrow {\cal X}_4\times {\cal Y}_4\times {\cal Z}_4\to {\cal X}_4\times {\cal Z}_4$, are defined by 
\begin{align}
\label{eq:D0 from L}
\acd{f}:&=
L^{-1}\star\diff\left(L\star f\star L^{-1}\right)\star L=
\diff{f}+\starcomm{ {U}^{(0)}}{{f}}\,,\\
\label{eq:D0t from L}
\tcd f:&=L^{-1}\star\diff\left(L\star f\star \pi(L^{-1})\right)\star \pi(L)=
\diff f+\picomm{ {U}^{(0)}}{f}
\,,
\end{align}
respectively.
It follows that 
\begin{align}
\label{eq:Dad as diffop}
\acd=&\diff+\Omega^{(0)\au\bu}Y_\au\partial^{(Y)}_\bu -i\Omega^{(0)\au\bu}\partial^{(Y)}_\au\partial^{(Z)}_\bu
\,,\\
\tcd=&
\diff
+\Omega_{(+)}^{(0)\au\bu}Y_\au\partial^Y_\bu-i\Omega_{(+)}^{(0)\au\bu}\partial^Y_\au\partial^Z_\bu
\nonumber\\&
-\frac{i}{2}\Omega_{(-)}^{(0)\au\bu}Y_\au Y_\bu-\Omega_{(-)}^{(0)\au\bu}Y_\au\partial^Z_\bu
+\frac{i}{2}\Omega_{(-)}^{(0)\au\bu}\partial^Y_\au\partial^Y_\bu+\frac{i}{2}\Omega_{(-)}^{(0)\au\bu}\partial^Z_\au\partial^Z_\bu
\,,
\end{align}
where
\begin{equation}
\Omega_{(\pm)}^{(0)\au\bu}Y_\au Y_\bu
:=
\frac{1}{2}(1\pm\pi)\Omega^{(0)\au\bu}Y_\au Y_\bu
\,.
\end{equation}

\paragraph{Symmetries.}
The background is left invariant under Cartan gauge transformations with rigid group elements $G^{(0)}$ obeying
\be D^{(0)} G^{(0)}=0\,,\qquad qG^{(0)}=0\,,\ee
that is,
\be G^{(0)}=L^{-1}\star G^{\prime(0)}\star L\,,\qquad dG^{\prime(0)}=0=G^{\prime(0)}\,,\ee
referred to as Cartan--Killing symmetries.

As for the linearized equations of motion (\ref{eq:qPhi1}--\ref{eq:DU1}),
they exhibit two types of symmetries.
Indeed, in addition to Cartan-Killing transformations
\be ({U}^{(1)},{V}^{(1)};\Phi^{(1)}\star\kappa_y)\to (G^{(0)})^{-1}\star ({U}^{(1)},{V}^{(1)};\Phi^{(1)}\star\kappa_y)\star G^{(0)}\,,\ee
they are also invariant under abelian linearized gauge transformations
\begin{equation} 
\label{eq:VE GIlin}
\delta_{\epsilon} {U}^{(1)}
= \acd\epsilon^{(1)}
\,,\qquad
\delta_{\epsilon} {V}^{(1)}
= q\epsilon^{(1)}
\,,\qquad
\delta_{\epsilon} {\Phi}^{(1)}
= 0 
\,,
\end{equation}
with unconstrained local parameters.
Thus, ${\Phi}^{(1)}$ is invariant under the abelian gauge transformations, while ${U}^{(1)}$ and ${V}^{(1)}$ decompose into sections and connections defined by the structure group.

\paragraph{Linearized solution spaces.}

Linearized solution spaces can be generated by applying finite gauge transformations with vacuum gauge function $L$ and (linearized) gauge function $H^{(1)}$ to a twisted-adjoint integration constant $C^{\prime(1)}$ for the Weyl zero-form $\Phi^{(1)}$, viz.\footnote{%
Note that, due to \eq{eq:qPhi1}, at first order $\Phi^{(1)}$ and $C^{(1)}$ defined in \eq{eq:phys proj sec 2} coincide. We shall therefore use these two notations interchangeably at first order.}
\begin{equation}
{U}^{(1)}=\acd{H}^{(1)}\,,\qquad  \Phi^{(1)}=C^{(1)}=L^{-1}\star C^{\prime(1)}\star \pi(L)\,,\qquad\end{equation}\begin{equation}  V^{(1)}=L^{-1}\star  V^{\prime(1)}\star L+q{H}^{(1)}\,,
\end{equation}
where $V^{\prime(1)}$ a particular solution to
\be  q V^{\prime(1)}+C^{\prime(1)}\star J=0\,,\qquad dV^{\prime(1)}=0\,.\ee 
The resulting classical solution space thus decomposes into Cartan gauge orbits which can be exhibited by defining
\be H^{(1)}=L^{-1}\star H^{\prime(1)}\star L\,,\qquad \epsilon^{(1)}=L^{-1}\star \epsilon^{\prime(1)}\star L\,,\ee
such that
\begin{equation}
{U}^{(1)}=L^{-1}\star dH^{\prime(1)}\star L\,,\qquad  \Phi^{(1)}=L^{-1}\star C^{\prime(1)}\star \pi(L)\,,\qquad\end{equation}\begin{equation}  V^{(1)}=L^{-1}\star(  V^{\prime(1)}+q{H}^{\prime(1)})\star L\,;
\end{equation}
the gauge orbits are then generated by 
\be
\delta_{\epsilon'} {H}^{\prime(1)}
= \epsilon^{\prime(1)}
\,,\qquad
\delta_{\epsilon'} {V}^{\prime(1)}
= 0
\,,\qquad
\delta_{\epsilon'} {C}^{\prime(1)}
= 0 \,.\ee
Thus, the solution space has the structure of proper gauge orbits over a linearized moduli space coordinatized by $C^{\prime(1)}$ and equivalence classes $[H^{\prime(1)}]$ defined such that $[H^{\prime(1)}]\sim [\widetilde H^{\prime(1)}]$ if $L^{-1}\star (\widetilde H^{\prime(1)}-H^{\prime(1)})\star L$ is a small gauge parameter.

\subsection{Twistor space decomposition and spacetime unfolded system}
\label{ssec:resolution}

\paragraph{Resolution operators and cohomology projectors.}
In order to embed unfolded Fronsdal fields into the gauge function, we decompose a form field $f$ given by a regular presentation in $\Omega({\cal U}_4)\otimes \Omega({\cal Z}_4|{\cal S})\otimes {\cal A}({\cal Y}_4)$, where ${\cal U}_4\subseteq {\cal X}_4$, as
\begin{equation}
\label{eq:sol qf=g}
f = q^* g   + q h + c \,,
\end{equation}
where
\begin{itemize}[label=---]
\item $g:=q f$ is the source of $f$;
\item $q^*$ is a resolution operator providing a particular co-source; 
\item $h$ is a gauge function (or form);
\item $c$ represents an element in the $q$-cohomology $H(q)\subset \Omega({\cal Z}_4|{\cal S})$ valued in $\Omega({\cal U}_4)\otimes {\cal A}({\cal Y}_4)$, that is,
\be c={\cal P}f\,,\qquad {\cal P}:\Omega({\cal Z}_4|{\cal S})\to H(q)\,,\qquad {\cal P}^2={\cal P}\,;\ee
\item the decomposition is compatible with the regular presentation, i.e.
\be f=q^\ast qf   + q h + {\cal P}f
\label{eq:sol qf=g detail}
\,,\ee
where $q^* q$ and ${\cal P}$ act on the Gaussian building blocks of $f$ \emph{prior to} performing the auxiliary integrals.
\end{itemize}
For ${\cal Z}_4\stackrel{\rm top}{\cong} S^2\times \overline{S}^2$, $H(q)$ is generated by $1$, $j_z$, $\bar\jmath_{\bar z}$ and $j_z\bar\jmath_{\bar z}$, and ${\cal P}f$ contains form fields on ${\cal U}_4$ in corresponding co-form degrees valued in ${\cal A}({\cal Y}_4)$. 

Thus, the projection of $(\Phi^{(1)},U^{(1)},V^{(1)})$ onto $H_{[0]}(q)$\footnote{%
Notice that $H_{[0]}(q)$ is the only cohomology that is relevant for the (duality unextended) Vasiliev system.
The cohomologies in degrees greater than one are activated in the Frobenius--Chern--Simons system \cite{Boulanger:2015kfa,Bonezzi:2016ttk} as well as in other extensions of the Vasiliev system \cite{Boulanger:2011dd,Vasiliev:2015mka} involving higher-degree forms.} 
yields a differential zero-form ${\cal P}\Phi^{(1)}$ and a spacetime one-form ${\cal P}U^{(1)}$ on ${\cal U}_4$;
while there is no cohomological part associated to $V^{(1)}\in\Omega^1({\cal Z}_4)$.

\paragraph{Twistor space gauges.}
Given a decomposition using $(q^{(A)\ast},{\cal P}^{(A)})$, we shall refer to the projection 
\begin{equation}
f^{(A)} := q^{(A)\ast} g +  c^{(A)}\equiv \left(q^{(A)\ast}q+{\cal P}^{(A)}\right)f\,,
\end{equation}
of $f$ obtained by setting $h^{(A)}$ to zero, as the twistor space $A$-gauge.
Two such gauges may be physically distinct, as the gauge function carries physical degrees of freedom (arising via boundaries or other topological defects). 

\paragraph{Projection to unfolded system on ${\cal X}_4$.} 

From Eq. \eqref{eq:qPhi1}, it follows that $\Phi^{(1)}$ is given by its cohomological part, that is,
\be\Phi^{(1)}={\cal P}\Phi^{(1)}\,,\ee
independently of the choice of ${\cal P}$, as the notation indicates.
From now on, we shall assume that $\Phi^{(1)}(x,Y)$ is analytic in $Y$ at $Y=0$, as is required for the interpretation \eqref{eq:comp Weyl} as Weyl tensor generating function\footnote{%
Strictly speaking, this requirement is necessary only at generic spacetime locations,
where the interpretation in terms of Fronsdal fields holds.
For example, this is not the case at the singular point of the black-hole-like solutions that are discussed in Section \ref{sec:examples}.
}.
Decomposing $V^{(1)}$ and $U^{(1)}$, respectively, using $(q^{(A)\ast},{\cal P}^{(A)})$ and $(q^{(B)\ast},{\cal P}^{(B)})$, it follows from Eqs. \eqref{eq:qV1} and \eqref{eq:qU1+DV1} that
\begin{align} {V}^{(1)}&=-q^{(A)\ast}(\Phi^{(1)}\star {J})+qh^{(1,A)}\,,\\
{U}^{(1)}&=q^{(B)\ast}\acd \left(q^{(A)\ast}(\Phi^{(1)}\star{J})-qh^{(1,A)}\right)+W^{(1,A,B)}\,,\end{align}
where the cohomological part 
\be W^{(1,B)}={\cal P}^{(B)}{U}^{(1)}\,,\ee
is a one-form field on ${\cal X}_4$ that does not depend on the $Z$ variables.
Thus, in the $A$-gauge, 
\begin{align}
\label{eq:VA}
{V}^{(1,A)}&=-q^{(A)\ast}(\Phi^{(1)}\star {J})
\,,\\
\label{eq:WAB}
{U}^{(1,A)}&=q^{(B)\ast}\acd q^{(A)\ast}(\Phi^{(1)}\star{J})+W^{(1,A,B)}
\,;
\end{align}
as the notation indicates, the choice of $(q^{(B)\ast},{\cal P}^{(B)})$ affects $W^{(1,A,B)}$ but not ${U}^{(1,A)}$.

The remaining linearized field equations, that is, Eqs. \eqref{eq:DPhi1} and \eqref{eq:DU1}, now read 
\begin{align}
\label{eq:COMST AB}
\acd W^{(1,A,B)}=&-(\acd q^{(B)\ast})(\acd q^{(A)\ast})(\Phi^{(1)}\star{J})\,,\\ D^{(0)}_{\rm tw} \Phi^{(1)}&=0
\,,
\end{align}
which constitute a free differential algebra on ${\cal X}_4$.
The Cartan integrability of the original system on ${\cal X}_4\times {\cal Z}_4$ implies that as the left-hand side of Eq. \eqref{eq:COMST AB} is $Z$-independent so is its right-hand side, whose normal and Weyl ordered forms are hence equal; for further details, see for example \cite{Didenko:2015cwv}. 

\paragraph{Resolution from homotopy contraction.}
\label{sec:homotopy}

A particular form of resolution operator on ${\cal V}_4\subset {\cal Z}_2$ can be obtained by choosing a vector field $\vec V$ on ${\cal V}_4$ such that every point on ${\cal V}_4$ is connected by a unique vector field flow to a base point $p_0$ (where thus $\vec V|_{p_0}=0$).
We then let
\be q^{(V)\ast}:= \int_{0}^1 \frac{dt}t t^{{\cal L}_{\vec V}} \imath_{\vec V}\,,\ee
denote the resolution operator given by homotopy contraction along $\vec V$.
For example, we may take $Z(p_0)=0$ and contract along the Euler vector field $\vec E:=Z^\au\vec\partial^Z_{\au}$. 
In particular, if $V^\au$ is $Z$-independent, then
\begin{equation}
\label{eq:rhoV}
q^{(E+V)\ast} g = \imath_{\vec E+\vec V} 
\int_0^1 \frac{\diff t}{t} g(x,tZ+(t-1)V;\dx,t\dZ;Y)
\,.
\end{equation}
The homotopy contractions square to zero. 
In trivial topology, the homotopy contraction $q^{(E+V)\ast}$ has the property that
\begin{equation}
{\cal P}^{(E+V)}:=1-qq^{(E+V)\ast}-q^{(E+V)\ast}q
\,,
\end{equation}
projects on the $q$ cohomology \cite{Didenko:2015cwv}.
This projector is the one appearing in Eq.\eqref{eq:sol qf=g detail},
as can be shown by acting on both sides of Eq.\eqref{eq:sol qf=g}.
Different linearized solution spaces can be constructed by using resolution schemes referring to different homotopy contracting vector fields in twistor space, as we shall exemplify next.
\paragraph{Standard homotopy contraction procedure.} Taking $q^{(A)\ast}=q^{(B)\ast}=q^{(E)\ast}$,  i. e.  resolving the $Z$-space equations for $V^{(1)}$ and $U^{(1)}$ using homotopy contraction along $\vec E$, yields 
\begin{equation}
\label{eq:COMST Vas}
\acd W^{(1,E,E)}=
-\frac{ib}{4}e^{\a\ad}e_{\a}^{\phantom\a\ad}\partial^\yb_\ad \partial^\yb_\ad \Phi^{(1)}(x;0,\yb)
-\frac{i\bb}{4}e^{\a\ad}e^{\a}_{\phantom\a\ad}\partial^y_\a \partial^y_\a \Phi^{(1)}(x;y,0)
\,,
\end{equation}
which provides a cocycle $\Sigma(e,e;\Phi^{(1)})$ that glues the twisted-adjoint zero-form module to the adjoint one-form module in a manifestly Lorentz covariant fashion in accordance with the COMST.

However, as we shall see in Sec. \ref{ssec:COMST from fact}, the above procedure amounts to imposing a gauge condition on the twistor space connection that can be relaxed without violating the COMST's requirements.

\paragraph{Weyl-ordered procedure.} Another possible choice is to resolve the $Z$-space equations for $V^{(1)}$ and $U^{(1)}$ using homotopy contraction along $\vec E+i\vec \partial_Y$ with $\vec \partial_Y:=\partial^{\au}_Y \vec\partial^Z_{\au}$,
\begin{equation}
\label{eq:rhoF}
q^{(E+i\partial_Y)\ast} g = \imath_{\vec E+i\vec \partial_Y} 
\int_0^1 \frac{\diff t}{t}\int\diff^4Y' g(x,tZ+i(t-1)\partial_Y;\dx,t\dZ;Y')
\delta^4\left(Y-Y'\right)
\,.
\end{equation}
This choice is equivalent to homotopy contracting in Weyl order using $q^{(E)\ast}$, viz.
\begin{equation}
q^{(E+i\partial_Y)\ast} = \hat\tau^{-1}q^{(E)\ast}\hat\tau 
\,,
\end{equation}
where 
\begin{equation}
\hat\tau f(Y,Z)
:=
\int\frac{\diff^4Y'\diff^4Z'}{(2\pi)^4}
\exp\left(-i(Y-Y')(Z-Z')\right)
f(Y',Z')
\,,
\end{equation}
maps symbols from normal to Weyl order, that is, if $f_N$ and $f_W$, respectively, are
the Weyl and normal ordered symbol of an operator, then $f_W=\hat\tau f_N$.

The advantage of homotopy contracting in Weyl order is the factorisation property
\begin{equation}
\label{eq:rhoF fact}
q^{(E+i\partial_Y )\ast}(f(Y)\star g(Z;\dZ))
=
f(Y)\star q^{(E)\ast} g(Z;\dZ)
\,,
\end{equation}
which, as shown in Appendix \ref{App:fact formal}, facilitates an explicit all order perturbative solution to the Vasiliev system provided that $q^{(E)\ast} j_z$ can be regularized\footnote{%
Applying $q^{(E)\ast}$ to $j_z$ yields the formal expression
$$q^{(E)\ast}j_z=\frac{ib}{2}\dz^\a\int_0^1\frac{dt}{t}z_\a\delta^2(z)\,,$$
which is a divergent integral multiplied by $0$ requiring regularization.}.
Following the regular scheme, we use the regular presentation \eqref{eq:kappaz param}, which yields 
\begin{align}
\label{eq:V1 unfact}
V^{(1,E+{i\partial_Y})}_\alpha
&=
-\frac{b}{2}
\frac{\partial}{\partial \rho^{\a}}
\int\frac{\diff^2u}{2\pi}
\Phi^{(1)}(u-z,\yb)e^{iy(z-u)}
\left.\int_{-1}^1
\frac{\diff s}{1+s}
\exp\left(
\tfrac{i}{2}\tfrac{1-s}{1+s}\, u{\cal D}u
+\tfrac{i}{1+s}\rho u
\right)
\right\vert_{\rho=0}
\,,
\\\label{eq:Ufact}
{U}^{(1,E+i\partial_Y)}&=W^{(1,E+{i\partial_Y},E+{i\partial_Y})}\,.
\end{align}
As reviewed in App. C, the all order completion of this solution gives the exact (particular) solution studied in \cite{Iazeolla:2011cb,Sundell:2016mxc,Aros:2017ror} (in the symmetric gauge).
\paragraph{$L$-rotated procedure.} An alternative scheme, which was used in \cite{Sezgin:2005pv}, consists of homotopy contracting in the primed gauge in normal order using $q^{(E)\ast}$ (prior to switching on the vacuum gauge function $L$).
This procedure is equivalent to using the resolution operator 
\begin{equation}
q^{(E+\widetilde{V})\ast}g
:=
L^{-1}\star q^{(E)\ast}\left(L\star g\star L^{-1}\right)\star L
\,,
\end{equation}
which defines the $Sp(4)$ spinor $\vec{\widetilde{V}}=\widetilde{V}^\au\partial^Z_\au$.
In particular, using the vacuum gauge function \eqref{eq:Lrot stereo} corresponding to stereographic coordinates, one has
\begin{align}
\tilde{v}_\a
:&=
2i\left(\partial^y_\a+\frac{1}{1-h}x_\a^{\phantom\a\ad}\partial^\yb_\ad\right)
\,,\qquad
\tilde{\bar{v}}_\ad
:=
2i\left(\partial^\yb_\ad+\frac{1}{1-h}\xb_\ad^{\phantom\ad\a}\partial^y_\a\right)
\,.
\end{align}

\subsection{Mapping between different resolution schemes}

Two linearized solution spaces obtained using decompositions\footnote{%
We assume that the $q$ cohomology is trivial on one-forms.} $(q^{(A)\ast};q^{(B)\ast},{\cal P}^{(B)})$ and $(q^{(A')\ast};q^{(B')\ast},{\cal P}^{(B')})$ are related by a gauge transformation with parameter ${H}^{(1,A\to A')}$, viz.
\begin{align}
\label{eq:HAB}
{V}^{(1,A')}&={V}^{(1,A)} + q {H}^{(1,A\to A')}
\,,\\
\label{eq:U1gt}
{U}^{(1,A')}
&=
{U}^{(1,A)}+\acd {H}^{(1,A\to A')}
\,,
\end{align}
and a map relating the initial data for the spacetime one-form induced by first replacing $q^{(B)\ast}$ by $q^{(B')\ast}$ in $U^{(1,A')}$, viz.
\begin{align}
{U}^{(1,A')}
&=
q^{(B)\ast}\acd q^{(A')\ast}(\Phi^{(1)}\star{J})+W^{(1,A',B)}\nonumber\\
&\equiv
q^{(B')\ast}\acd q^{(A')\ast}(\Phi^{(1)}\star{J})+W^{(1,A',B')}
\,,
\end{align}
and then performing the gauge transformation \eqref{eq:U1gt}, which yields the redefinition
\begin{equation}
\label{eq:wA'P}
W^{(1,A',B')}
= 
{\cal P}^{(B')}q^{(B)\ast}\acd q^{(A)\ast}(\Phi^{(1)}\star J)+{\cal P}^{(B')}\acd {H}^{(1,A\to A')}+W^{(1,A,B)}
\,.
\end{equation}

\paragraph{Twistor space gauge condition.}
If the $A'$-gauge is specified by a condition
\begin{equation}
{\cal O}_{A'}{V}^{(1,A')}=0
\,,
\end{equation}
rather than an explicit choice of $q^{(A')\ast}$, then the $A'$-gauge can be reached from the $A$-gauge by means of a gauge transformation with parameter $H^{(1,A\to A')}$ obeying
\begin{equation}
\label{eq:cdt HAA'}
{\cal O}_{A'}\left({V}^{(1,A)}+q{H}^{(1,A\to A')}\right)=0
\,,
\end{equation}
which fixes $H^{(1,A\to A')}$ up to a residual gauge parameter $h^{(1,A\to A')}\in{\rm ker}({\cal O}_{A'}q)$.

\paragraph{Preferred projection.}
We observe that a change of cohomology projector ${\cal P}^{(B)}$ induces a field redefinition of $W^{(1,A,B)}$ that,
if completely unconstrained, can be used to trivialise it.
Indeed, given a preferred integration constant $W^{(1)}$, one can always choose $(q^{(B')\ast},{\cal P}^{(B')})$ such that 
\be W^{(1,A,B')}=W^{(1)}\,,\label{eq:W1AB'}\ee
by rewriting Eq. \eqref{eq:WAB} as
\begin{equation}
{U}^{(1,A)}=q^{(B)\ast}\acd q^{(A)\ast}(\Phi^{(1)}\star{J})+W^{(1,A,B)}-W^{(1)}+W^{(1)}
\,,
\end{equation}
and defining
\be
q^{(B')\ast}g
:=
q^{(B)\ast}g+W^{(1,A,B)}-W^{(1)}
\,.
\ee
Hence it is necessary to choose a preferred projector that defines the gauge field,
the dynamics of which are then provided by the procedure of Section \ref{ssec:COMST from fact}.

\subsection{Relaxed twistor space gauge condition and COMST}
\label{ssec:COMST from fact}

In what follows, we shall give a family of \emph{relaxed} gauge conditions ${\cal O}_G$ and a projection ${\cal P}^{(G)}$ such that 
\begin{enumerate}[label=\alph*),ref=\alph*]
\item \label{it:O(Z^2)}
${\cal O}_G$ has an infinite-dimensional kernel that we shall employ in Sec. \ref{sec:NL} in imposing ALAdS boundary conditions;
\item\label{it:P^G}
$W^{(1,G,G)}$ is a generating functional for unfolded Fronsdal fields (embedded as in Eq.\eqref{eq:comp Fronsdal}) in accordance with the COMST;
\item\label{it:Ag analytic}
the relaxed gauge can be reached from any twistor space gauge $A$ in which $U^{(1,A)}$ and $V^{(1,A)}$ are real-analytic in $Z$ at $Z=0$\footnote{%
\label{ftn:VU reg}
In principle,
such a gauge $A$ can always be obtained from a preliminary gauge transformation, as the connection $U^{(1,A)}+V^{(1,A)}$ is sourced by $\Phi^{(1)}\star J$, itself regular in $Z$.
However, we do not discuss how to systematically perform this transformation.
}.
\end{enumerate}
The condition \eqref{it:P^G} amounts to using the following projector when solving the equation for $U^{(1)}$:
\be {\cal P}^{(G)}f:= f|_{Z=0}
\label{eq:proj Vgauge}
\,,\ee
where $Z$ is set to zero in normal order prior to performing all auxiliary integrals.
The existence of this projection itself requires $U^{(1,G)}$ to be analytic in $Z$ at $Z=0$.
The condition \eqref{it:P^G} moreover amounts to requiring $W^{(1,G,G)}\equiv 
{\cal P}^{(G)} U^{(1,G)}$ to be real analytic in ${\cal Y}_4$ at $Y=0$.
The gauge condition is taken to be the following relaxed version of the standard one\footnote{%
The standard procedure, that was recalled in Sec. \ref{sec:homotopy}, makes use of the stronger condition $H_2^{(1,G)}=0$.}:
\be 
\label{eq:ZV=O(Z2)}
{\cal O}_G V^{(1,G)}:= \imath_{\vec E} V^{(1,G)}- \vec E H_2^{(1,G)}=0\,,
\ee
where $H_2^{(1,G)}$ is an arbitrary symbol such that
\begin{equation}\label{eq:H2atZ=0}
{\cal P}^{(G)}\acd H_2^{(1,G)}=0
\,,
\end{equation}
i.e. any linearized gauge parameter that has no influence\footnote{%
Strictly speaking, it would be enough that it have no influence on the gauge curvature,
in which case the procedure of Section \ref{ssec:h1(Y)} would be modified by the introduction of the term ${\cal P}^{(G)}\acd H_2^{(1,(A,a)\to G)}$ in the definition of ${\cal O}^{(A,a)}$ in Eq.\eqref{eq:Wpart}.}
on the definition of the gauge connection $W^{(1,G,G)}$.
Since both $U^{(1,A)}$ and $U^{(1,G)}$ are assumed to be analytic in $Z$ at $Z=0$, this condition implies that $H_2^{(1,G)}$ is a $O(Z^2)$ function.
It follows that $H_2^{(1,G)}$ produces a deviation
of the solution satisfying the conditions (\ref{it:O(Z^2)}--\ref{it:Ag analytic}) from the standard one built using $q^{(E)\ast}$ \eqref{eq:rhoV}, viz.
\begin{align}
V^{(1,G)}&=-q^{(E)\ast}(\Phi^{(1)}\star J)+qH_2^{(1,G)}\,,\\
U^{(1,G)}&= q^{(E)\ast}\acd q^{(E)\ast}(\Phi^{(1)}\star J)+\acd H_2^{(1,G)}+W^{(1,G,G)}\,.
\end{align}
\paragraph{Reaching the relaxed gauge.}
Starting in a gauge $A$ satisfying condition \eqref{it:Ag analytic}, it follows that \eqref{eq:ZV=O(Z2)} is equivalent to 
\begin{align}
{H}^{(1,A\to G)}
&=
-\frac{1}{{\cal L}_{\vec E}}\imath_{\vec E} V^{(1,A)}
+h^{(1,A\to G)}
+H_2^{(1,A\to G)}
\nonumber\\&=
-\int_0^1\diff t\,Z^{\au} V^{(1,A)}_{\au}(x,tZ,Y) +h^{(1,A\to G)} +H_2^{(1,A\to G)}
\,,
\label{eq:H = int VL + h}
\end{align}
where $H_2^{(1,A\to G)}$ is a function satisfying Eq.\eqref{eq:H2atZ=0}
and where
$h^{(1,A\to G)}$ is homogeneous in degree zero in $Z$, i. e. $h^{(1,A\to G)}(x,Z;Y)$ depends on $Z$ only through ratios $Z^\au/Z^\bu$.
Because of the analyticity requirements, $h^{(1,A\to G)}$ cannot contain such ratios, that is
\be qh^{(1,A\to G)}=0\,.\ee
\paragraph{Generating function for unfolded Fronsdal fields.}
The generating function for the spacetime gauge fields is given by
\begin{align}
W^{(1,G,G)}
=& 
{\cal P}^{(G)}U^{(1,A)}
+
{\cal P}^{(G)}\acd H^{(1,A\to G)}
\nonumber\\=&
{\cal P}^{(G)}U^{(1,A)}
+i\Omega^{\bu\bu}\partial^Y_\bu{\cal P}^{(G)}\partial^Z_\bu \int_0^1\diff t\,Z^\au V^{(1,A)}_{\au}(x,tZ,Y)+\acd h^{(1,A\to G)}
\nonumber\\=&
\label{eq:def omega_phys}
{\cal P}^{(G)}U^{(1,A)}
+
i\Omega^{\au\bu}\partial^Y_\bu{\cal P}^{(G)} V^{(1,A)}_{\au}
+\acd h^{(1,A\to G)}
\,,
\end{align}
whose real-analyticity in ${\cal Y}_4$ at $Y=0$ fixes $h^{(1)}$ modulo its real-analytic part, as we shall analyze in more detail in Section \ref{ssec:h1(Y)}.
Formally, it follows that 
\begin{align}
\acd W^{(1,G,G)}
&=
\acd {\cal P}^{(G)}U^{(1,A)}
+
i\left(\left\{\Omega^{\au\bu}\partial^Y_\bu,\acd\right\}-\Omega^{\au\bu}\partial^Y_\bu\acd\right){\cal P}^{(G)} V^{(1,A)}_{\au}
\nonumber\\&=
\acd{\cal P}^{(G)}U^{(1,A)}
-i\Omega^{\au\bu}\partial^Y_\bu\acd{\cal P}^{(G)}V^{(1,A)}_{\au}
\nonumber\\&=
i\Omega^{\au\bu}\partial^Y_\au{\cal P}^{(G)}\left(\partial^Z_\bu U^{(1,A)}-\acd V^{(1,A)}_\bu\right)
+\Omega^{\au\bu}\partial^Y_\bu\Omega^{\cu\bu}\partial^Y_\bu{\cal P}^{(G)}\partial^Z_{[\cu} V^{(1,A)}_{\au]}
\label{eq:COMST eom step}\\
&=
-\Omega^{\au\bu}\partial^Y_\bu\Omega^{\cu\bu}\partial^Y_\bu {\cal P}^{(G)}(\Phi^{(1)}\star J)_{\cu\au}
\nonumber\\&=
-\frac{ib}{4}\Omega^{\a\bu}\partial^Y_\bu\Omega^{\c\bu}\partial^Y_\bu{\cal P}^{(G)}\left(\epsilon_{\c\a}\Phi^{(1)}(x;-z,\yb)e^{iyz}\right)
-\hc
\nonumber\\&=
\label{eq:COMST H1}
-\frac{ib}{4}e^{\a\ad}e_{\a}^{\phantom\a\ad}\partial^\yb_\ad \partial^\yb_\ad \Phi^{(1)}(x;0,\yb)
-\frac{i\bb}{4}e^{\a\ad}e^{\a}_{\phantom\a\ad}\partial^y_\a \partial^y_\a \Phi^{(1)}(x;y,0)
\,.
\end{align}
in agreement with the COMST (see App. D of \cite{Aros:2017ror}) albeit under the assumption that
\begin{itemize}
\item[i)] derivatives with respect to twistor variables commute, used to prove \eqref{eq:COMST eom step}.
\end{itemize}
This is not guaranteed for singular functions of the twistor variable,
as for example the factorised twistor space connection \eqref{eq:V1 unfact}.
The regular prescription ensures that property.
However, according to this prescription, the whole chain of derivatives should be taken prior to performing the homotopy integrals,
whereas the latter operation is in principle necessary in order to use the field equations after \eqref{eq:COMST eom step}.
Hence, we separately require the following assumption:
\begin{itemize}
\item[ii)] equations of motion hold inside a chain of twistor space derivatives.
\end{itemize}
In Sec. \ref{sec:examples}, we shall verify that the latter assumption holds within the regular scheme
in the case of massless particle and black hole solutions that have been studied in \cite{Didenko:2009td,Iazeolla:2011cb,Iazeolla:2012nf,Sundell:2016mxc,Iazeolla:2017vng}.

Because of Eq.\eqref{eq:COMST H1}, $W^{(1,G,G)}$ is in the same cohomology class as the generating function for unfolded fields.
Since the linearised Weyl zero-form $\Phi^{(1)}$ is real analytic in $Y$ at $Y=0$, it follows that the singular part of $W^{(1,G,G)}$, if it exists, is pure gauge.
Notice that, in the case where the starting point is a gauge $(\widetilde{L})$ such that $U^{(1,\widetilde{L})}=0$ and $V^{(1,\widetilde{L})}$ is real analytic in both $Y$ and $Z$ at $Y=Z=0$, then
\begin{equation}\label{3.70}
W^{(1,G,G)}=
i\Omega^{\au\bu}\partial^Y_\bu{\cal P}^{(G)} V^{(1,\widetilde{L})}_{\au}
\,,
\end{equation}
provides a compact generating function of Fronsdal fields.

\subsection{Real analyticity and refined gauge fixing}
\label{ssec:h1(Y)}
According to the COMST and to Eq.\eqref{eq:COMST H1},
if one can extract $\phi^{(1)}_{\mu(s)}$ from $W^{(1,G,G)}$ as in Eq.\eqref{eq:comp Fronsdal},
then they are actually Fronsdal fields on the mass shell.
This requires that $W^{(1,G,G)}$ be smooth at $Y=0$.
We claim that $h^{(1,A\to G)}$ can be used to reach a gauge where this condition is satisfied.
The solution will be given as in Eq.\eqref{eq:WGg}, by the sum of a completely gauge fixed part and of a residual gauge parameter that preserves both the analyticity in $Y$ and the gauge condition \eqref{eq:ZV=O(Z2)}.
\paragraph{Residual gauge parameters.}
Because of Eq.\eqref{eq:COMST AB} (resp. \eqref{eq:COMST H1}), $W^{(1,A,B)}$ (resp. $W^{(1,G,G)}$) is defined up to a pure gauge piece, viz.
\begin{equation}
\label{eq:WGg}
W^{(1,A,B)} = W^{(1,(A,a),B)}+\acd h^{(1,(A,a))}
\,,\qquad
W^{(1,G,G)} = W^{(1,(G,g),G)}+\acd h^{(1,(G,g))}
\,,
\end{equation}
where $(A,a)$ (resp. $(G,g)$) is a refined gauge fixing that includes the $A$-gauge condition (resp. the $G$-gauge condition \eqref{eq:ZV=O(Z2)})
and where $h^{(1,(A,a))}$ (resp. $h^{(1,(G,g))}$) is a $Z$-independent residual gauge parameter that does not break this gauge condition.
$W^{(1,(G,g),G)}$ is given by the following rewriting of Eq.\eqref{eq:def omega_phys}:
\begin{align}
W^{(1,(G,g),G)}
&=
{\cal P}^{(G)}U^{(1,(A,a))}
+
i\Omega^{\au\bu}\partial^Y_\bu{\cal P}^{(G)} V^{(1,(A,a))}_{\au}
+\acd h^{(1,(A,a)\to(G,g))}
\nonumber\\&=:
\label{eq:Wpart}
\Omega^{\au\bu}{\cal O}^{(A,a)}_{\au\bu}\Phi^{(1)}
+\acd h^{(1,(A,a)\to(G,g))}
\,.
\end{align}
where ${\cal O}^{(A,a)}_{\au\bu}$ is a field-independent operator defined by the latter equality.
For example, if one refines the initial $A$-gauge given by the factorised homotopy contraction \eqref{eq:rhoF} by imposing the additional condition 
\begin{equation}
\label{eq:W(1F0)=0}
W^{(1,(E+i\partial_Y,0),E+i\partial_Y)}=0
\,,
\end{equation}
then the resulting operator ${\cal O}^{(1,(E+i\partial_Y,0))}_{\au\bu}$ is given by
\begin{align}
&
{\cal O}^{(1,(E+i\partial_Y,0))}_{\au\bu}F(x;y,\yb)
\\&:=\nonumber
\frac{b}{2}
\int\frac{\diff^2w}{2\pi} 
\int_{-1}^{-1}\frac{\diff s}{(1+s)^2}
\exp\left(\tfrac{i}{2}\tfrac{1-s}{1+s}w{\cal D}w+iwy\right)
\begin{pmatrix}
-iw_\a w_\b&\tfrac{1}{2}w_\a\partial^\yb_\bd\\
\tfrac{1}{2}w_\b\partial^\yb_\ad&0
\end{pmatrix}
F(x;w,\yb)
+\hc
\,.
\end{align}
Since only real analytic residual transformations ($h^{(1)}$ valued in the higher-spin algebra $\mhs(2,3)$) are allowed in the end,
what we need is to find a particular relaxed Vasiliev gauge $(G,g)$ such that $W^{(1,(G,g),G)}$ is smooth in $Y$.
According to the choice of the $(A,a)$-gauge, from where one starts the procedure of Sec. \ref{ssec:COMST from fact}, $h^{(1,(A,a)\to(G,g))}=0$ may already be solution to that problem.
In the remaining part of the section, we discuss two different techniques to obtain a particular solution when it is not the case.

\paragraph{Cartan integration.}
One way to find $h^{(1)}:=h^{(1,(A,a)\to(G,g))}$ yielding a regular $W^{(1,(G,g),G)}$ is to embed \eqref{eq:Wpart} into the following system:
\begin{align}
\label{eq:COMST CI1}
\diff \Omega^{\au\bu}+\Omega^\au_{\phantom\au\cu}\Omega^{\cu\bu}
&=0
\,,\\\label{eq:COMST CI2}
\left(\diff+\omega^{\a\b}y_\a\partial^y_\b
+\bar\omega^{\ad\bd}\yb_\ad\partial^\yb_\bd
-ie^{\a\bd}(y_a\yb_\bd-\partial^y_\a\partial^\yb_\bd)
\right)
\Phi^{(1)}(y,\yb)&=0
\,,\\\label{eq:COMST CI3}
\left(\diff+\Omega^{\au\bu}Y_\au\partial^Y_\bu\right)W^{(1,(G,g),G)}
+\frac{ib}{4}
e^{\a\ad}e_{\a}^{\phantom\a\ad}\partial^\yb_\ad \partial^\yb_\ad \Phi^{(1)}(x;0,\yb)
-\hc
&=0
\,,\\\label{eq:COMST CI4}
\left(\diff+\Omega^{\au\bu}Y_\au\partial^Y_\bu\right)h^{(1)}(y,\yb)
-W^{(1,(G,g),G)}(y,\yb)
+\Omega^{\au\bu}({\cal O}^{(A,a)}_{\au\bu}\Phi^{(1)})(y,\yb)
&=0
\,,
\end{align}
If $\Phi^{(1)}(y,\yb)$, $W^{(1,(G,g),G)}(y,\yb)$ and $h^{(1)}(y,\yb)$ are seen as two infinite families of fields labelled by continous indices $Y_\au$, and if the Weyl zero-form $\Phi^{(1)}(y,\yb)$ is assumed to fall off at the boundary of $Y$ space, then this system is formally integrable and one can use Cartan's integration formula to solve it.
The only field allowed to live outside the space of regular functions is $h^{(1)}$, and the only operator that can make a field leave this class is ${\cal O}^{(A,a)}_{\au\bu}$.
Hence the subsystem (\ref{eq:COMST CI1},\,\ref{eq:COMST CI2},\,\ref{eq:COMST CI3}) will provide us with a regular solution for $W^{(1,(G,g),G)}$, while equation (\ref{eq:COMST CI4}) will give a (a priori not regular) solution for the gauge function $h^{(1,(A,a)\to(G,g))}$.
We leave the resolution of the above system for future work.

\paragraph{Homotopy integration.}
If $\Phi^{(1)}(x,Y)$ is analytic in $Y$, a solution for $W^{(1,(G,g),G)}$ is given by the action on the right hand side of \eqref{eq:COMST H1} of a resolution $\left(\acd\right)^{(g)\ast}$ for $\acd$ that squares to zero and preserves the  analyticity in $Y$, i. e. 
\begin{equation}
\label{eq:omega D*}
W^{(1,(G,g),G)}=
-\frac{ib}{4}\left(\acd\right)^{(g)\ast}\left(
e^{\a\ad}e_{\a}^{\phantom\a\ad}\partial^\yb_\ad \partial^\yb_\ad \Phi^{(1)}(x;0,\yb)
+\bb e^{\a\ad}e^{\a}_{\phantom\a\ad}\partial^y_\a \partial^y_\a \Phi^{(1)}(x;y,0)
\right)
\,.
\end{equation}
$h^{(1,A\to G)}$ can then be retrieved as a solution to \eqref{eq:Wpart}:
\begin{equation}
h^{(1,A\to G)}
=
-\left(\acd\right)^\ast\left(
\Omega^{\au\bu}{\cal O}^{(A,a)}_{\au\bu}\Phi^{(1)}
\right)\,.
\end{equation}
Given a resolution operator $\diff^{(g)\ast}$ for $\diff$ satisfying the aforementioned requirements,
a resolution operator $\left(\acd\right)^{(g)\ast}$ for $\acd$ is provided by equation \eqref{eq:D0 from L}:
\begin{equation}
\left(\acd\right)^{(g)\ast} f := L^{-1}\star\diff^{(g)\ast}\left(L\star f\star L^{-1}\right)\star L\,,
\end{equation}
which preserves the analyticity in $Y$ for $Z$-independent symbols, as it is equivalent to preserving the one in $Y^L$, that was defined in \eqref{eq:Lrot gen app}.
For example, the choice of homotopy
\begin{equation}
d^{(FS)\ast} g(x,Z;\dx,\dZ;Y) 
:= x^\mu\frac{\partial}{\partial \dx^\mu} 
\int_0^1 \frac{\diff t}{t} g(tx,Z;t\dx,\dZ;Y)
\,,
\end{equation}
yields
\begin{equation}
\label{eq:D0* FS}
\left(\acd\right)^{(FS)\ast} g(x,Z;\dx,\dZ;Y^L) 
:= x^\mu\frac{\partial}{\partial \dx^\mu} 
\int_0^1 \frac{\diff t}{t} g(tx,Z;t\dx,\dZ;Y^L)
\,.
\end{equation}
The particular spacetime connection
\begin{equation}
\label{eq:omega FS}
W^{(1,(G,FS),G)}=
-\frac{i}{4}\left(\acd\right)^{(FS)\ast}\left(
b\,e^{\a\ad}e_{\a}^{\phantom\a\ad}\partial^\yb_\ad \partial^\yb_\ad \Phi^{(1)}(x;0,\yb)
+\bb\,e^{\a\ad}e^{\a}_{\phantom\a\ad}\partial^y_\a \partial^y_\a \Phi^{(1)}(x;y,0)
\right)
\end{equation}
is hence regular in $Y$.
This gauge is the Fock-Schwinger gauge, characterised by
\begin{equation}
x^\mu\frac{\partial}{\partial \dx^\mu}W_{\rm part.}^{(1,FS)}=0
\,,
\end{equation}
as can be seen from
\begin{equation}
x^\mu\frac{\partial}{\partial \dx^\mu}\left(\acd\right)^{(FS)\ast}
=0
\,.
\end{equation}
The $Z$-independent part of the gauge parameter used to reach this gauge from $(A,a)$-gauge is
\begin{equation}
\label{eq:h1 FS}
h^{(1,(A,a)\to (G,FS))}(x;Y)
=
\int_0^1 \diff t\,
x^\mu\Omega_\mu^{\phantom\mu\au\bu}(tx)
\left[-
{\cal O}^{(A,a)}_{\au\bu}\Phi^{(1)}
\right]\left(tx;L(x)_\au^{\phantom\au\bu} L^{-1}(tx)_\bu^{\phantom\bu\cu}Y_\cu\right)\,.
\end{equation}

\section{Fronsdal fields carrying particle and black hole states}
\label{sec:examples}

We shall now apply the procedure described in Section \ref{sec:COMST} to massless particle \cite{Iazeolla:2017vng} and black hole \cite{Didenko:2009td,Iazeolla:2011cb} modes.
In Sections \ref{ssec:BH PT} and \ref{ssec:Weyl}, we review the construction of their Weyl tensors starting from initial data valued in an associative operator algebra ${\cal A}^{\prime(1)}({\cal Y}_4)\subset{\cal A}^{(1)}({\cal Y}_4)$.
In Sections \ref{ssec:PTBH fact} to \ref{ssec:DV}, we shall map those Weyl zero-forms to linearised master fields in the relaxed Vasiliev gauge, 
in particular extracting gauge field generating functions that properly encode Fronsdal fields in accordance with the COMST.
The starting point will be either initial data for both particles and black holes in factorised gauge (\ref{eq:V1 unfact},\,\ref{eq:Ufact}), or alternatively black hole states in Didenko-Vasiliev gauge \cite{Didenko:2009td}.


\subsection{Algebra of zero-form integration constants} \label{ssec:BH PT}


In what follows, we describe the construction of linearized particle and black hole initial data as elements of the associative algebra ${\cal A}^{\prime(1)}({\cal Y}_4)$ of (generally complex) endomorphisms of the extended supersingleton Fock space.
More precisely, we consider an initial datum $\Phi^{\prime(1)}$ in ${\cal A}^{\prime(1)}({\cal Y}_4)$ given by an expansion of the form \eqref{Wexp} where the coefficient $f_{\ell,\bar\ell}$ are endomorphisms of the (anti-)supersingleton Fock space obeying additional reality conditions.
The algebra ${\cal A}^{\prime(1)}({\cal Y}_4)$ itself is left invariant under the transformation \eqref{eq:PTvsBH} that exchanges particle and black hole states \cite{Iazeolla:2017vng} (which is reminescent of a Tannaka-Krein duality transformation).
We then provide regular presentations for such elements, that will be useful in constructing the aforementioned map to properly unfolded Fronsdal fields.

\paragraph{Scalar and spinor singleton endomorphisms.}
The massless particle and black hole states of the bosonic model are unified into the even subalgebra 
\be {\cal A}^{\prime(1)}({\cal Y}_4):=\frac12(1+\pi\bar{\pi}){\rm End}({\cal F})\ee
of the algebra of endomorphisms of the extended Fock space 
\be {\cal F}=\bigoplus_{\sigma=\pm} {\cal F}^{(\sigma)}\,,\ee
where ${\cal F}^{(+)}$ and ${\cal F}^{(-)}$, respectively, are the Fock and anti-Fock spaces of two sets of harmonic oscillators given by linear combinations 
\be y^\pm_i:=y_i^{\pm\underline{\alpha}}Y_{\underline \alpha}
\,,\qquad
\left[y_i^{\epsilon},\,y_j^{\epsilon'}\right]_\star =
\delta_{ij}\epsilon^{\epsilon\epsilon'}
\,,\qquad  y_i^{-}=(y^+_i)^\dagger
\,,\qquad i=1,2\,,
\ee
where $\epsilon^{-+}=-\epsilon^{+-}=1$\,. The $y_i^\pm$  can be extracted from the $Y$ oscillators \cite{Iazeolla:2011cb} by means of the spin-frame $(u^\pm_\a,\ub^\pm_{\ad})$ as
\begin{align}
y_1^+&=\frac12\left(y^++i\yb^-\right)
\,,\qquad
y_2^+=\frac12\left(-y^-+i\yb^+\right)
\,,\label{eq:ypm1}\\
y_1^-&=\frac12\left(\yb^+-iy^-\right)
\,,\qquad
y_2^-=\frac12\left(-\yb^--iy^+\right)
\,.\label{eq:ypm2}
\end{align}
Thus, the Weyl-ordered number operators 
\be w_i:=\frac12(y^+_i\star y^-_i+y^-_i\star y^+_i)
\,,\qquad 
\left[w_i,y_j^\epsilon\right]_\star =\epsilon\delta_{ij}
\,,
\ee
span a compact Cartan subalgebra of $\mso(2,3)$ (see \eq{eq:w(E,J)}), 
while $w_i$ is the compact Cartan generator of the $\mathfrak{osp}(1|2)_i$ algebra generated by $(y^+_i,y^-_i)$; accordingly, 
\be {\cal F}^{(\pm)}\downarrow_{\mathfrak{osp}(1|2)}\cong S^{(\pm)}(\pm1/4)_1 \otimes S^{(\pm)}(\pm1/4)_2\,,\ee
where $S^{(+)}(1/4)$ and $S^{(-)}(-1/4)$, respectively, are lowest and highest weight superquartions (see \cite{Sorokin:1993ua} and references therein). 
These decompose under $\mathfrak{sp}(2)$ as
\be 
S^{(\pm)}(\pm1/4)\downarrow_{\mathfrak{sp}(2)}=\mD^{(\pm)}(\pm 1/4)\oplus \mD^{(\pm )}(\pm 3/4)\,,\ee
where $\mD^{(\pm)}(\pm 1/4)$ and $\mD^{(\pm )}(\pm 3/4)$ are the lowest and highest weight quartions of $\mathfrak{sp}(2)$.
Thus, 
\be {\cal A}^{\prime(1)}({\cal Y}_4)=\bigoplus_{\epsilon=\pm} {\cal A}^{\prime(1)}_{\epsilon}({\cal Y}_4)\,,\qquad 
{\cal A}^{\prime(1)}_{\pm }({\cal Y}_4)= {\cal F}_\pm\otimes ({\cal F}_\pm)^\ast\,,\ee
where 
\be 
{\cal F}_-\cong \mD^{(+)}(1/2;(0))\oplus \mD^{(-)}(-1/2;(0))\,,\qquad 
{\cal F}_+\cong \mD^{(+)}(1;(1/2))\oplus \mD^{(-)}(-1;(1/2))\,,\ee
respectively, are the extended scalar and spinor singletons.

\paragraph{Mixing of particle and black hole states under extended HS transformations.}
Letting 
\be K:=\kappa_y\star \bar{\kappa}_{\bar y}\,,\ee
which obeys $K\star K=1$ and $K^\dagger=K$, one can show that\footnote{%
As can be recovered from a limit of Eq.\eqref{eq:gauss vs stargauss}, one has $K=\exp_\star (2i\pi E)$ \cite{Aros:2019pgj}.}
\be \Pi_\e\star {\cal A}^{\prime(1)}_{\epsilon'}({\cal Y}_4)=\delta_{\e,\e'}{\cal A}^{\prime(1)}_{\epsilon'}({\cal Y}_4)\,,\qquad \Pi_\pm=\frac12(1\pm K)\,.\ee
Thus, ${\cal A}^{\prime(1)}_{\pm }({\cal Y}_4)$ are two decoupled algebras each consisting of massless particle and black hole states.
Each one of these two subalgebras decompose further under the (unextended) Weyl algebra into four subsectors
\be {\cal A}^{\prime(1;\sigma,\sigma')}_{\pm }({\cal Y}_4)= {\cal F}^{(\sigma)}_\pm\otimes ({\cal F}^{(\sigma')}_\pm)^\ast\ ,\ee
which are mixed under general extended higher spin transformations, as $\kappa_y\star {\cal F}^{(\pm)}\cong {\cal F}^{(\mp)}$.
To exhibit these transformations, we define 
\be h_+:=\frac{1}{2}(\kappa_y+\bar{\kappa}_{\bar y})=\kappa_y\star \Pi_+\,,\qquad h_-=\frac{i}{2}(\kappa_y-\bar{\kappa}_{\bar y})=i\kappa_y\star \Pi_-\,,\ee
it follows from 
\be (h_\pm)^\dagger=h_\pm\,,\qquad h_\pm \star {\cal F}_\mp=0\,,\qquad h_\pm\star h_\pm=\pm \Pi_\pm\,,\ee
that
\begin{align} g_+:=e_\star^{ih_+\theta_+}&=\left(\cos \theta_+ + i\kappa_y\sin \theta_+\right)\star\Pi_++\Pi_-\,,\\
g_-:=e_\star^{ih_-\theta_-}&=\left(\cosh \theta_- - \kappa_y\sinh \theta_-\right)\star\Pi_-+\Pi_+\,,\end{align}
that is, the extended higher spin group contains a subgroup
\be G_+\times G_-\cong SO(2)\times SO(1,1)\,,\ee
that is represented in the extended Fock space as
\begin{align}\rho_{{\cal F}_+\oplus {\cal F}_-}(g_+,g_-)=g_+\Pi_+ + g_-\Pi_-\,.
\end{align}
Thus, the extended scalar and spinor singletons decompose under $G_+\times G_-$ into doublets, and hence ${\cal A}^{\prime(1)}_{\pm }({\cal Y}_4)$ decompose under the adjoint action of $G_+\times G_-$ into singlets and triplets arising in the anti-symmetric and symmetric direct products, respectively. 

In other words, the notion of particle, anti-particle and higher spin black hole states at the linearized level, as defined above, is not left invariant under (rigid) extended higher spin symmetry transformations.
On the other hand, these transformations factor out from the adapted traces used to form higher spin invariants provided that one follows the regular scheme (and that this scheme gives a finite result at the leading order).
Under these conditions, these invariants can be interpreted physically (for example, as higher spin amplitudes) in a given duality frame of particle, anti-particle and black holes states provided that this frame can be fixed globally on the base manifold, which we shall assume in what follows.

\paragraph{Massless particle modes in compact weight basis.}
The particle modes are obtained from initial data expanded over non-polynomial function $T_{e;(s)}$ of the $\mso(2,3)$ generators that are enveloping-algebra realizations of AdS${}_4$ massless particle states $|e,(s)\rangle$: as such, they have definite eigenvalues under the twisted-adjoint action of the compact Cartan generators $E:=P_0$ and $J:=M_{12}$ and of the quadratic Casimir $\frac12 M^{rs}\star M_{rs}$ of $\mso(3)$, 
\bea
&[E,T_{e;(s)}]_\pi=\{E,T_{e;(s)}\}_\star  =eT_{e;(s)}\,,&\\ 
&\frac12 [M^{rs},[M_{rs},T_{e;(s)}]_\pi]_\pi =\frac12 [M^{rs},[M_{rs},T_{e;(s)}]_\star]_\star =s(s+1)T_{e;(s)}\,,&
\eea
where each $T_{e;(s)}$ is a $(2s+1)$-plet with elements distinguished by the eigenvalue $j_s$ of $J$, $j_s = -s, -s+1,\ldots, s-1, s $,
and they span lowest-weight  modules (highest-weight modules for the anti-particle states) built via the action of energy-raising (lowering) operators $L^+_r$ ($L^-_r$) on a lowest-weight (highest-weight) state $T_{e_0;(s_0)}$ ($T_{-e_0;(s_0)}$), 
\be [L^-_r,  T_{e_0;(s_0)}]_\pi =L^-_r \star  T_{e_0;(s_0)}-T_{e_0;(s_0)}\star L^+_r =0
\,,\qquad{\rm for}\quad 
e_0 = s_0+1 
\,.
\label{eq:LWcond}\ee
All massless particle and anti-particle states have
\be |e|>s\,,\ee
and can in fact be built via $\mhs(2,3)$ action on the $\mD(1,0)$ massless scalar particle lowest weight state \cite{Iazeolla:2008ix} $T_{1;(0)} \leftrightarrow |1,(0)\rangle$ and on the $\mD(-1,0)$ massless scalar anti-particle highest weight state $T_{-1;(0)} \leftrightarrow |-1,(0)\rangle$, projectors that admit the realization ($\epsilon:=\pm1$)
\be T_{\epsilon;(0)}\equiv{\cal P}_{\epsilon} =4\exp(-4\epsilon E) =4\exp(\epsilon y\sigma_0\yb)  
\,. 
\label{eq:vac proj} 
\ee

The $Y$-space elements built that way actually diagonalize the separate left and right action of the compact Cartan generators: in fact, from the point of view of the left and right action of $\mso(2,3)$, the $T_{e;(s)}$ correspond to enveloping algebra realizations of (anti-)singleton states \cite{Iazeolla:2008ix}. This is a reflection, at an operatorial level, of the compositeness of massless particle states. Thus, each element $T_{e;(s)}$ corresponds to a specific linear combination of operators on the (anti-)singleton Hilbert space, i.e., is an enveloping-algebra realization of the specific tensor product of singleton states corresponding to any specific massless particle state according to the Flato-Fronsdal theorem \cite{Flato:1978qz}. For example,
\be
E\star T_{1;(0)} 
=
\frac12\, T_{1;(0)} =T_{1;(0)}\star E 
\,,\qquad 
L^-_r\star T_{1;(0)} =0 =T_{1;(0)}\star L^+_r 
\,,
\ee
thus
\be
T_{1;(0)} =|\ft12;(0)\rangle \langle\ft12;(0)| 
\qquad\leftrightarrow\qquad  
|\ft12;(0)\rangle_1 |\ft12;(0)\rangle_2 = | 1;(0)\rangle  
\,,\ee
where $ |\ft12;(0)\rangle$ is the singleton lowest-weight state. The massless particle lowest-weight states of spin $s$ can be analogously encoded in the element 
\be
T_{s+1;(s)} = 
{\cal N}_s \sum_{k=0}^{s}f_{s;k}(-1)^{s-k}  L^+_{\{r_1} \star \ldots \star L^+_{r_k}\star T_{1;(0)}\star L^-_{r_{k+1}} \ldots L^-_{r_s\}} 
\,,
\label{eq:LWSs}
\ee
where the normalization constant is given in \cite{Iazeolla:2008ix},
\be f_{s;k} ={s \choose k} \frac{(\ft12-s)_k}{(\ft12)_k} 	\,,\ee
with $(a)_n$ denoting the Pochhammer symbol, and the curly brackets around the indices denote symmetric and traceless projection. The relative coefficients between the terms forming the linear combination \eq{eq:LWSs} are fixed by the lowest-weight condition \eq{eq:LWcond}. For instance, the spin-$1$ lowest-weight element is
\be T_{2;(1)}  \ \propto \ L^+_r\star e^{-4E}+e^{-4E}\star L^-_r \ \propto \ M_{0r}e^{- 4E}\,. \label{eq:s1LWS}\ee

\paragraph{Massless particle modes in Fock-space basis.} The particle states can thus be reflected into operators on the (anti-)singleton Hilbert space \cite{Iazeolla:2008ix}. 
All such operators can be obtained from the ground state projectors \eq{eq:vac proj} and realized in terms of linear combinations of operators $P_{\mathbf{m}|\mathbf{n}} = \left\vert m_1,m_2\right\rangle\left\langle n_1,n_2\right\vert$ on a two-dimensional (anti-)Fock-space, with all $m_i$ and $n_j$ being strictly positive (negative) half-integers.
In the following, the set of such $(\mathbf{m},\mathbf{n})$ will be referred to as $\mathscr{N}$.
$P_{\mathbf{m}|\mathbf{n}}$ is an eigenfunction under the left and right star-product actions, respectively, of the number operators $w_i$ with eigenvalues $m_i$ and $n_i$, viz.
\begin{equation}
(w_i-m_i)\star P_{\mathbf{m}|\mathbf{n}} =0 =P_{\mathbf{m}|\mathbf{n}} \star (w_i-n_i)   \,.
\end{equation}
These number operators are related to the energy $E$ and spin $J$ as \cite{Iazeolla:2011cb}
\begin{equation}
\label{eq:w(E,J)}
w_1:= E-J
\,,\qquad
w_2:= E+J \,.
\end{equation}
The $e$ and $j_s$ eigenvalue under twisted-adjoint action can thus be obtained from the left and right eigenvalues of $P_{\mathbf{m}|\mathbf{n}}$ as
\be e =\frac{m_1+m_2+n_1+n_2}{2} 
\,,\qquad 
j_s =\frac{m_2-m_1+n_1-n_2}{2}  
\,.
\ee
The operators $ P_{{\mathbf m}|{\mathbf n}}$ satisfy
\be P_{{\mathbf m}|{\mathbf n}}=\pi\bar\pi(P_{{\mathbf m}|{\mathbf n}})\,,\ee
and 
\be P_{{\mathbf m}|{\mathbf m}'} \star P_{{\mathbf n}|{\mathbf n}'} =\delta_{{\mathbf m}',{\mathbf n}} P_{{\mathbf m}|{\mathbf n}'} \,. \label{eq:projalg}\ee
The lowest-weight and highest-weight projectors \eq{eq:vac proj} correspond to ${\cal P}_\e\equiv  P_{\ft{\e}{2},\ft{\e}{2} | \ft{\e}{2},\ft{\e}{2}}$, and all (anti-)particle modes can be written as appropriate linear combinations of
\begin{align}
P_{\mathbf{m}|\mathbf{n}} 
&=
P_{m_1|n_1}(y^+_1,y^-_1)\star P_{m_2|n_2}(y^+_2,y^-_2) = P_{m_1|n_1}(y^+_1,y^-_1)P_{m_2|n_2}(y^+_2,y^-_2)
\nn\\&\propto
\left(y_1^{\epsilon}\right)^{\star|m_1|-\tfrac{1}{2}}\star
\left(y_2^{\epsilon}\right)^{\star|m_2|-\tfrac{1}{2}}\star
{\cal P}_\epsilon\star
\left(y_1^{-\epsilon}\right)^{\star|n_1|-\tfrac{1}{2}}\star
\left(y_2^{-\epsilon}\right)^{\star|n_2|-\tfrac{1}{2}} 
\,, 
\label{eq:Pmn0}
\end{align}
where $y_i^\pm$ are the creation and annihilation operators (\ref{eq:ypm1},\,\ref{eq:ypm2}), $w_i=y^+_i y^-_i$, and 
\begin{align}
y_i^{-\epsilon}\star{\cal P}_{\e} =0 ={\cal P}_{\e}\star y_i^{\epsilon}
\,.
\end{align}
In Weyl-ordered form, each factor $P_{m_i|n_i}$ in \eq{eq:Pmn0} can be rewritten as\footnote{%
Note that the expressions (\ref{eq:Pmn1},\,\ref{eq:Pmn2}) include the cases when $m_i,n_i<0$, since via Kummer's transformation $e^{-2w}L^{(m-n)}_{n-1/2}(4w) = \frac{\sin(n-1/2)\pi}{\sin(-m-1/2)\pi}  e^{2w}L^{(m-n)}_{m-1/2}(-4w)$.} \cite{Aros:2019pgj}
\be
P_{m_i|n_i}
\propto
(y^+_i)^{m_i-n_i} L^{(m_i-n_i)}_{n_i-1/2}(4w_i)\,e^{-2w_i} 
\,,
\label{eq:Pmn1}
\ee
when $m_i\geq n_i$ and 
\be 
P_{m_i|n_i}
\propto 
(y^-_i)^{n_i-m_i} L^{(n_i-m_i)}_{m_i-1/2}(4w_i)\,e^{-2w_i} 
\,,  
\label{eq:Pmn2}
\ee
for $n_i\geq m_i$, where $L^{(a)}_k(x)$ are generalized Laguerre polynomials. 

The generalized Fock-space operators  $P_{\mathbf{m}|\mathbf{n}}$ correspond to the elements of the $\mso(3)$ multiplets $T_{e;(s)}$ with definite $e$ and $j_s$ engenvalues. For instance, the two terms in \eq{eq:s1LWS}, degenerate in the eigenvalues $e=2$ and $s=1$, split with respect to $J$ into the elements
\begin{align}
L^+_r\star e^{-4E} 
&=
\underbrace{(P_{\ft12,\ft52|\ft12,\ft12}}_{j_s=1},\underbrace{P_{\ft32,\ft32|\ft12,\ft12}}_{j_s=0},\underbrace{P_{\ft52,\ft12|\ft12,\ft12}}_{j_s=-1}) 
\,,\\
e^{-4E}\star L^-_r 
&=
\underbrace{(P_{\ft12,\ft12|\ft52,\ft12}}_{j_s=1},\underbrace{P_{\ft12,\ft12|\ft32,\ft32}}_{j_s=0},\underbrace{P_{\ft12,\ft12|\ft12,\ft52}}_{j_s=-1}) 
\,.
\end{align}

\paragraph{Black hole states.}

As was previously mentioned, ${\cal A}^{\prime(1)}({\cal Y}_4)$ also includes what we call black hole states,
that is to say operators that mix the singleton and the anti-singleton sectors.
Each black hole mode is associated to a particle mode through the chiral $Y$-space Fourier transform
\begin{equation}
\label{eq:PTvsBH}
\Psi_{\rm bh}':=\Psi_{\rm pt}'\star\k_y
\,,
\end{equation}
In other words, we consider an expansion of the Weyl zero-form of type
\be 
\label{eq:PTBHmn expansion}
\Phi'^{(1)} 
=\sum_{(\mathbf{m},\mathbf{n})\in\mathscr{N}}
\left( \m_{{\mathbf m}|{\mathbf n}}P_{{\mathbf m}|{\mathbf n}}+\n_{{\mathbf m}|{\mathbf n}}P_{{\mathbf m}|{\mathbf n}}\star \k_y \right)
\,,
\ee
generalizing to arbitrary spin the expansion studied in \cite{Iazeolla:2017vng}. 
It turns out that the constraints imposed by the reality condition \eqref{eq:RC} on the deformation parameters $\m_{\mathbf{m}|\mathbf{n}}$ and $\n_{\mathbf{m}|\mathbf{n}}$ are different \cite{Iazeolla:2017vng},
the condition \eqref{eq:RC} being incompatible with the duality \eqref{eq:PTvsBH}.
As $\k_y\star w_1=-w_2\star\k_y$, from the point of view of the eigenvalues of the twisted-adjoint action of the Cartan generators $E$ and $J$ the twisted Fock-space operator $P_{\mathbf{m}|\mathbf{n}}\star\k_y$ behaves as
\be P_{m_1,m_2|n_1,n_2}\star \k_y \ \propto \ P_{m_1,m_2|-n_2,-n_1} \,,\ee 
which implies that black-hole states fill the wedge in weight space in between particle and anti-particle modules, i.e. satisfy
\be |e| \leq s \,.\ee
In particular, the operator that maps the singleton and anti-singleton ground states into one another can be computed from \eqref{eq:vac proj} as 
\begin{equation}
{\cal P}_\e\star\k_y=8\pi\delta^2(y-i\e\sigma_0\yb)
\,.
\end{equation}
The delta function appearing in that expression means in particular that it does not make any sense to look at the Lorentz tensor components of the Weyl zero-form at the unfolding point $x^a=0$.  However, as we shall see after turning on the spacetime dependence  (see also \cite{Iazeolla:2011cb,Iazeolla:2017vng}), $\Phi^{(1)}(x;Y)$ is a regular function of $Y$ at generic locations (i.e., away from $x^a=0$), and the conventional interpretation in terms of Lorentz tensor fields is therefore restored.

%
\paragraph{Regular presentation.} The enveloping-algebra realisation of the (anti-)particle lowest-weight (highest-weight) state  \eq{eq:vac proj} ensures their idempotency, but results in a divergent product between ${\cal P}_1$ and ${\cal P}_{-1}$.
As shown in \cite{Iazeolla:2011cb,Iazeolla:2017vng}, the latter can be regularised to 0 by representing ${\cal P}_\e$ with an integral presentation
\begin{equation}
{\cal P}_{\epsilon}
=2\epsilon\oint_{C(\epsilon)}
\frac{d\eta}{2\pi i}\,\frac{\eta+\epsilon}{\eta-\epsilon}
\,e^{\eta y\sigma_0\yb}
\,,\label{eq:regP}
\end{equation}
where $C(\epsilon)$ is a complex contour that encircles $\epsilon$ in the complex plane, and prescribing that all star-product computations be performed before evaluating the contour integral (see Section \ref{ssec:reg}).  One simple way of encoding all the other massless particle modes, such as \eq{eq:LWSs}, is by means of a generating function, as
\begin{equation}
T_{e;(s)} =\Pi_{e;(s)}\left(\frac{\partial}{\partial X_\au}\right) 2\epsilon\left.
\oint_{C(\epsilon)}
\frac{d\eta}{2\pi i}\,\frac{\eta+\epsilon}{\eta-\epsilon}
\,\exp\left(\eta y\sigma_0\yb+\chi y+\bar\chi\yb\right)
\right\vert_{X=0}
\,,\label{eq:regHS}
\end{equation}
where $X_\au=(\chi_\a,\bar\chi_\ad)$ are polarisation spinors, $\Pi_{e;(s)}(\frac{\partial}{\partial X_\au})$ are  differential operators in $(\chi_\a,\bar\chi_\ad)$ endowed with the appropriate projections onto the irreducible $\mso(3)$-irrep corresponding to a given massless particle mode as well as normalization factors\footnote{%
Alternatively, one may consider integral presentations also for the polynomials in $Y$ that dress the exponential for the lowest- (highest-) weight (anti-)particle state, as done in specific examples in \cite{Iazeolla:2011cb,Iazeolla:2017vng,Aros:2017ror,Aros:2019pgj}. Such fully integral presentation is in general crucial in order to go beyond the first order in perturbation theory and satisfy \eq{eq:projalg}. As the analysis of the present paper is purely linear, we shall use the above simpler, mixed presentation \eq{eq:regHS}, and we shall not fix the normalization factors as they will not be necessary.}, and $\epsilon={\rm sign}(e)$. For instance, $\Pi_{1;(0)}(\frac{\partial}{\partial X_\au})=1$, while
\be T_{2;(1)}  
\ \propto \ \left.
\left[(\s_{0r})^{\a\b}\frac{\partial^2}{\partial \chi^\a \partial\chi^\b}+(\bar \s_{0r})^{\ad\bd}\frac{\partial^2}{\partial \bar\chi^{\ad} \partial\bar\chi^{\bd}}\right]\oint_{C(1)}
\frac{d\eta}{2\pi i}\frac{\eta+1}{\eta-1}
\exp\left(\eta y\sigma_0\yb+\chi y+\bar\chi\yb\right)
\right\vert_{X=0} 
\,.\ee
A similar operation can be performed on the black hole modes.
The expansion \eqref{eq:PTBHmn expansion} can hence be rewritten
\begin{align}
\Phi'^{(1)}(Y) 
=
\sum_{(\mathbf{m},\mathbf{n})\in\mathscr{N}}
&\left(
\m_{{\mathbf m}|{\mathbf n}}\Pi_{{\mathbf m}|{\mathbf n}}\left(\frac{\partial}{\partial X^{\au}}\right) {\cal O}_{\eta,\epsilon} \Phi_{\rm pt}'^{(1)}(Y;\eta;X)
\right.\nn\\&\quad\left.
+
\nu_{{\mathbf m}|{\mathbf n}}\left.\Pi_{{\mathbf m}|{\mathbf n}}\left(\frac{\partial}{\partial X^{\au}}\right) {\cal O}_{\eta,\epsilon} \Phi_{\rm bh}'^{(1)}(Y;\eta;X)\right)\right\vert_{X=0}
\,, 
\end{align}
where we have denoted 
\be
{\cal O}_{\eta,\epsilon} :=\oint_{C(\epsilon)}
\frac{d\eta}{2\pi i}\frac{\eta+\epsilon}{\eta-\epsilon}\,,
\ee
 $\Pi_{{\mathbf m}|{\mathbf n}} $ is the operator that differentiates with respect to $X_{\au}$ as to reproduce the appropriate polynomial characterizing each $P_{{\mathbf m}|{\mathbf n}} $ according to (\ref{eq:Pmn1},\,\ref{eq:Pmn2}), and we have defined the generating functions
\begin{align}
\label{eq:Phi' pt}
\Phi_{\rm pt}'^{(1)}(Y;\eta;X) 
&= 
\exp\left(\eta y\sigma_0\yb+\chi y+\bar\chi\yb\right)
\,,\\
\label{eq:Phi' bh}
\Phi_{\rm bh}'^{(1)}(Y;\eta;X) 
&=
2\pi\delta^2(y-i\eta\sigma_0\yb+i\chi)\exp(\bar\chi\yb)\,.
\end{align}
encoding the initial data for particle and black hole modes, respectively. 

Since we are expanding the spacetime-independent equations \eqref{eq:VE'} around the trivial vacuum $V^{'(0)}=\Phi^{\prime(0)}=0$,
the field $\Phi^{\prime(1)}$ can be identified with its integration constant $C^{\prime(1)}$ in \eqref{eq:Psi' vs Phi'}.
In particular, one has
\begin{align}
\label{Psipt}
\Psi_{\rm pt}^{\prime(1)}(Y;\eta;X)
:&=
\Phi_{\rm pt}^{\prime(1)}(Y;\eta;X)\star\k_y
=
2\pi\delta^2(y-i\eta\sigma_0\yb+i\chi)\exp(\bar\chi\yb)
\,,\\
\Psi_{\rm bh}^{\prime(1)}(Y;\eta;X)
:&=
\Phi_{\rm bh}^{\prime(1)}(Y;\eta;X)\star\k_y
=
\exp\left(\eta y\sigma_0\yb+\chi y+\bar\chi\yb\right)
\,.
\end{align}
The spacetime dependence can then be obtained by computing the star products in \eq{eq:def Phi'} and using \eqref{eq:Lrot stereo} on the adjoint quantity $\Psi^{\prime(1)}$.

\subsection{Weyl zero-form}
\label{ssec:Weyl}

Using a gauge function $L$ and applying \eqref{eq:Lrot stereo} to \eqref{eq:Phi' bh}, the Weyl zero-form constant $C'^{(1)}=\Psi'^{(1)}\star\kappa_y$ can be mapped to a Weyl zero-form 
\begin{equation}
\label{eq:def Phi'}
\Phi^{(1)}=L^{-1}\star C'{}^{(1)}\star \pi(L)=\Psi^{(1)}\star \k_y\,,\qquad \Psi^{(1)}= L^{-1}\star\Psi'\star L 
\,.
\end{equation}
In what follows, we shall use the vacuum gauge function $L$ \eq{eq:L gf stereo} defined on the (inner) stereographic coordinate chart\footnote{%
The metric $$ds^2=\frac{4dx^2}{(1-x^2)^2}\,,\qquad x^2\neq 1\,,$$
provides a global cover of proper AdS spacetime, with $x^2=1$ serving as a two-sheeted boundary, while the stereographic gauge function $L$ is defined for $x^2<1$.}.

\paragraph{Massless particle states.}
The adjoint action of the gauge function $L$ on \eq{Psipt} gives
\be
\Psi_{\rm pt}^{(1)}(x;Y;\eta;X) 
=2\pi\delta^2(Ay+B\yb+i\chi)\exp(\bar\chi\yb^L)
\,,\\
\label{eq:PsiLpt}
\ee
from which it follows that the spacetime-dependent massless particle mode generating function is
\begin{align}
\Phi_{\rm pt}^{(1)}(x;Y;\eta;X) 
&=
\Psi_{\rm pt}^{(1)}(x;Y;\eta;X) \star \k_y
\nn\\&=
\frac{1}{\det A}\exp\left(
iyM\yb-yA^{-1}\chi-\tfrac{i}{h}\bar\chi\xb A^{-1}\chi+\tfrac{1}{h}\bar\chi(1-\xb M)\yb
\right)
\,,
\label{eq:PhiLpt}
\end{align} 
where we have defined
\be
 A_\a^{\phantom\a\b} :=\frac{1}{h}\left(\epsilon_\a^{\phantom\a\b}-i\eta\sigma_{0\a}^{\phantom{0\a}\ad}\xb_\ad^{\phantom\ad\b}\right)
=: \bar{A}_{\phantom\b\a}^{\b}
\,,\qquad 
B_\a^{\phantom\a\bd}:=\frac{1}{h}\left(x_\a^{\phantom\a\bd}-i\eta\sigma_{0\a}^{\phantom{0\a}\bd}\right)
=:\bar{B}^\bd_{\phantom\bd\ad}
\,, 
\ee
with
\be \det A =\frac{1-2i\eta x_0+\eta^2 x^2}{1-x^2} \,,\ee
and 
\be M_\a^{\phantom\a\bd} :=A^{-1}_\a{}^\b B_\b{}^{\bd} =f_1(x,\eta)x_\a{}^{\bd}-if_2(x,\eta)(\s_0)_\a{}^{\bd} 
=:\bar{M}^\bd_{\phantom\bd\a}
\,, \ee
\be f_1 :=\frac{1-2i\eta x_0+\eta^2}{1-2i\eta x_0 + \eta^2 x^2} \,, \qquad f_2 
:=\eta \,\frac{1-x^2}{1-2i\eta x_0 + \eta^2 x^2}   \label{eq:A-1B}\,.\ee

For $\chi_{\a}=\bar \chi_{\ad}=0$ the generating function \eq{eq:PhiLpt} reduces to the one for massless rotationally-invariant scalar field modes already studied in \cite{Iazeolla:2017vng}.

\paragraph{Black hole states.} Analogously, the expansion of the Weyl zero-form and its dual over black hole states are based on the $x$-dependent generating functions
\begin{align}
\label{eq:PhiL bh}
\Phi_{\rm bh}^{(1)}(x;Y;\eta;X)
=&
-\frac{i}{\sqrt{\eta^2} r}
\exp\left(
-\tfrac{1}{2\eta}y(\vark^L)^{-1}y
+iy(\vark^L)^{-1}v^L\yb
-\tfrac{i}{\eta}y(\vark^L)^{-1}\chi^L
\right)\\\nonumber&\times 
\exp\left(
\tfrac{\eta}{2}\yb(\varkb^L-\bar{v}^L(\vark^L)^{-1}v^L)\yb
+\tfrac{1}{2\eta}\chi^L(\vark^L)^{-1}\chi^L
+\yb\bar{v}^L(\vark^L)^{-1}\chi^L+\bar\chi^L\yb
\right)
\,,
\end{align}
and
\be
\label{eq:Psi BH}
\Psi_{\rm bh}^{(1)}(x;Y;\eta;X)
\ = \
\exp\left(
\tfrac{\eta}{2}y\vark^Ly
+\eta yv^L\yb
+\tfrac{\eta}{2}\yb\varkb^L\yb
+\chi^Ly+\bar\chi^L\yb
\right)
\,,\ee
where
\be
\chi^L:=\frac{1}{h}\left(\chi-x\bar\chi\right)
\,,\qquad
\bar\chi^L:=\frac{1}{h}\left(\bar\chi-\xb\chi\right)
\,,\ee
\be
\vark^L:=
\frac{1}{h^2}\left(\sigma_0\xb-x\bar\sigma_0\right)
\,,\qquad
\varkb^L:=
\frac{1}{h^2}\left(\bar\sigma_0x-\xb\sigma_0\right)
\,,\qquad
v^L:=
\frac{1}{h^2}\left(\sigma_0-x\bar\sigma_0x\right)
\,,
\ee
\be (\vark^L)^{-1}_{\a\b} =\frac{\vark^L_{\a\b}}{r^2} \,,  \qquad  \det \vark^L_{\a\b} = \frac12\vark^{\a\b}\vark_{\a\b} = -r^2\ee
(analogously for the hermitian conjugate, with $\det \vark^L_{\a\b}=\det \varkb^L_{\a\b}$). 
The Weyl zero-form is indeed divergent at the point $r=0$, which is one of the motivations for the black-hole interpretation \cite{Didenko:2009td,Iazeolla:2011cb}. For $\chi_{\a}=\bar \chi_{\ad}=0$ the generating function \eq{eq:PhiL bh} reduces to the one for spherically-symmetric higher-spin black hole states studied in \cite{Iazeolla:2011cb,Iazeolla:2017vng}.

\subsection{Connections in factorized gauge}
\label{ssec:PTBH fact}

In what follows, we shall start from the Weyl zero-forms (\ref{eq:PhiLpt},\,\ref{eq:PhiL bh}) for particle and black-hole modes, and we give the associated $Z$-space and spacetime one-form connections in factorized gauge using (\ref{eq:V1 unfact},\,\ref{eq:Ufact}).
For the ground states, i.e. when the polarizations are turned off, such linear solutions correspond to the first-order solutions given in \cite{Iazeolla:2017vng} with $n=\epsilon$.

\paragraph{Particle states.}
Applying (\ref{eq:V1 unfact},\,\ref{eq:Ufact}) to equation \eqref{eq:PsiLpt} gives
\begin{align}
U_{\rm pt}^{(1,E+i\partial_Y)}
&=
W_{\rm pt}^{(1,{E+i\partial_Y},{E+i\partial_Y})}
\,,\\
V_{\rm pt}^{(1,E+i\partial_Y)}
&=
\label{eq:Vpt(C)}
\frac{ib}{2}
(\dz{\cal D}\ytp)
\exp\left(
i\ytp z
\right)
\int_{-1}^1\frac{\diff s}{(1+s)^2}
\exp\left(
\tfrac{i}{2}\tfrac{1-s}{1+s}\ytp{\cal D}\ytp
\right)
\left.\Phi_{\rm pt}^{(1)}\right\vert_{y=0}
-\hc
\,,
\end{align}
where $\ytp$ was defined as
\begin{equation}
\ytp_\a := y_\a + M_\a{}^{\ad}\yb_\ad +i(A^{-1})_\a^{\phantom\a\b}\chi_\b
\,.
\end{equation}
\paragraph{Black hole states.}
The corresponding connection in factorised gauge is given by the application of (\ref{eq:V1 unfact},\,\ref{eq:Ufact}) on \eqref{eq:PhiL bh}
\begin{align}
&
U_{\rm bh}^{(1,E+i\partial_Y)}
=
W_{\rm bh}^{(1,{E+i\partial_Y},{E+i\partial_Y})}
\,,\\
&
V_{{\rm bh}}^{(1,E+i\partial_Y)} 
\nonumber\\&=
-\frac{b}{2}
\left.\Phi_{\rm bh}^{(1)}\right\vert_{y=0}
\,
(\dz \partial^\rho)
\int_{-1}^1\frac{\diff s}{(1+s)\sqrt{\det G}}
\exp\left(
\tfrac{1}{2}\ytb\,G^{-1}\ytb 
+\tfrac{i}{1+s}\rho\left(
1+i\tfrac{1-s}{1+s}G^{-1}{\cal D}
\right)z
\right)\times\nonumber\\&\quad\left.\times\exp\left(
-\tfrac{1}{1+s}\rho\,G^{-1}\ytb 
-\tfrac{1-s}{1+s}z{\cal D}G^{-1}\ytb
+\tfrac{i}{2}\tfrac{1-s}{1+s}
z\left(
{\cal D}+i\tfrac{1-s}{1+s}{\cal D}G^{-1}{\cal D}
\right)z
\right)
\right\vert_{\rho=0}
-\hc
\label{eq:V1 BH}
\,,
\end{align}
where the following definitions were introduced
\begin{align}
\ytb:&=
y+(\vark^L)^{-1}v^L\yb-\frac{1}{\eta}(\vark^L)^{-1}\chi^L
\,,\qquad
G:=\frac{1}{\eta r^2}\vark^L-i\frac{1-s}{1+s}{\cal D}
\,,
\end{align}
and where the inverse is meant in the sense of NW-SE contraction \eqref{eq:NWSE}.
More details about the relevant computations are given in Appendix \ref{App:BH}.

\subsection{COMST from factorised solution}

In this subsection we show that one can start from the factorised particle and black hole linearised solutions (\ref{eq:Vpt(C)},\,\ref{eq:V1 BH}) and apply the procedure of Sec. \ref{ssec:COMST from fact} to get the COMST.
As mentioned in that section, the regular prescription allows to get all the way to \eqref{eq:COMST eom step},
and what we will show in the following is that performing the subsequent steps is compatible with said prescription. 
In other words, we shall now show explicitly that assumption ii) below \eqref{eq:COMST H1} is verified for the massless particle and black hole solutions, starting from the factorised gauge. 

\paragraph{Spacetime dependence of internal connection.}
The vanishing of the first term in Eq.\eqref{eq:COMST eom step} is compatible in both cases with the regular prescription.
This comes from the fact that the solution is factorised as Eqs.(\ref{eq:V^m fact},\ref{eq:U^m fact}), that $\Phi^{(1)}$ was built using Eq.\eqref{eq:def Phi'}
and that neither of those observations need the parametric integral to be performed.
Then, one has
\begin{align}
\acd V^{(1,E+i\partial_Y)}
&=
\acd\left(L^{-1}\star\Psi'{}^{(1)}\star L\star v_1(z)\right)+\hc 
\nonumber\\&=
\acd\left(L^{-1}\star\Psi'{}^{(1)}\star L\right)\star v_1(z)+\hc 
\nonumber\\&=
L^{-1}\star d\Psi'{}^{(1)}\star L\star v_1(z)+\hc 
=0
\,,
\end{align}
prior to doing any (parametric or contour) integral.

\paragraph{Particle states.}
After ensuring the vanishing of the first term in Eq.\eqref{eq:COMST eom step}, using Eq.\eq{eq:Vpt(C)} one has
\begin{align}
\acd W_{\rm pt}^{(1,G,G)}
&=
\Omega^{\au\bu}\Omega^{\cu\bu}\partial^Y_\bu\partial^Y_\bu\left(\partial^Z_{\left[\cu\right.}V^{(1,E+i\partial_Y)}_{{\rm pt}\,\left.\au\right]}\right)_{Z=0}
\nonumber\\&=
-\frac{b}{4}
\int_{-1}^1\frac{\diff s}{(1+s)^2}
\Omega^{\a\bu}\Omega_\a^{\phantom\a\bu}
\partial^Y_\bu\partial^Y_\bu
(\ytp{\cal D}\ytp)
\exp\left(
\tfrac{i}{2}\tfrac{1-s}{1+s}\ytp{\cal D}\ytp
\right)
\left.\Phi_{\rm pt}^{(1)}\right\vert_{y=0}
+\hc
\nonumber\\&=
-\frac{b}{4}
\int_{-1}^1\frac{\diff s}{(1+s)^2}
(\ytp{\cal D}\ytp)
\exp\left(
\tfrac{i}{2}\tfrac{1-s}{1+s}\ytp{\cal D}\ytp
\right)
e^{\a\bd}e_\a^{\phantom\a\bd}
\partial^\yb_\bd\partial^\yb_\bd
\left.\Phi_{\rm pt}^{(1)}\right\vert_{y=0}
+\hc
\nonumber\\&=
-\frac{ib}{4}
e^{\a\bd}e_\a^{\phantom\a\bd}
\partial^\yb_\bd\partial^\yb_\bd
\left.\Phi_{\rm pt}^{(1)}\right\vert_{y=0}
+\hc
\,.
\label{eq:COMST pt}
\end{align}
The final line was obtained using Eq.\eqref{eq:int exp=1/z2}.
The one before comes from the following procedure.
First one does the change of variables $(y,\yb)\to(\ytp,\yb)$ and writes the derivatives as
\begin{align}
&\Omega^{\a\bu}\Omega_\a^{\phantom\a\bu}
\partial^Y_\bu\partial^Y_\bu
\\&=\nonumber
\left(
\left(\omega^{\a\b}-e^{\a\ad}M^{\b}{}_{\ad}
\right)\partial^{\ytp}_\b
+e^{\a\ad}\partial^\yb_\ad
\right)
\left(
\left(\omega_\a^{\phantom\a\b}-e_\a^{\phantom\a\ad}M^{\b}{}_{\ad}\right)\partial^{\ytp}_\b
+e_\a^{\phantom\a\ad}\partial^\yb_\ad
\right)
\,.
\end{align}
What one has to show is then that the action of respectively one and two $\ytp$-derivatives is trivial on
\begin{equation}
v
:=
\int_{-1}^1\frac{\diff s}{(1+s)^2}
(\ytp{\cal D}\ytp)
\exp\left(
\tfrac{i}{2}\tfrac{1-s}{1+s}\ytp{\cal D}\ytp
\right)
\,,
\end{equation}
with the prescription that all the derivatives are taken one the integrand and that the $s$-integral is performed as the very last step.
Using standard integration tools as well as Eqs. (\ref{eq:delta'},\,\ref{eq:int exp=1/z2},\,\ref{eq:lem int(1+...)exp})\,,
one finds respectively
\begin{align}
\partial^\ytp_\a v
&=
2\pi({\cal D}\ytp)_\a\delta^2(\ytp)=0
\\
\partial^\ytp_\a\partial^\ytp_\a v
&=
2\pi\partial^\ytp_\a(({\cal D}\ytp)_\a)\delta^2(\ytp)
+2\pi({\cal D}\ytp)_\a\partial^\ytp_\a(\delta^2(\ytp))
=0
\,,
\end{align}
concluding the proof.

\paragraph{Black-hole states.}
Starting again from the second term in \eqref{eq:COMST eom step}, this time for the $Z$-space connection \eq{eq:V1 BH}, one has
\begin{align}
\acd W_{\rm bh}^{(1,G,G)}
&=
\Omega^{\au\bu}\Omega^{\cu\du}\partial^Y_\bu\partial^Y_\du
\left(\partial^Z_{[\cu}V_{{\rm bh}\,\au]}^{(1,E+i\partial_Y)}\right)_{z=0}
\nonumber\\&=-\frac{ib}{4}
\Omega^{\a\bu}\Omega_\a^{\phantom{\a}\du}\partial^Y_\bu\partial^Y_\du
\left(v(\ytb)
\left.\Phi_{\rm bh}^{(1)}\right\vert_{y=0}
\right)-\hc
\nonumber\\&=-\frac{ib}{4}
e^{\a\bd}e_\a^{\phantom{\a}\dd}\partial^\yb_\bd\partial^\yb_\dd
\left.\Phi_{\rm bh}^{(1)}\right\vert_{y=0}
-\hc
\,,
\label{eq:bh COMST fact}
\end{align}
where $v(\ytb)$ was defined as
\begin{equation}
v(\ytb):=
\int_{-1}^1\frac{\diff s}{(1+s)\sqrt{\det G}}
\left[
\Tr\left(1+i\frac{1-s}{1+s}G^{-1}{\cal D}\right)
+i\frac{1-s}{1+s}\ytb G^{-1}{\cal D}G^{-1}\ytb
\right]
\exp\left(\tfrac{1}{2}\ytb G^{-1}\ytb
\right)
\,,
\end{equation}
and where the conclusion comes from 
doing the change of variables $(y,\yb)\to(\ytb,\yb)$
and then noticing that
\begin{equation}
v(\ytb)=1\,,\qquad 
\partial^\ytb_\a v(\ytb)=0\,,\qquad
\partial^\ytb_\a\partial^\ytb_\a v(\ytb)=0\,.
\end{equation}
The latter statement is compatible with the regular prescription, as we will now show.
First, using the lemmas collected in Appendix \ref{App:BH}, it is possible to check that 
\begin{align}
\label{eq:lem comst bh}
v(\ytb)
&=
\left[-\frac{1}{\sqrt{\det G}}\frac{1-s}{1+s}
\exp\left(\tfrac{1}{2}\ytb G^{-1}\ytb\right)\right]_{-1}^1
\,.
\end{align}
The evaluation of the upper boundary term uses the fact that $G$ tends to become $s$-independent for $s\to 1$,
hence one has a regular prefactor multiplying $(1-s)$, which then gives 0 in the limit.
In the case of the lower boundary, one has
\begin{equation}
\label{eq:G at s=-1}
G= -\frac{2i}{1+s}{\cal D}+O(1)
\,,\qquad 
\det G = \frac{4}{(1+s)^2}
+O\left(\frac{1}{1+s}\right)
\,,
\end{equation}
implying in turn
\begin{equation}
v(\ytb) = \lim_{s\to-1}\frac{1-s}{2} =1\,.
\end{equation}
The action of $\ytb$-derivatives is to take down powers of $G^{-1}$, that behaves as a constant toward the upper boundary of the integration domain,
and falls off towards the lower one,
thereby proving \eqref{eq:lem comst bh}.
This concludes the proof.

\subsection{Master fields in relaxed Vasiliev gauge}

In this Section, we shall give the linearised master fields for massless particles and black hole states
that satisfy the relaxed Vasiliev gauge discussed in Section \ref{ssec:COMST from fact}, reaching it from the factorised gauge (\ref{eq:V1 unfact},\ref{eq:Ufact}), and more precisely from the refined gauge \eqref{eq:W(1F0)=0}.
This procedure gives the gauge function in the relaxed Vasiliev gauge.
Indeed, since the spacetime connection $U$ is trivial in factorised gauge,
it is characterised by the AdS gauge function $L$, up to a regular residual transformation.
The gauge function \eqref{eq:def gF} in relaxed Vasiliev gauge is hence given perturbatively by
\begin{equation}
G = L\star\left(
1+H^{(1,E+i\partial_Y\to G)}
\right)
+G^{(\geq2)}
\end{equation}
\paragraph{Particle states.}
In the particle case, the gauge parameter needed to go from factorised to the relaxed Vasiliev gauge is given by Eq. \eqref{eq:H = int VL + h}
\begin{align}
H^{(1,E+i\partial_Y\to G)}_{\rm pt}
=&
-
\frac{ib}{2\det A}
(z\,{\cal D}\ytp)
\exp\Big(-\tfrac{i}{h}\bar\chi\xb A^{-1}\chi+\tfrac{1}{h}\bar\chi(1-\xb M)\yb
\Big)
\int_0^1\diff t\,\exp\left(it\ytp z\right)
\nonumber\times\\&\times
\left.
\int_{-1}^1\frac{\diff s}{(1+s)^2}
\exp\left(
\tfrac{i}{2}\tfrac{1-s}{1+s}\ytp{\cal D}\ytp
\right)\right\vert_{\rho=0}
+h^{(1,E+i\partial_Y\to G)}_{\rm pt}
+H^{(1,E+i\partial_Y\to G)}_{\rm pt,2}
\,.
\end{align}
Because of the lemma \eqref{eq:int exp=1/z2}, it is clear that it is singular at the point $\ytp=0$.
From there, the particular spacetime connection \eqref{eq:Wpart} is
\begin{align}
W_{\rm pt}^{(1,(G,g),G)}
:=&
\left.\acd H^{(1,(E+i\partial_Y,0)\to (G,g))}_{\rm pt}\right\vert_{z=0}
\nonumber\\=&
\frac{ib}{2\det A}
\exp\left(
-\tfrac{i}{h}\bar\chi\xb A^{-1}\chi+\tfrac{1}{h}\bar\chi(1-\xb M)\yb
\right)
\int_{-1}^1\frac{\diff s}{1+s}
\exp\left(
\tfrac{i}{2}\tfrac{1-s}{1+s}\ytp{\cal D}\ytp
\right)
\nonumber\times\\&\times
\left[
\Tr((\omega-e\bar{M}){\cal D})
+i\frac{1-s}{1+s}\ytp{\cal D}(\omega-e\bar{M}){\cal D}\ytp
+\frac{1}{h}\bar\chi(1-\xb M)\bar{e}{\cal D}\ytp
\right]
\nonumber\\&
-\hc + \acd h^{(1,(E+i\partial_Y,0)\to (G,g))}_{\rm pt}
\,.
\label{eq:Wphys pt}
\end{align}
Because again of the same lemma, we see that, although Eq.\eqref{eq:COMST pt} shows that it would anyway give rise to COMST,
the solution $
W_{\rm pt}^{(1,(G,0),G)}$ with $h^{(1,i\partial_Y\to G)}_{\rm pt}=0$ would not be a genuine generating function for unfolded Fronsdal fields.
Such a generating function can be constructed choosing the Fock-Schwinger gauge $(G,FS)$ given by \eqref{eq:h1 FS}, viz.
\begin{align}
h^{(1,(E+i\partial_Y,0)\to (G,FS))}_{\rm pt}
=&
\frac{b}{2}\int_0^1dt\int_{-1}^1\frac{ds}{(1+s)^2}
\left(
m_1
+\ytp M_2\ytp
+\bar\chi M_3\ytp
\right)
\times\nonumber\\&\times
\exp\left(
\ytp M_4\ytp + \bar\chi M_5\ytp +\bar\chi\yb^L
\right)
+\hc
\,,
\end{align}
where we have defined
\begin{align}
m_1:&=
i\eta\Tr\left(x^a\sigma_{0a}D\right)
\,,\\
M_2:&=
\eta\frac{1-s}{1+s}\bar{A}(1-it\eta\sigma_0\xb)
D(x^a\sigma_{0a})D
(1-it\eta x\bar\sigma_0)A
\,,\\
M_3:&=
-\frac{(\xb-it\eta x^2\bar\sigma_0)D(1-it\eta x\bar\sigma_0)A}{1-2it\eta x_0+t^2\eta^2x^2}
\,,\\
M_4:&=
-\frac{i}{2}\frac{1-s}{1+s}
\bar{A}(1-it\eta\sigma_0\xb)D(1-it\eta x\bar\sigma_0)A
\,,\\
M_5:&=
-t\frac{\xb-it\eta x^2\bar\sigma_0}{1-2it\eta x_0+t^2\eta^2 x^2}
\,,\\
\bar{A}
:&=
-\frac{1}{h}(1-i\eta x\bar\sigma_0)
\,.
\end{align}
One fact that is crucial for this construction is that $A\ytp=-\ytp\bar{A}=y^L-i\eta\sigma_0\yb^L+i\chi$ is not affected by the homotopy integral.

\paragraph{Black hole states.}
The gauge function can be found again from Eq. \eqref{eq:H = int VL + h}, and in this case is 
\begin{align}
H^{(1,E+i\partial_Y\to G)}_{\rm bh}
=&
\frac{b}{2}
\left.\Phi_{\rm pt}^{(1)}\right\vert_{y=0}
\int_0^1dt\int_{-1}^1\frac{ds}{\sqrt{\det G}(1+s)^2}
\left(
zG^{-1}\ytb+t\frac{1-s}{1+s}zG^{-1}{\cal D}z
\right)
\times\nonumber\\&\times
\exp\left(
\tfrac{1}{2}\ytb{\cal D}\ytb 
-t\tfrac{1-s}{1+s}z{\cal D}G^{-1}\ytb+\tfrac{t^2}{2}\tfrac{1-s}{1+s}z\left({\cal D}+i\tfrac{1-s}{1+s}{\cal D}G^{-1}D\right)z
\right)
\nonumber\\&
-\hc
+h^{(1,E+i\partial_Y\to G)}_{\rm bh}
+H^{(1,E+i\partial_Y\to G)}_{\rm bh,2}
\,.
\end{align}
The  $O(Z^2)$ piece $H^{(1,E+i\partial_Y\to G)}_{\rm bh,2}$ maybe be redefined so as to give
\begin{align}
H^{(1,E+i\partial_Y\to G)}_{\rm bh}
=&
\frac{b}{2}
\left.\Phi_{\rm pt}^{(1)}\right\vert_{y=0}
\int_0^1dt\int_{-1}^1\frac{ds}{\sqrt{\det G}(1+s)^2}
\left(
zG^{-1}\ytb
\right)
\times
\phantom{+t\frac{1-s}{1+s}zG^{-1}{\cal D}z}
\nonumber\\&\times
\exp\left(
\tfrac{1}{2}\ytb{\cal D}\ytb 
-t\tfrac{1-s}{1+s}z{\cal D}G^{-1}\ytb+\tfrac{t^2}{2}\tfrac{1-s}{1+s}z\left({\cal D}+i\tfrac{1-s}{1+s}{\cal D}G^{-1}D\right)z
\right)
\nonumber\\&
-\hc
+h^{(1,E+i\partial_Y\to G)}_{\rm bh}
+\widetilde{H}^{(1,E+i\partial_Y\to G)}_{\rm bh,2}
\,.
\end{align}
The determined part of this expression is analytic around $\ytb=0$ for a generic spacetime point.
Indeed, it has been argued in \cite{Iazeolla:2017vng} that $\det G$ (given by Eq.\eqref{eq:detG}) has no pole inside the interval $[-1,1]$. 
Towards $s=-1$, Eq.\eqref{eq:G at s=-1} allows to show that the integrand stays finite.
The spacetime connection \eqref{eq:Wpart} is then given by
\begin{align}
W_{\rm bh}^{(1,(G,g),G)}
:=&
\left.\acd H^{(1,(E+i\partial_Y,0)\to (G,g))}_{\rm bh}\right\vert_{z=0}
\nonumber\\=&
\frac{ib}{4}
\left.\Phi^{(1)}_{\rm bh}\right\vert_{y=0}
\int_0^1dt\int_{-1}^1\frac{ds}{\sqrt{\det G}(1+s)^2}
\exp\left(
\tfrac{1}{2}\ytb G^{-1}\ytb
\right)
\times\nonumber\\&\times
\left[
m_6
+
\ytb M_7\ytb
+m_8
\right]
-\hc + \acd h^{(1,(E+i\partial_Y,0)\to (G,g))}_{\rm bh}
\,,
\end{align}
where
\begin{align}
m_6&=\Tr\left((\omega+e\bar{v}^L(\kappa^L)^{-1})G^{-1}\right)
\,,\\
M_7&=G^{-1}(\omega+e\bar{v}^L(\kappa^L)^{-1})G^{-1}
\,,\\
m_8&=\left(
-\eta\yb(\kappa^L)^{-1}-\bar\chi^L(\kappa^L)^{-1}v^L+\bar\chi^L
\right)\bar{e}G^{-1}\ytb
\,.
\end{align}
It is regular, because the particular part of $H^{(1,i\partial_Y\to G)}_{\rm bh}$ is.
This makes $h^{(1,(E+i\partial_Y,0)\to (G,g))}_{\rm bh}=0$ an acceptable choice.

\subsection{Black-hole solutions in Didenko-Vasiliev gauge}
\label{ssec:DV}
In this section, we discuss another way to resolve the $z$-dependence in the case of black-hole initial data,
which provides us with a example of a gauge $(DV)$ where $U^{(1,DV)}$ is non trivial,
and from which we perform the procedure of Section \ref{ssec:COMST from fact}.
It corresponds to the polarised generalisation of the linearisation of the solution studied in \cite{Didenko:2009td},
and we hence refer to this resolution as the Didenko-Vasiliev, or $(DV)$, gauge.
The construction of the exact solution of \cite{Didenko:2009td} requires that $\Psi^{(1)}_{\rm bh}$ be a projector,
which is the case when $\eta=\pm1$ and $X_\au=0$
.
Since here the aim is simply to give an example of the linearised procedure where $U^{(1,DV)}$ is non-trivial, we restrict ourselves to the first order analysis.
One property that is crucial in this construction is
\begin{equation}
\Psi^{(1)}_{\rm bh}\star f(Z)
=
\Psi^{(1)}_{\rm bh}
\int\frac{d^4U}{(2\pi)^2}f(A-U)
\exp\left(-\frac{1}{2\eta}\left(
u\vark^Lu + \ub\varkb^L\ub + 2uv^L\ub
\right)\right)
\,,
\end{equation}
where $A_\au=(a_\a,\ab_\ad)$ is defined as
\begin{align}
a:&=z+i\eta\vark^Ly+i\eta v^L\yb-i\chi^L
=z+i\eta\vark^L\ytb
\,,\\
\ab:&=\zb+i\eta\vb^Ly+i\eta\varkb^L\yb-i\bar\chi^L
\,.
\end{align}
This allows in particular to rewrite the source in equation \eqref{eq:qV1} as
\begin{equation}
-\Phi^{(1)}_{\rm bh}\star J
=
-\Psi^{(1)}_{\rm bh}\star
j_z
+\hc
=
\frac{b}{4r\sqrt{\eta^2}}
\Psi^{(1)}_{\rm bh}
\exp\left(
-\frac{1}{2\eta}a(\vark^L)^{-1}a
\right)
dz^\a dz_\a
+\hc
\,.
\end{equation}
One choice of resolution is the homotopy contraction along $a_\a$:
\begin{equation}
V^{(1,a)}_{{\rm bh},\a}
=
-\frac{b}{4r\sqrt{\eta^2}}
\Psi^{(1)}_{\rm bh}
a_\a\int_0^1d\tau\exp\left(
-\frac{\tau}{2\eta}a(\vark^L)^{-1}a
\right)
\,.
\end{equation}
The Didenko-Vasiliev gauge for the internal connection is defined as
\begin{equation}
V^{(1,DV)}_{{\rm bh},\a}
=
-\frac{b}{2r\sqrt{\eta^2}}
\Psi^{(1)}_{\rm bh}
a^+_{E,\a}\int_0^1d\tau\exp\left(
-\frac{\tau}{2\eta}a(\vark^L)^{-1}a
\right)
\,,
\end{equation}
where $a^+_{E,\a}$ is defined
as
\begin{equation}
a^{\pm}_{E,\a}
:=
\frac{1}{2}\left(
\epsilon_\a^{\phantom\a\b}\pm\frac1r(\vark^L)_\a^{\phantom\a\b}
\right)a_\b
=
\pm u^\mp_{E,\a}u^{\pm\b}_Ea_\b
\end{equation}
in terms of the $E$-adapted spin-frame (\ref{eq:uE}).
That $V^{(1,DV)}_{{\rm bh},\a}$ and its antiholomorphic counterpart solve \eqref{eq:qV1} comes from the following key property
\begin{equation}
\partial^a_{[\a}\left(
(a^+_{E}-a^-_{E})_{\b]}
f(a\vark^La)
\right)
=0
\,,
\end{equation}
for any function $f$.
The spacetime connection in Didenko-Vasiliev gauge is defined as
\begin{align}
U^{(1,DV)}_{\rm bh}
=&
\frac{b}{4\sqrt{\eta^2}r^2}
d\left(\frac{\vark^L}r\right)^{\a\b}a^+_{E,\a}a^+_{E,\b}
\int_0^1d\tau(1-\tau)\exp\left(-\frac{\tau}{2\eta}a(\vark^L)^{-1}a\right)
-\hc
\nonumber\\&
+W^{(1,DV,DV)}_{\rm bh}
\label{eq:U1DV}
\,.
\end{align}
It solves \eqref{eq:qU1+DV1} because of the property that
\begin{equation}
\label{eq:DV qU1+DV1}
\acd V^{(1,DV)}_{{\rm bh},\a} -\partial^z_\a U^{(1,DV)}_{\rm bh}
=
\int_{0}^1d\tau\frac{d}{d\tau}\left(
\tau(\tau-1)
\exp\left(-\frac{\tau}{2\eta}a(\vark^L)^{-1}a\right)
\right)
=0\,.
\end{equation}
The latter result is a consequence of the identity
\begin{equation}
\acd\left(
\Psi^{(1)}_{\rm bh}
f(x,a)
\right)
=
\Psi^{(1)}_{\rm bh}
\left(
dx^\mu\partial^x_\mu-\frac{\eta}{2}d(\vark^L)^{\a\b}\partial^a_\a\partial^a_\b
\right)
f(x,a)
\,,
\end{equation}
where $\partial^x_\mu$ only acts on the explicit $x$-dependence of $f(x,a)$, and not on the one in $a$.

\paragraph{COMST.}
We start from \eqref{eq:COMST eom step}.
The first term vanishes, since the application of $Y$-derivatives on the middle hand side of \eqref{eq:DV qU1+DV1} can only bring down positive powers of $\tau$,
hence not spoiling the vanishing.
The second term works as in \eqref{eq:bh COMST fact}, where here $v(\ytb)$ should be replaced by
\begin{equation}
w(\ytb)
:=
\left[t\exp\left(
\frac{\eta}{2}(1-t)\ytb\vark^L\ytb
\right)\right]_{t=0}^1
=1
\,.
\end{equation}
The vanishing after taking derivatives is again consistent with the prescription of taking them before evaluating the boundary term,
as each $\ytb$-derivative will bring down positive powers of $(1-t)$.

\paragraph{Relaxed Vasiliev gauge.}
The gauge parameter is constructed using Eq.\eqref{eq:H = int VL + h}
\begin{align}
H^{(1,DV\to G)}_{\rm bh}
=&
\frac{b}{4\sqrt{\eta^2} r}
\Psi^{(1)}_{\rm bh}
\int_0^1d\tau\int_0^1dt
\left(
\frac{t}{r}z\varkappa^Lz
+i\eta z(r+\varkappa^L)\ytb
\right)
\times\nonumber\\&\times
\exp\left(
-\tfrac{\tau}{2\eta}(tz-i\eta\ytb\vark^L)(\vark^L)^{-1}(tz+i\eta\varkappa^L\ytb)
\right)
\nonumber\\&
+h^{(1,DV\to G)}_{\rm bh}
+H^{(1,DV\to G)}_{\rm bh,2}
\,.
\end{align}
It is regular as the integrand and the volume on which one integrates are.
The particular connection contains an additional contribution with respect to the previous one:
\begin{equation}
W_{\rm bh,part}^{(1)}
:=
\left.\acd H^{(1,DV\to G)}_{\rm bh}\right\vert_{z=0}
+\left.U^{(1,DV)}\right\vert_{z=0}
\,.
\end{equation}
The first term benefits from $H^{(1,DV\to G)}_{\rm bh}$ being analytic around $\ytb=0$ while the second one inherits from the regularity of the particular part of $U^{(1,DV)}$ that was given in Eq.\eqref{eq:U1DV}.
This makes again $h^{(1,DV\to G)}_{\rm bh}=0$ an acceptable choice.

\section{A proposal for asymptotically anti-de Sitter geometries}
\label{sec:NL}

In what follows, we shall examine the following aspects of the Fefferman--Graham-like expansions of AAdS higher spin geometries:
\begin{enumerate}[label=\roman*),ref=\roman*]
\item\label{it:asymp bc} \emph{AAdS boundary conditions}: taking ${\cal X}'_4$ as in Eq.\eqref{eq:topcalx4}, the full master fields are required to reduce to the AdS vacuum plus free master fields consisting of properly unfolded Fronsdal fields in a tubular neighborhood of $\partial{\cal X}'_4\times {\cal Z}_4$ following the perturbative procedure consisting of steps (\ref{it:step AAdS}--\ref{it:step h}) on page \pageref{it:step AAdS}, which determines perturbatively defined equivalence classes of gauge functions and zero-form integration constants as in Eqs. (\ref{eq:AsGlu Phi}--\ref{eq:AsGlu U});
\item\label{it:max sub}
\emph{Maximal subtraction scheme}: we propose to organize the perturbative expansion obtained in (\ref{it:asymp bc}) by taking the asymptotically free master fields to be given by the linearized master fields, which determines the equivalence classes of sub-leading data in terms of the linearized data as in Eq.\eqref{eq:max sub CH};
\item\label{it:dual cond} \emph{Dual boundary condition}: the requirement that the Weyl zero form belongs to ${\cal E}$ may require further conditions narrowing down the classes of gauge functions, i. e. the Cartan gauge group ${\cal G}$, thereby making it possible to implement (ii) successfuly and for classical observables to be class functions;
\item\label{it:fct class Phi}
\emph{Finite on-shell action}: it may be required to further constrain zero-form integration constants $\Phi^{\prime(n)}$ so that the on-shell action \eqref{eq:freeE} is finite (viewed as a functional of $\Phi^{\prime(1)}$).
\end{enumerate}
To spell out the above scheme in more detail, we start in Section \ref{ssec:NL} by pushing the procedure of Section \ref{sec:COMST} to interacting orders in perturbation.
We then impose the AAdS boundary condition (i) in Section \ref{ssec:AAdS}, after which we turn to the maximal substraction scheme (ii) in Section \ref{ssec:maxsubscheme}.
In Section \ref{ssec:freeE}, we examine (iv) and comment on the potential role played by (iii).

\subsection{Perturbatively defined solution spaces}
\label{ssec:NL}

Spaces of solutions to Eqs.(\ref{eq:VE Phi(x,z)},\,\ref{eq:VE UV}) can be obtained by expanding 
perturbatively around the AdS${}_4$ vacuum \eqref{eq:AdS}, viz.
\begin{align}
\Phi
&=
\sum_{n=1}^\infty\Phi^{(n)}
\,,&
V
&=
\sum_{n=1}^\infty V^{(n)}
\,,&
U
&=
\Omega + \sum_{n=1}^\infty U^{(n)}
\,.
\end{align}
which leads to the following perturbatively defined equation systems:
\begin{align}
\label{eq:qPhin}
q\Phi^{(n)}
&=-
\sum_{k=1}^{n-1}\picomm{V^{(k)}}{\Phi^{(n-k)}}
\,,\\
\label{eq:DPhin}
\tcd\Phi^{(n)}
&=-
\sum_{k=1}^{n-1}\picomm{U^{(k)}}{\Phi^{(n-k)}}
\,,\\
\label{eq:qVn}
qV^{(n)}
+\Phi^{(n)}\star J
&=-
\sum_{k=1}^{n-1}V^{(k)}\star V^{(n-k)}
\,,\\
qU^{(n)}
+\acd V^{(n)}
&=-
\sum_{k=1}^{n-1}\staracomm{V^{(k)}}{V^{(n-k)}}
\,,\\
\label{eq:DUn}
\acd U^{(n)}
&=-
\sum_{k=1}^{n-1} U^{(k)}\star U^{(n-k)}
\,.
\end{align}
\paragraph{$n^{\rm th}$-order solution space.}
Assuming the solution is known up to order $n-1$ in classical perturbation theory, a particular solution $(\Phi^{(n,A)}_{\rm l.o.},\,V^{(n,A)}_{\rm l.o.},\,U^{(n,A)}_{\rm l.o.})$ to the $n^{\rm th}$ order equation system can be constructed from the moduli of orders $n'<n$.
Therefore, as was discussed in Section \ref{sec:COMST}, the $n^{\rm th}$ order solution space is given by
\begin{align}
\Phi^{(n)}
&=
C^{(n,A)}+\Phi^{(n,A)}_{\rm l.o.}
\,,\\
V^{(n)}
&=
{\cal V}^{(n,A)}[C^{(n,A)}]+qH^{(n,A)}+V^{(n,A)}_{\rm l.o.}
\,,\\
U^{(n)}
&=
{\cal U}^{(n,A)}[C^{(n,A)}]+\acd H^{(n,A)}+U^{(n,A)}_{\rm l.o.}
\,,
\end{align}
where $C^{(n,A)}$ are homogeneous solutions annihilated by both $q$ and $\tcd$,
$H^{(n,A)}$ is a gauge function\footnote{%
The part of $U^{(n,A)}$ that is in the $q$-cohomology, corresponding to $W^{(1,A,B)}$ in Eq.\eqref{eq:WAB} has been split between a particular solution included in ${\cal U}^{(n,A)}[C^{(n,A)}]$ and a fluctuating pure gauge part in $\acd H^{(n,A)}$.
},
and ${\cal V}^{(n,A)}$ and ${\cal U}^{(n,A)}$ are linear functionals obeying
\begin{align}
q{\cal V}^{(n,A)}[f^{(n)}]+f^{(n)}\star {J}
&=0
\,,&
q{\cal U}^{(n,A)}[f^{(n)}]+\acd{\cal V}^{(n,A)}[f^{(n)}]
&=0
\,,&
\acd{\cal U}^{(n,A)}[f^{(n)}]
&=0\,,
\end{align}
for any symbol $f^{(n)}$ such that $qf^{(n)}=\tcd f^{(n)}=0$.
The perturbative\footnote{%
As the geometry is non-commutative, the moduli of the full theory also include a non-trivial flat connection \eqref{eq:non trivial flat connection},
that is here already fixed by the choice of the vacuum $V^{(0,A)}=0$.
} moduli are hence $H^{(n,A)}$ and $C^{\prime(n,A)}$, or equivalently $H^{(n,A)}$ and $\Psi^{\prime(n,A)}$, where $C^{\prime(n,A)}$ and $\Psi^{\prime(n,A)}$ are defined as
\begin{align}
\label{eq:def Psi'n}
C^{(n,A)}&=:L^{-1}\star C^{\prime(n,A)}\star\pi(L)
\,,&
\Psi^{\prime(n,A)}:&=C^{\prime(n,A)}\star\k_y
\,,&
dC^{\prime(n,A)}&=d\Psi^{\prime(n,A)}=0
\,,
\end{align}
where $L$ is the AdS gauge function \eqref{eq:AdS L}.
In the above, $A$ labels an infinite sequence of (possibly different) perturbatively defined resolution operators (rather than a single resolution operator).
\paragraph{Mapping between different resolution schemes.}

Starting from the $A$-gauge, in which
\begin{align}
\label{eq:Phin A}
\Phi^{(n,A)}
&=
C^{(n,A)}+\Phi^{(n,A)}_{\rm l.o.}
\,,\\
V^{(n,A)}
&=
{\cal V}^{(n,A)}[C^{(n,A)}]+V^{(n,A)}_{\rm l.o.}
\,,\\
\label{eq:Un A}
U^{(n,A)}
&=
{\cal U}^{(n,A)}[C^{(n,A)}]+U^{(n,A)}_{\rm l.o.}
\,,
\end{align}
we may reach another gauge, $G$ say, characterized by an infinite sequence of resolution operators as well, by means of a perturbatively-defined gauge transformation \eqref{eq:gF}, viz.
\begin{align}
M^{(A\to G)}
&=
1+\sum_{n=1}^\infty H^{(n,A\to G)}
\,,\\
\label{eq:Phi GfromA}
\Phi^{(G)}
&=
\left(M^{(A\to G)}\right)^{-1}\star \Phi^{(A)}\star\pi\left(M^{(A\to G)}\right)
\,,\\
V^{(G)}
&=
\left(M^{(A\to G)}\right)^{-1}\star V^{(A)}\star M^{(A\to G)}
+\left(M^{(A\to G)}\right)^{-1}\star qM^{(A\to G)}
\,,\\
\label{eq:U GfromA}
U^{(G)}
&=
\left(M^{(A\to G)}\right)^{-1}\star U^{(A)}\star M^{(A\to G)}
+\left(M^{(A\to G)}\right)^{-1}\star dM^{(A\to G)}
\,.
\end{align}
The solution $G$ is still general, although we only have modified the gauge function and not the integration constants $\Psi^{\prime(n,G)}$.
Indeed, a change of initial data $C^{(n,G)}$ is equivalent to a modification of $C^{(n,A)}$, that was so far kept arbitrary\footnote{%
A solution with independent $n$th order homogeneous integration constants for the Weyl zero-form can be obtained from an all-order solution defined in terms of a single initial datum $C^{(1)}$ by replacing $C^{(1)}$ by $\sum_{n=1}^\infty C^{(n)}$ and re-expanding perturbatively.}
The perturbative expansion of the solution $G$ can be obtained by plugging Eqs.(\ref{eq:Phin A}--\ref{eq:Un A}) into Eqs.(\ref{eq:Phi GfromA}--\ref{eq:U GfromA}), which yields
\begin{align}
\label{eq:Phin GfromA}
\Phi^{(n,G)}
&=
C^{(n,A)}+\Phi^{(n,G)}_{\rm l.o.}
\,,\\
\label{eq:Vn GfromA}
V^{(n,G)}
&=
{\cal V}^{(n,A)}[C^{(n,A)}]+qH^{(n,A\to G)}+V^{(n,G)}_{\rm l.o.}
\,,\\
\label{eq:Un GfromA}
U^{(n,G)}
&=
{\cal U}^{(n,A)}[C^{(n,A)}]+\acd H^{(n,A\to G)}+U^{(n,G)}_{\rm l.o.}
\,,
\end{align}
where $\Phi^{(n,G)}_{\rm l.o.}$, $V^{(n,G)}_{\rm l.o.}$ and $U^{(n,G)}_{\rm l.o.}$ form a particular solution that only depends on the lower order moduli contained in $C^{(n'<n,A)}$ and $H^{(n'<n,A\to G)}$.
The perturbative moduli of this solution $G$ are hence $C^{(n,A)}$ and $H^{(n,A\to G)}$, that can be constraint to obey boundary conditions, as will be discussed in Section \ref{ssec:AAdS} for the conditions discussed on page \pageref{it:asymp bc}.

\paragraph{Gauge function.}
If the spacetime connection $U^{(L)}$ is uncorrected in the $A$-gauge, which we denote in this case by $(L)$, viz.
\begin{align}
\Phi^{(L)}&=L^{-1}\star \Phi'\star\pi(L)
\,,&
V^{(L)}&=L^{-1}\star V'\star L
\,,&
U^{(L)}&=\Omega
\,,
\end{align}
where $(\Phi',V')$ is a solution to Eq.\eqref{eq:VE'} and $L$ is an AdS gauge function \eqref{eq:AdS L};
then it follows that the field configuration in the physical gauge $G$ is given by
\begin{align}
\Phi^{(G)}&=G^{-1}\star \Phi'\star\pi(G)
\,,&
V^{(G)}&=G^{-1}\star V'\star G+G^{-1}\star qG
\,,&
U^{(G)}&=G^{-1}\star dG
\,,
\end{align}
with the full gauge function 
\begin{equation}
\label{eq:GfromL}
G=L\star\left(1+ \sum_{n\geqslant 1} H^{(n,L\to G)}\right)
\,.
\end{equation}

\subsection{AAdS boundary conditions}
\label{ssec:AAdS}   
Letting $r$ be a coordinate on ${\cal X}'_4$ such that $r(N)=\infty$, we expand the full master fields $(\Phi,U,V)$ in powers of $1/r$ in a tubular neighbourhood of $\partial{\cal X}_4\times{\cal Z}_4$ using a basis ${\cal B}$ for ${\cal A}({\cal Y}_4)$ that is of $O(1)$.
The AAdS boundary conditions in Eq. \eqref{it:asymp bc} on page \pageref{it:asymp bc} are then equivalent\footnote{%
It is a priori not equivalent to asking
\begin{align*}
&&
q{\Phi}^{(G)}
&=O_{\cal B}(1/r)
\,,&
\tcd{\Phi}^{(G)}
&=O_{\cal B}(1/r)
\,,\\
q{V}^{(G)}+{\Phi}^{(G)}\star {J}
&=O_{\cal B}(1/r)
\,,&
q{U}^{(G)}+\acd{V}^{(G)}
&=O_{\cal B}(1/r)
\,,&
\acd{U}^{(G)}
&=O_{\cal B}(1/r)
\,.
\end{align*}
These relations are consequences of Eqs.(\ref{eq:AsVE Phi},\,\ref{eq:AsVE UV}) if and only if the basis $\cal B$ is such that the derivatives $q$, $\acd$ and $\tcd$ act faithfully on the sub-leading part, viz.
\begin{align*}
q:O_{\cal B}(1/r)\to O_{\cal B}(1/r)
&\,,&
\acd:O_{\cal B}(1/r)\to O_{\cal B}(1/r)
&\,,&
\tcd:O_{\cal B}(1/r)\to O_{\cal B}(1/r)
&\,.
\end{align*}
} 
to writing
\begin{align}
\label{eq:AAdS bc}
\Phi^{(G)}&=\widetilde{\Phi}^{(G)}+O_{\cal B}(1/r)
\,,&
V^{(G)}&=\widetilde{V}^{(G)}+O_{\cal B}(1/r)
\,,&
U^{(G)}&=\widetilde{U}^{(G)}+O_{\cal B}(1/r)
\,,
\end{align}
where $O_{\cal B}(1/r)$ stands for forms on the tubular neighbourhood of $\partial{\cal X}_4\times{\cal Z}_4$ that are sub-leading in the $1/r$ expansion; and $\widetilde{\Phi}^{(G)}$, $\widetilde{V}^{(G)}$ and $\widetilde{U}^{(G)}$ form a solution to the linearized Vasiliev equations (\ref{eq:qPhi1}--\ref{eq:DU1})
\bea
\label{eq:AsVE Phi}
&
q\widetilde{\Phi}^{(G)}
= 0
\,, \qquad \quad
\tcd\widetilde{\Phi}^{(G)}
= 0  
\,,&\\
\label{eq:AsVE UV}
& q\widetilde{V}^{(G)}+\widetilde{\Phi}^{(G)}\star {J}
=0
\,,\qquad
q\widetilde{U}^{(G)}+\acd\widetilde{V}^{(G)}
=0
\,,\qquad 
\acd\widetilde{U}^{(G)}
=0\,, &
\eea
that encode free unfolded Fronsdal fields.
Expanding the AAdS boundary conditions in Eq. \eqref{eq:AAdS bc} in classical perturbation theory and using (\ref{eq:Phin GfromA}--\ref{eq:Un GfromA}), one has
\begin{align}
\label{eq:AsGlu Phi}
\widetilde\Phi^{(n,G)}
&=
C^{(n,A)}+\Phi^{(n,G)}_{\rm l.o.}
+O_{\cal B}(1/r)
\,,\\
\widetilde{V}^{(n,G)}
&=
{\cal V}^{(n,A)}[C^{(n,A)}]+qH^{(n,A\to G)}+V^{(n,G)}_{\rm l.o.}
+O_{\cal B}(1/r)
\,,\\
\label{eq:AsGlu U}
\widetilde{U}^{(n,G)}
&=
{\cal U}^{(n,A)}[C^{(n,A)}]+\acd H^{(n,A\to G)}+U^{(n,G)}_{\rm l.o.}
+O_{\cal B}(1/r)
\,.
\end{align}
where thus $(\widetilde\Phi^{(n,G)},\widetilde{V}^{(n,G)},
\widetilde{U}^{(n,G)})$ obey the free master field equations (\ref{eq:AsVE Phi},\,\ref{eq:AsVE UV}) for each separate value of $n$.

\paragraph{Perturbative implementation scheme.} Assuming that the AAdS boundary conditions are satisfied up to order $m$ in classical perturbation theory with $H^{(m,A\to G)}$ containing a term $H_2^{(m,A\to G)}$ that is an arbitrary solution to Eq.\eqref{eq:H2atZ=0}, we propose to 
\begin{enumerate}[label=\arabic*),ref=\arabic*]
\item\label{it:step AAdS} 
Use $H_2^{(m,A\to G)}$ and $C^{(m+1,A)}$ to impose the AAdS boundary conditions\footnote{%
The dual boundary condition (\ref{it:fct class Phi}) holds provided that the right hand side of Eq.\eqref{eq:Phin GfromA} belongs to ${\cal E}({\cal X}_4)$ as well.} at order $m+1$, that is, to ensure Eq. (5.25), or, equivalently, that the right hand sides of (\ref{eq:AsGlu Phi}--\ref{eq:AsGlu U}) obey the linearized field equations (\ref{eq:AsVE Phi}--\ref{eq:AsVE UV}).
In particular, one has
\begin{align}
\Phi^{(m+1,G)}_{\rm l.o.}
=
L^{-1}\star\Psi^{\prime(m+1,G)}_{l.o.}\star\k_y\star\pi(L)
+{\cal O}_{{\cal B}}(1/r)
\,,\\
d\Psi^{\prime(m+1,G)}_{l.o.}=
q\Psi^{\prime(m+1,G)}_{l.o.}=0
\,;
\end{align}
\item\label{it:step UVreg}
Use $H^{(m+1,A\to G)}$ to eliminate the possible singular part in $Z$ of $\widetilde{V}^{(m+1,G)}$ and $\widetilde{U}^{(m+1,G)}$;
\item\label{it:step ZV}
Use the part of $H^{(m+1,A\to G)}$ that is analytic in $Z$ to impose the relaxed gauge condition \eqref{eq:ZV=O(Z2)}, that is,
\be 
\imath_{\vec E}\widetilde{V}^{(m+1,G)}
\in
\vec E \ker\left({\cal P}^{(G)}\acd\right)\,,
\ee
which fixes $H^{(m+1,A\to G)}$ up to $h^{(m+1,A\to G)}\in \ker q$ and $H_2^{(m+1,A\to G)}\in\ker\left({\cal P}^{(G)}\acd\right)\,$;
\item\label{it:step h}
Use $h^{(m+1,A\to G)}$ from (3) to ensure the regularity in $Y$ of the asymptotically defined connection
\begin{equation}
\widetilde{W}^{(m+1,G)}:={\cal P}^{(G)}\widetilde{U}^{(m+1,G)}\,.
\end{equation}
\end{enumerate}
Provided that step \eqref{it:step AAdS}, which is non-trivial, can be taken, the success of steps (\ref{it:step UVreg} -- \ref{it:step h}) is guaranteed\footnote{%
The success of step \eqref{it:step UVreg} is in principle guaranteed by the one of step \eqref{it:step AAdS}, see footnote \ref{ftn:VU reg}.}, as they are equivalent to the procedure detailed in Section \ref{ssec:COMST from fact}.
Finally, the fact that $H_2^{(m+1,A\to G)}$ remains undetermined at order $m+1$ allows the procedure to be repeated at order $m+2$.

\subsection{Maximal subtraction scheme} \label{ssec:maxsubscheme}

Asssuming that the above procedure has been carried out, one may adopt a \emph{maximal subtraction scheme} by requiring the leading order in the $1/r$ expansion to coincide with the first order in classical perturbation theory around the AdS background as follows\footnote{%
\label{ftn:ho FG}Imposing this boundary condition amounts to adopting a Fefferman-Graham-like scheme in which each order in classical perturbation theory would be sub-leading in the $O(1/r)$ expansion with respect to the previous order (and not just with respect to the leading order).}:
\begin{enumerate}[label=\alph*),ref=\alph*]
\item The background is the AdS${}_4$ solution \eqref{eq:AdS};
\item The linearised configuration is obtained from a given initial datum $C^{(1)}$ following the procedure of Section \ref{sec:COMST};
\item At order $n>1$, the configurations are given by the method above, where in step \eqref{it:step AAdS}, Eq.\eqref{eq:AAdS bc} is replaced by the following stronger condition:
\begin{align}
\label{eq:maxsub bc}
\Phi^{(n,G)}&=O_{\cal B}(1/r)
\,,&
V^{(n,G)}&=O_{\cal B}(1/r)
\,,&
U^{(n,G)}&=O_{\cal B}(1/r)
\,.
\end{align}
\end{enumerate}
If the maximal subtraction scheme has been implemented up to a given order $m$ in classical perturbation theory, the admissibility of \eqref{eq:maxsub bc} at order $m+1$ is guaranteed by that of the AAdS boundary conditions in Eq. \eqref{eq:AAdS bc} at the same order.
To show this, we consider a particular solution $\{C^{(m+1,A)}_{\rm part.},\,H^{(m+1,A\to G)}_{\rm part.},\,H^{(1,A\to G)}_2\}$ 
with corresponding asymptotic fields $\{\widetilde{\Phi}^{(m+1,G)}_{\rm part.},\,\widetilde{V}^{(m+1,G)}_{\rm part.},\,\widetilde{U}^{(m+1,G)}_{\rm part.}\}$ properly encoding unfolded Fronsdal fields.
Since both $\{\widetilde\Phi^{(m+1,G)}_{\rm part.},\,\widetilde{V}^{(m+1,G)}_{\rm part.},\,\widetilde{U}^{(m+1,G)}_{\rm part.}\}$ and $\{\widetilde\Phi^{(m+1,G)}_{\rm part.},\,{\cal V}^{(m+1,A)}[\widetilde\Phi^{(m+1,G)}_{\rm part.}],\,{\cal U}^{(m+1,A)}[\widetilde\Phi^{(m+1,G)}_{\rm part.}]\}$ solve the linearized equations (\ref{eq:AsVE Phi},\,\ref{eq:AsVE UV}), these two solutions are related by a gauge parameter $\widetilde{H}^{(m+1,A\to G)}_{\rm part.}$, that is,
\begin{align}
\widetilde{V}^{(m+1,G)}_{\rm part.}
&=
{\cal V}^{(m+1,A)}[\widetilde\Phi^{(m+1,G)}_{\rm part.}]
+q\widetilde{H}^{(m+1,A\to G)}_{\rm part.}
\,,\\
\widetilde{U}^{(m+1,G)}_{\rm part.}
&=
{\cal U}^{(m+1,A)}[\widetilde\Phi^{(m+1,G)}_{\rm part.}]
+\acd\widetilde{H}^{(m+1,A\to G)}_{\rm part.}
\,.
\end{align}
Therefore, the configuration 
\begin{align}
\label{eq:max sub CH}
C^{(m+1,A)}_{\rm max.sub.}
&=
C^{(m+1,A)}_{\rm part.}-\widetilde{\Phi}^{(m+1,A)}_{\rm part.}
\,,&
H^{(m+1,A\to G)}_{\rm max.sub.}
&=
H^{(m+1,A\to G)}_{\rm part.}-\widetilde{H}^{(m+1,A\to G)}_{\rm part.}
\,,
\end{align}
satisfy the condition \eqref{eq:maxsub bc}, as can be seen by plugging it into Eqs.(\ref{eq:AsGlu Phi}--\ref{eq:AsGlu U}), which concludes the proof.
The maximal subtraction \eqref{eq:max sub CH} translates to the integration constant $\Psi^{\prime(m+1,A)}$ as
\begin{equation}
\Psi^{\prime(m+1,A)}=-\Psi^{\prime(m+1,G)}_{l.o.}
\,,
\end{equation}
thereby completely fixing it.
By applying, this relation iteratively, one may express $\Psi^{\prime(n,A)}$ as an $n$-linear functional of $\Psi^{\prime(1)}$ (valued in ${\cal A}^{(n)}({\cal Y}_4)$).
However, $H^{(m+1,A\to G)}$ is only determined by Eq.\eqref{eq:max sub CH} up to a piece $H^{(m+1,A\to G)}_{1/r}$ that decays fast enough so as to preserve eq.\eqref{eq:maxsub bc}.

\subsection{On-shell action}  
\label{ssec:freeE}

\paragraph{Topological vertex operators.} 

Following the adaptation of the AKSZ approach to non-commutative quasi-topological field theories \cite{Boulanger:2012bj}, the FCS formulation of HSG gives rise to a partition function given in the saddle-point approximation by a sum \eqref{QTFTZ} over classical field configurations weighted by $\exp(iK)$, where $K$ is an on-shell action.
The latter is a classical observable that can be resolved off-shell as a functional whose total variation vanishes on-shell, referred to as a topological vertex operator \cite{Sezgin:2011hq}, which can be added to the BV master action of AKSZ type without ruining the nilpotency of the BRST operator.
Thus, for higher spin representations in which $K$ is a (positive) definite function on the classical moduli space (at least to the leading order in classical perturbation theory), $e^{iK}$ regularizes the AKSZ partition function such that the QTFT functor can assign infinite-dimensional state spaces to boundaries. 

\paragraph{On-shell action from twistor space Chern class.}
In the case of the FCS model, the topological vertex operators are (duality-extended) Chern classes \cite{Boulanger:2015kfa}.
On ${\cal X}_4\times {\cal Z}_4$ of topology $S^1 \times S^3\times S^2\times \overline{S}^2$, a natural on-shell action is the second Chern class on ${\cal Z}_4$, 
\begin{align}
\label{eq:freeE}
K&=
\oint_{{\cal Z}_4} {\rm Tr}_{{\cal A}({\cal Y}_4)}\left(F\star F\right)
\,,&
F:&=\diffhat A+A\star A
\,.
\end{align}
Upon using the field equations \eqref{eq:VE}, the identities (\ref{eq:pifromJ},\,\ref{eq:J^2}), the reality condition \eqref{eq:RC}, and the relation \eq{eq:d2z} between $dz^\a dz_\a$ in $J$ and the integration measure $d^2z$ as well as their analogues in the anti-holomorphic sector, the resulting on-shell action is a bilinear function of the Weyl zero-form $\Phi$, viz.
\begin{align}
K
&=
\oint_{{\cal Z}_4} {\rm Tr}_{{\cal A}({\cal Y}_4)}\left(
\Phi\star\Phi^{\dagger}\star J^{\star2}\right)
\\&=
-\frac{1}{4}
\int d^4Z\Tr_{{\cal A}({\cal Y}_4)}\big(
(\Phi\star\kappa)\star(\Phi\star\kappa)\star(\k\star\bar\k)\big)\ ,
\end{align}
that is positive definite for particle states.
This functional is a particular case of the Wilson loop observables, or zero-form charges,  discussed in Section \ref{ssec:OWL}.
More precisely, in the language of Eq.\eqref{eq:def ZFC}, it reads
\begin{equation}
K=
-\pi^2
I_{2,1}(0)
\,.
\end{equation} 
In particular, when applied on localizable states, it has a series of properties, discussed in Section \ref{ssec:OWL}, that are expected from an on-shell action.
As a zero-form charge, it is constructed to be independent of the gauge function $M$ in Eq.\eqref{eq:gF}, hence depending only on the integration constants $\Psi^{\prime(n,A)}$ for a perturbative solution submitted to the AAdS boundary conditions.
In particular, if the starting solution $A$ is the factorised solution (\ref{eq:Phi^m fact}--\ref{eq:U^m fact}), the on-shell action reads 
\begin{equation}
K=
-\pi^2\sum_{n,n'}
\Tr_{{\cal A}({\cal Y}_4)}\big(
\Psi^{\prime(n,A)}\star\Psi^{\prime(n',A)}\star(\k_y\star\bar\k_y)
\big)
\,.
\end{equation}
Under the maximally subtractive boundary condition \eqref{eq:maxsub bc}, it is given by
\begin{equation}
\label{eq:K from lo}
K=
-\pi^2\sum_{n,n'}
\Tr_{{\cal A}({\cal Y}_4)}\big(
\Psi^{\prime(n,G)}_{\rm l.o.}\star\Psi^{\prime(n',G)}_{\rm l.o.}\star(\k_y\star\bar\k_y)
\big)
\,,
\end{equation}
where we recall that $\Psi^{\prime(n,G)}_{\rm l.o.}$ are $n$-linear functionals of $\Psi^{\prime(1)}$ fixed by the maximal subtraction scheme. 
Hence, if the maximal subtraction scheme triggers non-trivial integration constants $\Psi^{\prime(n,A)}$ of arbitrarily high orders, then $K$, despite being a bilinear function of $\Phi$, will have an indefinite perturbative expansion in $\Psi^{\prime(1)}$.

\paragraph{Finiteness.}
We expect that the AAdS boundary conditions determine equivalence classes $\left[H^{(n,A\to G)}\right]$ and $[\Psi^{\prime(n,A)}]$ of gauge functions and integration constants, respectively.
Although the maximal subtraction scheme \eqref{eq:max sub CH} trivialises the latter, the choice of representative for $[H^{(n,A)}]$ influences $\Psi^{\prime(n,A)}$ through its dependence on the lower order gauge functions $H^{(n',A\to G)}$ with $n'<n$.

Thus, in order for the on-shell action functional to be well-defined it must be a class function.
To achieve this, further reduction of the equivalence classes may be required for which the dual boundary condition (\ref{it:dual cond}) on page \pageref{it:fct class Phi} may serve a purpose.

\section{Wilson loop observables}
\label{sec:obs}

In this Section, we discuss the observables that are insensitive to any (small or large) gauge function.
In Section \ref{ssec:OWL}, we first review their construction as open Wilson lines in ${\cal Z}_4$, after which we re-interpret them as closed Wilson lines with an insertion of a transition function.
In Section \ref{ssec:OWL pert}, we discuss their perturbative computation in the factorised gauge described in Appendix \ref{App:fact}.
In particular, we isolate the first sub-leading correction, which is thus due to the interactions in $Z$-space, and verify that, just like the leading contribution, it is independent of the spin-frame \eqref{eq:spin frame} used to build the solution.
This result supports the usage of the regular prescription discussed in Section \ref{ssec:reg}.

We would like to remark that, as discussed in Section \ref{sec:VE}, the set of classical observables also contains functionals that are only invariant under small gauge transformations.
These can be constructed by choosing a structure group that is a proper subgroup of ${\cal G}$, as discussed in \cite{Sezgin:2011hq}, after which observables can be defined as integrals of globally defined (hence central) and on-shell closed composite spacetime $p$-forms built from the ingredients of the model\footnote{%
In \cite{Sezgin:2011hq}, it was shown that taking the structure group to be generated by $\pi$-odd parameters, the resulting de Rham cohomology consists of one complex element in every even form degree.
Although not stressed in \cite{Sezgin:2011hq}, also trivial elements can be used to construct classical observables, provided that they are integrated over open cycles and that boundary conditions are imposed to make them finite.}.
An alternative construction was provided in \cite{Didenko:2015pjo}, based on a modification of the Vasiliev system via the introduction of additional central forms in spacetime.
It would be interesting to study the relation between these two approaches.

\subsection{Construction}
\label{ssec:OWL}
The classical observables that are least sensitive to the topology of ${\cal X}_4$ are Wilson loops in ${\cal Z}_4$.
Using the regular computation scheme, these objects, which are zero-forms on ${\cal X}_4$, are invariant under (large) Cartan gauge transformations, hence on-shell de Rham closed \cite{Sezgin:2005pv}.
They are furthermore interpretable as extensive variables \cite{Colombo:2010fu,Iazeolla:2011cb}, thus serving as natural building blocks for the on-shell action and thereby also the higher spin amplitudes \cite{Colombo:2012jx}.
For these reasons, they are referred to interchangeably as Wilson loop observables, zero-form charges and quasi-amplitudes\footnote{%
The existence of the infinite tower of conserved zero-form charges reflects the formal integrability of Vasiliev's theory.
In principle, there is nothing preventing the existence of analogous conserved quantities for gravity with a finite cosmological constant; for further discussions, see \cite{Boulanger:2008up}.}.

\paragraph{Holonomy and closed Wilson loop.} To construct the Wilson loops, one starts from an oriented curve 
\be C_{p_0\to p_1}(M)=\left\{p(t)\in {\cal X}_4\times {\cal Z}_4|\,t\in [0,1]\right\}\,,\qquad p(0)=p_0\,,\quad p(1)=p_1\,,\ee
which consists of its projection ${\rm Proj}_{{\cal X}_4}(C_{p_0\to p_1}(M))$ to ${\cal X}_4$ (which is an ordinary open curve on a commuting space), and its projection ${\rm Proj}_{{\cal Z}_4}(C_{p_0\to p_1}(M))$ to ${\cal Z}_4$, given by a \emph{classical curve}
\be C(M)=\left\{Z^\au_{C}(t)|t\in [0,1]\right\}\,,\qquad Z^\au_{C}(0)=0\,,\qquad Z^\au_{C}(1)=M^\au\,,
\ee
emanating from a \emph{non-commutative base point}, that we may take to be ${\rm Proj}_{{\cal Z}_4}(p_0)$, to a commutative chiral point\footnote{%
Strictly speaking, we need to extend the hermitian conjugation map so as to act on $M^\au$ in accordance with real form of ${\cal Z}_4$.} $M^\au=(\mu^\alpha,-\bar\mu^{\dot\alpha})\in\Real^2\times \Comp^2$,  viz.
\be z^\alpha(p(t)):=z^\alpha+z^\alpha_{C}(t)\,,\qquad z^\alpha:=z^\alpha(p_0)\,.\ee
The resulting path ordered integral 
\be 
\label{eq:def holon}
H_{C_{p_0\to p_1}(M)}[A]={\rm P} \exp_\star \int_{C_{p_0\to p_1}(M)} A\,,\ee
defines a holonomy element\footnote{%
If the path passes from one chart of ${\cal X}_4$ to another, the corresponding transition function is inserted into the trace.} in ${\cal G}$, which transforms under Cartan gauge transformations as 
\be H_{C_{p_0\to p_1}(M)}\left[A^{(G)}\right]=(G|_{p_0})^{-1}\star H_{C_{p_0\to p_1}(M)}[A]\star (G|_{p_1})\,.\ee
Taking $p_0=p_1=p$, the resulting closed Wilson loop
\be W_{C(0)}[V]:= \int_{Z(p)\in {\cal Z}_4}\frac{d^4Z}{(2\pi)^2} {\rm Tr}_{{\cal A}({\cal Y}_4)}\,H_{C_{p\to p}(0)}[A]\,,\ee
is invariant under (large) Cartan gauge transformations, and hence independent of ${\rm Proj}_{{\cal X}_4}(p)$, as indicated above, provided that it can be evaluated using the regular scheme.
Moreover, from the fact that $\imath_{\vec v} F=0$ for $\vec v$ tangent to ${\cal X}_4$, it follows that $W_{C(M)}$ is independent of ${\rm Proj}_{{\cal X}_4}(C_{p\to p}(0))$, hence $U$, as also indicated above.

The closed Wilson loop does not encode any non-commutative structure in any non-trivial fashion.
On a non-commutative symplectic space, the exists, however, an alternative closed Wilson loop.

\paragraph{Wigner deformed oscillators.}
The symplectic non-commutativity of ${\cal Z}_4$ implies that
\be V^\# := T^\#\star (V+q)\star T^\#=- V-idZ^\au Z_\au\,,\qquad T^\#:=\kappa\bar\kappa
\,.\ee
is a gauge equivalent connection; thus, 
\be S:=i(V^\#-V)=dZ^{\au}(Z_\au-2i V_\au)\,,\ee
is an adjoint quantity, obeying 
\be S^\#=T^\#\star S\star T^\#=-S\,.\ee
From Vasiliev's equations it follows that $S$ generate a Wigner deformed oscillator algebra with (adjoint) deformation parameters
\be \chi:=\Phi\star\kappa\,,\qquad \bar\chi:=\Phi\star\bar{\kappa}\,,\ee
and central elements $dz^2$ and $d\bar z^2$, viz.\footnote{%
The deformations are non-trivial \cite{Vasiliev:1990en,Vasiliev:1990vu} since $(1-\chi)^{\star(-1/2)}\star (dz^\alpha S_\alpha) \notin \Omega_{[1]}({S}^2|j_z)$, as $dz^\alpha \kappa_z$ is excised from $\Omega_{[1]}({S}^2|j_z)$; see footnote \ref{ftn:chiral forms}.}
\be S\star S= i dz^2(1-b\chi)+i d\bar z^2(1-\bb\bar\chi)\,,\qquad [S,\chi]_\star=0\,;\ee
the remainder of the equations amount to the covariant constancy conditions
\be dS+[U,S]_\star=0\,,\qquad d\chi+[U,\chi]_\star=0\,.\ee

\paragraph{Open Wilson lines in ${\cal Z}_4$.}
Since the geometry is non-commutative,
 more zero-form observables can be constructed as traces of new adjoint quantities.
Indeed, using the lemma
\begin{equation}
f(Z)\star e^{iM^\au Z_\au}= e^{iM^\au Z_\au}\star f(Z-2M)
\,.
\end{equation}
and chosing a path $C_{p_0\to p_1}(M)$
such that ${\rm Proj}_{{\cal X}_4}(p_0)={\rm Proj}_{{\cal X}_4}(p_1)$ while ${\rm Proj}_{{\cal Z}_4}(C_{p_0\to p_1}(M))$ is left open, one has
\be H_{C_{p_0\to p_1}(M)}\left[A^{(G)}\right]\star e^{iM^\au Z_\au/2} =(G|_{p_0})^{-1}\star H_{C_{p_0\to p_1}(M)}[A]\star e^{iM^\au Z_\au/2}\star (G|_{p_0})\,.\ee
Thus, the open Wilson line 
\be W^{(\rm open)}_{C(M)}[A]:= \int_{Z(p)\in {\cal Z}_4}\frac{d^4Z}{(2\pi)^2} {\rm Tr}_{{\cal A}({\cal Y}_4)}\left( H_{C_{p_0\to p_1}(M)}[A]\star e^{iM^\au Z_\au/2}\right)\,,\ee
is invariant under large Cartan gauge transformations and independent of ${\rm Proj}_{{\cal X}_4}(C_{p_0\to p_1}(M))$, hence $U$, provided that it can be evaluated using the regular scheme. 

The contour $C(M)$ in the holonomy element $H_{C_{p_0\to p_1}(M)}$ can be deformed (while keeping $H_{C_{p_0\to p_1}(M)}$ fixed) at the expense of inserting curvature corrections along the path-ordered integral.
As was argued in \cite{Gross:2000ba}, this can be used to rewrite 
\be W^{(\rm open)}_{C(M)}[A]= \int_{Z(p)\in {\cal Z}_4}\frac{d^4Z}{(2\pi)^2} {\rm Tr}_{{\cal A}({\cal Y}_4)}\left( O_{C(M)}[A]\star H_{L_{p_0\to p_1}(M)}[A]\star e^{iM^\au Z_\au/2}\right)\,,\ee
where the holonomy is computed along a straight line $L_{p_0\to p_1}(M)$ in ${\cal Z}_4$ with classical piece
\begin{equation}
\label{eq:def L(M)}
L(M):=\{tM^\au\vert t\in[0,1]\}
\,,
\end{equation}
and the adjoint decoration
\be O_{C(M)}[V]:=\sum_{m,n,p} \Sigma_{n,p}^{\au(m)}[C(M)]  \left(S_\au\right)^{\star m}\star \chi^{\star n}\star \left(\k\kb\right)^{\star p} \,,\ee
encodes the geometry of $C(M)$.
As was shown in \cite{Bonezzi:2017vha}, 
\be 
\label{eq:OWL=e^iMS}
H_{L_{p_0\to p_1}(M)}[A]\star e^{iM^\au Z_\au/2}=e_\star^{iM^\au S_\au(p_0)/2}\,.\ee
It follows that
\be W_{C(M)}[A]=\sum_{m,n,p} \Sigma_{n,p}^{\au(m)}[C(M)] \int_{{\cal Z}_4} \frac{d^4Z}{(2\pi)^2} \Tr_{{\cal A}
({\cal Y}_4)}\Big(\left(S_\au\right)^{\star m}\star {{\cal V}}_n(M/2)\star \left(\k\kb\right)^{\star p}
\Big)
\,,\ee
where 
\be 
{{\cal V}}_n(M):=\chi^{\star n}\star
H_{L_{p_0\to p_1}(2M)}[A]\star e_\star^{iM^\au Z_\au}
\,,\qquad n=0,1,\dots\,.
\ee
Thus, the quantities
\be
\label{eq:def ZFC}
{\cal I}_{n,p}(M)
:=
\int_{{\cal Z}_4}  \frac{d^4Z}{(2\pi)^2} \Tr_{{\cal A}({\cal Y}_4)}
\Big(
{{\cal V}}_n(M/2)\star
\left(\k\kb\right)^{\star p}
\Big)
\,,\qquad 
p\in\{0,1\}
\,,\ee
and their derivatives with respect to $M$ form a basis of the ring of zero-form charges.

\paragraph{Closing the straight Wilson line.}
The decorated straight Wilson line \eqref{eq:def ZFC} can be rewritten in a way that can be straightforwardly generalised to more general differential Poisson manifolds, where the translation operator cannot be generated by means of star products.

To this end, one observes that $H_{L_{p_0\to p_1}(M)}[V]$ can be combined with 
\be T^\#\star H_{{L_{p_1\to p_0}(M)}}[V]\star T^\# =H_{{L_{p_1\to p_0}(M)}}[V^\#]
\,,\ee
to form a non-trivial closed Wilson loop
\be W_{L(M)\cup \overline{L(M)}}[V]:= \int_{{\cal Z}_4} \frac{d^4Z}{(2\pi)^2} {\rm Tr}_{{\cal A}({\cal Y}_4)}\left( H_{L_{p_0\to p_1}(M)}[V]\star H_{{L_{p_1\to p_0}(M)}}[V^\#]\right)\,,\ee
with classical closed path in ${\cal Z}_4$ given by $L(M)\cup \overline{ L(M)}$, which is invariant under large Cartan gauge transformations and independent of ${\rm Proj}_{{\cal X}_4}(p_0)$, provided that it can be evaluated using the regular scheme.
In other words, $W_{L(M)\cup \overline{L(M)}}$ is the trace of the holonomy resulting from parallel transporting an object from $Z$ to $Z+M$ along $L(M)$ using $V$, and then back along the same path using $V^\#$. 
The non-triviality of the closed Wilson loop follows from 
\begin{align}
& H_{L_{p_0\to p_1}(M)}[V]\star H_{{L_{p_1\to p_0}(M)}}[V^\#]=
H_{L_{p_0\to p_1}(M)}[V]\star\left(H_{L_{p_0\to p_1}(M)}[V^\#]\right)^{-1}
\nonumber\\&=
e_\star^{iM^\au S_\au(p_0)/2}\star e_\star^{-iM^\au Z_\au/2}\star \left(e_\star^{iM^\au S^\#_\au(p_0)/2}\star e_\star^{-iM^\au Z_\au/2}\right)^{-1}=e_\star^{iM^\au S_\au(p_0)}\,,
\end{align}
using $S^\#=-S$, from which it follows that
\be W^{(\rm open)}_{L(2M)}[V]=W_{L(M)\cup \overline{L(M)}}[V]\,,\ee
that is, the open Wilson line equals a folded closed Wilson loop consisting of two holonomy segments with two gauge equivalent connections, one in the opposite direction of the other. 

\subsection{Perturbative expansion}
\label{ssec:OWL pert}
The observables \eqref{eq:def ZFC} can be evaluated on solutions that are built perturbatively around the AdS${}_4$ vacuum \eqref{eq:AdS}.
Their vacuum value is
\begin{equation}
{\cal I}_{n,p}^{(0)}=\delta_{n,0}\Tr_{{\cal A}({\cal Y}_4)}\big((\k_y\bar\k_y)^{\star p}\big)
\,.
\end{equation}
The perturbative expansion of ${\cal I}_{n,p}$ actually starts at order $n$, 
where ${\cal I}_{n,p}^{(n)}$ is clearly a function of the initial datum $\Phi^{(1)}$ only.
The resolution used to solve the $Z$-space dependence of the fields are only expected to have an effect in sub-leading corrections to the zero-form charges.
Because they are constructed as traces of adjoint quantities, those observable are in principle invariant under all transformations of the form \eqref{eq:GIi}.
Although singular gauge functions can affect the cyclicity of the trace,
this is prevented if one uses the regular computational scheme presented in Section \ref{ssec:reg}.
In particular, the zero-form charges are independent of the gauge function $M$ in Eq.\eqref{eq:gF},
and can be written only in term of the virtual master fields $\Phi'$ and $V'$; hence only in terms of the virtual moduli $\Psi'$ and $\theta'$.
This means that the only perturbative moduli that they depend on are the zero-form integration constants $\Psi^{\prime(n)}$.

\paragraph{Localised states and extensive nature of zero-form charges.} In \cite{Colombo:2010fu,Iazeolla:2011cb}, it was argued that for initial data of the form $\Psi^{\prime(1)}=\sum_\xi \Psi^{\prime(1)}_\xi$, where  $\Psi^{\prime(1)}_\xi$ are linearized Weyl zero-forms that are \emph{localized} at mutually well-separated points $x_\xi\in {\cal X}_4$ (as can be achieved for particle and black hole states), a zero-form charge ${\cal I}_n$ evaluated on the corresponding full solution has a perturbative multi-body expansion of the form
\be {\cal I}_n(\Psi^{\prime(1)})=\sum_\xi {\cal I}_n(\Psi^{\prime(1)}_\xi)+ \sum_{\xi<\eta} \Delta(\Psi^{\prime(1)}_\xi,\Psi^{\prime(1)}_\eta)+\cdots\,,\ee
where ${\cal I}_n(\Psi^{\prime(1)}_\xi):={\cal I}_n(\Psi^{\prime(1)})|_{\forall\eta\neq \xi,\,\Psi^{\prime(1)}_\eta=0}$ is the full one-body zero-form charge, and the two-body corrections obey
\be \left|\Delta(\Psi^{\prime(1)}_\xi,\Psi^{\prime(1)}_\eta)\right|\ll \left|{\cal I}_n(\Psi^{\prime(1)}_\xi)\right|\,,\quad \left|{\cal I}_n(\Psi^{\prime(1)}_\eta)\right|\,.\ee
More generally, it is expected that the zero-form charges cluster decompose over well-separated subsets of localizable states.
Thus, when evaluated on a dilute gas of localizable states, the zero-form charges are extensive variables, hence natural building blocks for an on-shell action in the micro-canonical ensemble.

Thus, for localizable states, as first proposed in \cite{Engquist:2005yt}, the zero-form charges are expected to have classical perturbative expansions in terms of building blocks for higher spin amplitudes, referred to as quasi-amplitudes \cite{Colombo:2010fu}.
Indeed, the above properties have been verified in the Type A as well as Type B models \cite{Colombo:2012jx,Didenko:2012tv} for particle states, whose quasi-amplitudes reproduce the correlation functions of the free $U(N)$ vector model in the leading order.
In \cite{Bonezzi:2017vha}, this correspondence was enhanced to cyclic structures, supportive of the underlying topological open string picture \cite{Engquist:2005yt}.
\paragraph{Factorised solutions.}
The Wilson loop observables can be evaluated on the factorised solution (\ref{eq:Phi^m fact}--\ref{eq:U^m fact}).
In particular, one can write
\begin{align}
V^{(E+i\partial^Y)}
&=
\left.
\sum_{k=1}^\infty \sum_{\bar{k}=1}^\infty
\frac{1}{k!\bar{k}!}
\Psi^{\star k}\star
\bar\Psi^{\star \bar{k}}\star 
\left(\partial_\nu\right)^k\left(\partial_{\bar\nu}\right)^{\bar{k}}
V_{\nu,\bar{\nu}}\right\vert_{\nu=\bar\nu=0}
\nonumber\\
&=
\left.
\sum_{k,\bar{k}}
\frac{1}{k!\bar{k}!}
\Psi^{\star(k+\bar{k})}\star\left(\kappa_y\kb_{\yb}\right)^{\star\bar{k}}\star 
\left(\partial_\nu\right)^k\left(\partial_{\bar\nu}\right)^{\bar{k}}
V_{\nu,\bar{\nu}}
\right\vert_{\nu=\bar\nu=0}
\,,\\
V_{\nu,\bar{\nu}}(Z)
:&=
\sum_{k=1}^\infty \nu^{k} v_{k}(z)
+
\sum_{\bar{k}=1}^\infty \bar\nu^{\bar k} \bar{v}_{\bar{k}}(\zb)
\,.
\end{align}
The advantage of this formulation is that it allows to rewrite the holonomies \eqref{eq:def holon} as
\begin{align}
&
H_{C_{p_0\to p_1}(M)}\left[U+V\right]
\nonumber\\&=
\left.
\sum_{k=0}^\infty
\sum_{\bar k=0}^\infty
\frac{1}{k!\bar{k}!}
\Psi^{\star(k+\bar{k})}\star\left(\kappa_y\kb_{\yb}\right)^{\star\bar{k}}\star 
\left(\partial_\nu\right)^k\left(\partial_{\bar\nu}\right)^{\bar{k}}
H_{C_{p_0\to p_1}(M)}\left[U+V_{\nu,\bar{\nu}}\right]
\right\vert_{\nu=\bar\nu=0}
\,,
\end{align}
and in turn the zero-form charges \eqref{eq:def ZFC} as
\begin{align}
{\cal I}_{n,p}(M)
=&
\label{eq:ZFC fact}
\sum_{N=n}^\infty\sum_{P=0}^1
\alpha_{n,p;N,P}\Tr_{{\cal A}({\cal Y}_4)}\left(
\Psi^{\star N}\star\left(\kappa_y\kb_{\yb}\right)^{\star P}\right)
\\=&
\label{eq:ZFC fact'}
\sum_{N=n}^\infty\sum_{P=0}^1
\alpha_{n,p;N,P}\Tr_{{\cal A}({\cal Y}_4)}\left(
\Psi^{\prime\star N}\star\left(\kappa_y\kb_{\yb}\right)^{\star P}\right)
\,,\\
\label{eq:ZFC fact a}
\alpha_{n,p;N,P}
:=&
\sum_{\substack{0 \leqslant\,\bar{k}\,\leqslant\, N-n\\\bar{k}\equiv P-p\mod2}}
B_{n,p;N-n-\bar k,\bar k}
\,,\\
\label{eq:ZFC fact b}
B_{n,p;k,\bar k}
:=&
\frac{\left(\partial_\nu\right)^k\left(\partial_{\bar\nu}\right)^{\bar{k}}}{k!\bar{k}!}
\left.
\int_{{\cal Z}_4}  \frac{d^4Z}{(2\pi)^2}
H_{L_{p_0\to p_1}(M)}\left[V_{\nu,\bar{\nu}}\right]\star e^{iM^\au Z_\au/2}\star\k_z^{\star(p+n)}\star\kb_z^{\star p}
\right\vert_{\nu=\bar\nu=0}
\,.
\end{align}
The equivalence between the forms \eqref{eq:ZFC fact} and \eqref{eq:ZFC fact'} is in principle guaranteed by the regular presentation of the gauge functions.
Such an equivalence is moreover ensured for the traces (\ref{eq:Tr*},\,\ref{eq:Tr reg}) by the fact that their respective definitions are invariant under linear transformations of the $Y$ variables, such as the one in Eq.\eqref{eq:Lrot gen app}.

The lesson that comes out of this result is that for a given initial datum $\Psi^\prime$, the Wilson loop observables of the full solution are proportional to the ones of the linearised solution.
In particular, if the initial data are bulk-to-boundary propagators of particle states, and the solution is the factorised one,
those observables give correlators of the dual CFT to all orders in perturbation theory.
Because of the regular presentation of the gauge function $H^{(1,E+i\partial_Y\to G)}$ that relates them to a solution satisfying the COMST,
there is in principle a solution satisfying COMST and whose observables are the correlators of the dual CFT to all orders in perturbation theory.
However, imposing boundary conditions on the solution, such as the dual boundary conditions discussed in Section \ref{sec:NL}, can have a non trivial effect on the zero-form charges if they require to activate the higher order initial data $\Psi^{\prime(n,A)}$ for $n>1$.
\paragraph{Protection.}
Some of the coefficients $B_{n,p;k,\bar{k}}$ in Eq.\eqref{eq:ZFC fact b} are protected, in the sense that the contribution they get from terms that are sub-leading in perturbation theory vanish identically.
To show that, one starts by factorising them using Eq.\eqref{eq:OWL=e^iMS}
\begin{align}
B_{n,p;k,\bar k}
&=
\b_{p+n;k}\bar\b_{p,\bar k}
\,,\\
\b_{m;k}
:&=
\frac{\left(\partial_\nu\right)^k}{k!}
\left.
\int_{S^2}\frac{d^2z}{2\pi}
e_\star^{\tfrac{i}{2}\mu S_\nu}
\star\k_z^{\star m}
\right\vert_{\nu=0}
\,,\\
\bar\b_{p,\bar k}
:&=
\frac{\left(\partial_{\bar\nu}\right)^{\bar{k}}}{\bar{k}!}
\left.
\int_{\bar{S}^2}\frac{d^2\zb}{2\pi}
e_\star^{\tfrac{i}{2}\bar\mu \bar{S}_{\bar\nu}}
\star\kb_z^{\star p}
\right\vert_{\bar\nu=0}
\,,
\end{align}
in terms of the quantities
\begin{align}
S_{\nu,\a}(z)
:&=
z_\a-2i
\sum_{k=1}^\infty \nu^{k}\frac{\partial}{\partial dz^\a}v_{k}(z)
\,,&
\bar{S}_{\bar\nu,\ad}(\zb)
:&=
\zb_\ad-2i
\sum_{\bar{k}=1}^\infty \bar\nu^{\bar k}\frac{\partial}{\partial d\zb^\ad}\bar{v}_{\bar{k}}(\zb)
\,,
\end{align}
that star-commute with each other.
The protection argument concerns $\b_{1;k}$ for $k\geqslant1$ and $\bar{\b}_{1;\bar{k}}$ for $\bar{k}\geqslant1$, 
and is based on the bosonic projection \eqref{eq:BP}, that yields in particular
\begin{align}
\pi(S_{\nu,\a})
&=
-S_{\nu,\a}
\,,&
\bar\pi(\bar{S}_{\nu,\ad})
&=
-\bar{S}_{\nu,\ad}
\,.
\end{align}
The coefficients $\b_{1;k}$ (resp. $\bar\b_{1;\bar{k}}$) can be expanded in powers of $\mu$ (resp. $\bar\mu$) and read
\begin{align}
\b_{1;k}
&=
\delta_{k,0}
\,,&
\bar\b_{1;\bar{k}}
&=
\delta_{\bar{k},0}
\,.
\end{align}
While the value of the $M$-independent part is straightforward to compute,
the proof of the $\mu$-independence of $\b_{1;k}$ relies on the cyclicity of the trace and on Eqs.(\ref{eq:k^2=1},\,\ref{eq:pifromk}) as follows
\begin{align}
\int_{S^2}d^2z (\mu S_\nu)^{\star N}\star\k_z
=
-\int_{S^2}d^2z (\mu S_\nu)^{\star N-1}\star\k_z\star(\mu S_\nu)
=
-\int_{S^2}d^2z (\mu S_\nu)^{\star N}\star\k_z
\,,
\end{align}
and similarly for the $\bar\mu$-independence of $\bar\b_{1;\bar{k}}$.
\paragraph{Spin-frame independence.}
By their definition \eqref{eq:ZFC fact a}, the coefficients $\a_{n,p;N,P}$ might a priori depend on the specific solution for the auxiliary connection $V$.
However, they cannot depend on $V$, as ensured by their independence on the gauge function,
itself resulting from the regular prescription introduced in Section \ref{ssec:reg}. 
A test of said prescription would be to evaluate the zero-form charges on different solutions with the same Weyl tensors $C$,
and verify their matching.
In particular, one can compute them on the family of solutions derived in Appendix \ref{App:fact param},
and verify their independence on the spin-frames $u^{\pm}$ and $\ub^{\pm}$.
Asking for this independence to be manifest, this would mean that the $M$-dependence of the coefficients should be given by
\begin{equation}
\label{eq:sf indep}
\a_{n,p;N,P}(M)=
\a_{n,p;N,P}^{(0,0)}
+\a_{n,p;N,P}^{(1,0)}\delta^2(\mu)
+\a_{n,p;N,P}^{(0,1)}\delta^2(\bar\mu)
+\a_{n,p;N,P}^{(1,1)}\delta^4(M)
\,.
\end{equation}

The result (\ref{eq:ZFC fact}--\ref{eq:ZFC fact b}) is compatible with the leading part of the observables being completely determined by the initial datum $\Psi^\prime$, as the holonomy factor would be the one of the trivial connection $V_{0,0}=0$.
More precisely, the leading order results are given by
\begin{align}
\a_{2n,0;2n,0}
&=
64\pi^2\delta^4(M)
\,,&
\a_{2n+1,0;2n+1,0}
&=
8\pi\delta^2(\bar\mu)
\,,\\
\a_{2n,1;2n,1}
&=
1
\,,&
\a_{2n+1,1;2n+1,1}
&=
8\pi\delta^2(\mu)
\,,
\end{align}
the other coefficients being trivial.
To explore to first sub-leading order, one starts by using Eqs.(\ref{eq:def holon},\,\ref{eq:def L(M)},\,\ref{eq:V1z param}) to write
\begin{align}
&
H_{L_{p_0\to p_1}(M)}^{(1)}(V_{\nu,\bar\nu})
\\&=\nonumber
-\frac{b\nu}{2}\mu^\a\partial^\rho_\a
\left.
\int_0^1dt\int_{-1}^1\frac{ds}{1+s}
\exp\left(
\tfrac{i}{2}\tfrac{1-s}{1+s}(z+t\mu){\cal D}(z+t\mu)
+\tfrac{i}{1+s}\rho(z+t\mu)
\right)\right\vert_{\rho=0}
-\hc
\,.
\end{align}
As can be shown using Eqs.(\ref{eq:Tr exp param},\,\ref{eq:int exp=1/z2}), the non-trivial coefficients at this order are:
\begin{align}
\a_{2n,0;2n+1,0}
&=
b(2\pi)^2\delta^2(\mu)
\,,&
\a_{2n+1,0;2n+2,1}
&=
\bb
\,,\\
\a_{2n,0;2n+1,1}
&=
\bb(2\pi)^2\delta^2(\bar\mu)
\,,&
\a_{2n+1,1;2n+2,1}
&=
b
\,.
\end{align}
These results show that the spin frame independence \eqref{eq:sf indep} is reached at least at the first two orders in perturbation theory,
thereby supporting the regular prescription.

\section{Conclusions and outlook}
\label{sec:ccl}

In this paper, we have revisited Vasiliev's equations at the linearized level, and demonstrated explicitly in the cases of particle and black hole states how to use gauge functions to go from the holomorphic gauge given in Weyl order, suitable for integrating the system to all orders, to a relaxed Vasiliev gauge given in normal order, suitable for identifying Fronsdal fields on-shell in accordance with the COMST and imposing asymptotically locally anti-de Sitter boundary conditions.
More specifically, we have found that working in normal order, the symbols for $\Phi^{(1)}$ and $U^{(1)}\equiv D^{(0)}H^{(1)}$ are indeed regular in twistor space at $Z=0$, where they serve as generating functions for linearized Weyl tensors and spacetime gauge fields, respectively, even though the linearized gauge function $H^{(1)}$ has singularities in twistor space. 
The latter, however, have no impact on the spacetime gauge fields, as the singular part of $U^{(1)}|_{Z=0}$ is cohomologically trivial with respect to the AdS covariant derivative $D^{(0)}$.
In other words, the $D^{(0)}$-cohomology provides a splitting of $U^{(1)}|_{Z=0}$ into a regular part, which thus defines the unfolded Fronsdal fields on-shell, and a singular part that can be cancelled by choosing the $Z$-independent part of $H^{(1)}$, as we have done explicitly in the case of particle states.

The construction of $H^{(1)}$ involves replacing Vasiliev's original gauge condition on the twistor space connection \cite{Vasiliev:1992av,Sezgin:2002ru}\footnote{%
In particular, in \cite{Boulanger:2015ova} it was shown that combining the original Vasiliev gauge with a computational scheme in which the master fields are expanded in formal power series in $Y$ \emph{prior to} performing star products leads to ill-defined second-order corrections to the resulting Fronsdal field equations.
To our best understanding, it remains an open problem whether relaxing the original Vasiliev gauge condition or switching to the regular computation scheme, or a combination of the two, may yield finite results.} with a relaxed version that only fixes $H^{(1)}$ up to terms of $O(Z^2)$ (in normal order), as the latter do not affect the COMST.
Instead, we argue that this freedom in $H^{(1)}$ is of use in imposing asymptotically (locally) AdS boundary conditions.
To this end, we have proposed a Fefferman--Graham-like expansion in which the full master fields linearize asymptotically.
The perturbative corrections to the bulk master fields involve star-product interactions that may affect their leading order in the asymptotic expansion, in which case corrections to the zero-form initial data as well as to the gauge function will be required, that we plan to report on elsewhere. 

The analysis is facilitated by the usage of a \emph{regular computation scheme}, whereby master fields and gauge functions are presented by means of parametric integrals with integrands given by regular Gaussian functions in twistor space.
We have proposed a prescription for how to nest auxiliary integrals and star products so as to maintain associativity in perturbative computations.
We have tested this scheme at the linearized level in the aforementioned mapping between gauges, and at the fully nonlinear level in constructing perturbatively exact solutions in the holomorphic gauge in Weyl order \cite{Iazeolla:2017vng}.

Following the QTFT approach to HSG advocated in \cite{Boulanger:2015kfa,Bonezzi:2016ttk,Arias:2017bvi}, the HSG partition function is given by a sum over classical field configurations weighted by an on-shell action.
Starting from first principles, the simplest such on-shell action is given by the zero-form charge obtained from the second Chern-class on twistor space.
In the leading order, this quantity has been shown to produce physically meaningful holographic two-point functions in the case of a perturbative expansion around anti-de Sitter spacetime with boundary conditions corresponding to free holographic CFTs.
We plan to report elsewhere on whether the sub-leading terms in the perturbative expansion of the on-shell action due to the aforementioned corrections to the zero-form initial data will induce the physically desired $n$-point functions as well. 
We have verified, however, that the first sub-leading corrections to a related set of classical observables, given by Wilson loops in twistor space, also known as zero-form charges, are well-defined in the sense that they do not depend on an auxiliary spin-frame structure introduced in the holomorphic gauge.

The formalism advocated in this paper shifts the focus of HSG away from the issue of finding classes of non-local vertices for Fronsdal fields in quasi-Riemannian spacetimes to that of constructing  star-product local functionals for curvatures of non-commutative higher spin geometries.
In this spirit, we have proposed a set of dual boundary conditions on Vasiliev's nonlinear master fields for higher spin geometries containing asymptotically locally anti-de Sitter spacetimes as well as a globally defined on-shell action functional given by an integral over twistor space.
If successful, the resulting framework may open up a dual approach to quantum field theory in which each particle is described by an infinite collections of topological fields making up a horizontal form on a (non-commutative) fibered space, rather than a finite set of propagating fields on a (commutative) quasi-Riemannian manifold.

\paragraph{Acknowledgements.} We are grateful to C. Arias, R. Aros, R. Bonezzi, O. A. Gelfond, A. V. Korybut, E. Skvortsov, C. Sleight, M. Taronna, B. Vallilo, Y. Yin and especially to N. Boulanger, V. E. Didenko, E. Sezgin and M. A. Vasiliev for many useful conversations. 
D.D.F. is a Research Fellow at the F.R.S.-FNRS (Belgium).
He is thankful to G. Marconi University (Rome, Italy) and Universidad Andres Bello (Santiago, Chile) for hospitality at various stages of this work.
The work of C.I. was supported in part by the Russian Science Foundation grant 14-42-00047 in association with the Lebedev Physical Institute in Moscow. 
P.S. acknowledges the support of Conicyt grant DPI 2014-0115, Fondecyt Regular grants 1151107 and 1140296, and the NSF China Contracts No. 11775110 and 11690034.
He is thankful to Edna Cheung for kind hospitality during the finalization of this project.
Finally, the Authors would like to thank the Erwin Schr\"odinger Institute in Vienna for hospitality within the program `Higher Spins and Holography' during the final stages of this work.

\appendix

\section{Conventions}
\label{App:conv}
\paragraph{Indices.}
Throughout the paper, we use the following types of indices:
\begin{itemize}[label=---]
\item
Greek letters from the end of the alphabet are world indices;
\item
Lower case letters from the beginning of the alphabet are Lorentz indices;
\item
Lower case letters from the end of the alphabet are $so(3)$ indices;
\item
(Possibly dotted) greek letters from the beginning of the alphabet are $Sp(2)$ spinor indices;
\item
Underlined greek letters from the beginning of the alphabet are $Sp(4)$ spinor indices.
\end{itemize}
The Lorentz metric $\eta_{ab}$ is taken as ${\rm diag}(-+++)$.
Every $Sp(4)$ spinor $A_\au$ consist in two $Sp(2)$ spinors $(a_\a,\bar{a}_\ad)$.
The symplectic matrices 
\begin{equation}
C_{\au\bu}=\begin{pmatrix}
\epsilon_{\a\b}&0\\0&\epsilon_{\ad\bd}
\end{pmatrix}
\,,\qquad
C_{\au\bu}=\begin{pmatrix}
\epsilon_{\a\b}&0\\0&\epsilon_{\ad\bd}
\end{pmatrix}
\,,
\end{equation}
are defined with the convention that
\begin{equation}
\epsilon^{\a\b}\epsilon_{\a\c} = \d^\b_\c
\,,\qquad
\epsilon^{\ad\bd}\epsilon_{\ad\cd} = \d^\bd_\cd
\,.
\end{equation}
They are used to raise, lower and contract all the spinor indices with the so-called NW-SE convention.
For example, for two $Sp(2)$ spinors $\lambda_\a$ and $\mu_\a$, one has
\begin{equation}
\label{eq:NWSE}
\lambda^\a=\epsilon^{\a\b}\lambda_\b 
\,,\qquad
\lambda_\a=\lambda^\b\epsilon_{\b\a} 
\,,\qquad 
\lambda\mu = \lambda^\a\mu_\a = -\mu\lambda 
\,.
\end{equation}
The convention is the same for dotted and underlined indices.
The convention is often used implicitely to omit all spinorial indices.
Another related convention is that the same symbol can be used for different indices in the same expression 
to indicate that they must be completely symmetrized (with weight one). 
Let us define a spin-frame $(u^\pm_\a,\ub^\pm_\ad)$ such that $u^{+\a}u^-_\a=\ub^{+\ad}\ub^-_\ad=1$.
That is to say the convention for the symplectic metric reads in this basis
\begin{equation}
\epsilon_{\a\b}:=u^-_\a u^+_\b - u^+_\a u^-_\b
\,,\qquad
\epsilon_{\ad\bd}:=\ub^-_\ad \ub^+_\bd - \ub^+_\ad \ub^-_\bd
\,.
\end{equation}
We usually use that property to expand spinors in that basis as
\begin{align}
\lambda^\pm\:=u^{\pm\a}\lambda_\a
\,,\quad
\lambda_\a = u^-_\a\lambda^+-u^+_\a\lambda^-
\,,\quad 
\bar\lambda^\pm:=\ub^{\pm\ad}\bar\lambda_\ad 
\,,\quad
\bar\lambda_\ad = u^-_\ad\bar\lambda^+-u^+_\ad\bar\lambda^-
\,.
\end{align}
Vector and spinor indices are related by the Van der Waerden symbols, 
that have the following properties
\begin{align}
\s_a\sb_b
&=
\eta_{ab}+\s_{ab}
\,,&
\s_{ab}:=\frac12(\s_a\sb_b-\s_b\sb_a)
\,,\\
\sb_a\s_b
&=
\eta_{ab}+\sb_{ab}
\,,&
\sb_{ab}:=\frac12(\sb_a\s_b-\sb_b\s_a)
\,,\\
&&
(\s^a)_{\a\ad}(\s_{a})_{\b\bd} 
=-2\epsilon_{\a\b}\epsilon_{\ad\bd} 
\,.
\end{align}
For a Lorentz vector $v^a$, one has
\begin{equation}
v_{\a\ad}:=v^a(\s_a)_{\a\ad}
=:\vb_{\ad\a}
\,,\qquad 
v^a=-\frac12(\s^a)_{\a\ad}v^{\a\ad}
\,.
\end{equation}
The Van der Waerden symbols can be realised in terms of the spin-frame as
\begin{align}
(\s_0)_{\a\ad}
&=
u^+_\a\ub^+_\ad+u^-_\a\ub^-_\ad
=
(\sb_{0})_{\ad\a}
\,,\label{eq:sigmau1}\qquad&
(\s_1)_{\a\ad}
&=
u^+_\a\ub^-_\ad+u^-_\a\ub^+_\ad
=
(\sb_{1})_{\ad\a}
\,,\\
(\s_2)_{\a\ad}
&=
i(u^-_\a\ub^+_\ad-u^+_\a\ub^-_\ad)
=
(\sb_{2})_{\ad\a}
\,,\qquad&
(\s_3)_{\a\ad}
&=
u^+_\a\ub^+_\ad-u^-_\a\ub^-_\ad
=
(\sb_{3})_{\ad\a}
\,.\label{eq:sigmau2}
\end{align}

\paragraph{AdS${}_4$ generators and connection.}
We use the following convention for the isometry algebra of AdS${}_4$
\begin{equation}
[M_{ab},M_{cd}]=4i\eta_{[d\vert[a}M_{b]\vert c]}
\,,\qquad
[M_{ab},P_c]=2i\eta_{c[b}P_{a]}
\,,\qquad 
[P_a,P_b]=iM_{ab}
\,,
\end{equation}
which corresponds to giving to the cosmological constant the constant value -3.
We usually redefine them into the compact generators $M_{rs}$ and $E$ and the ladder operators $L^{\pm}_r$, with the definitions
\begin{equation}
E:=P_0
\,,\qquad 
L^{\pm}_r:=M_{0r}\mp iP_r
\,.
\end{equation}
The Cartan connection of AdS${}_4$ is the pull-back of a Maureer-Cartan form, that is to say it is an algebra-value 1-form field $U^{(0)}$ satisfying
\begin{align}
dU^{(0)}+\frac12[U^{(0)},U^{(0)}]=0\,.
\end{align}
It can be splitted as
\begin{equation}
U^{(0)} := -i \left(e^{(0)a} P_a +  \tfrac{1}{2}\omega^{(0)ab}M_{ab}\right)
\,,
\end{equation}
where $e^a$ is the vierbein and $\omega^{ab}$ is te spin connection, satisfying
\begin{equation}
d\omega^{(0)ab} + \omega^{(0)a}_{\phantom{(0)a}c}\omega^{(0)cb} + e^{(0)a} e^{(0)b}=0
\,,\qquad 
de^{(0)a} + \omega^{(0)a}_{\phantom{(0)a}b}e^{(0)b}=0
\,.
\end{equation}
The algebra can be realized in terms of the star-product \eqref{eq:NOprod}
with the following definition of the generators
\begin{equation}
\label{eq:genAdS4}
M_{ab}=
-\frac18(\s_{ab})^{\a\a}y_\a y_\a
-\frac18(\sb_{ab})^{\ad\ad}\yb_\ad \yb_\ad
\,,\qquad
P_a=\frac14(\s_a)^{\a\ad}y_\a\yb_\ad
\,.
\end{equation}
In this language, the Cartan connection becomes Eq.\eqref{eq:AdS U}, viz.
\begin{equation}
U^{(0)}
=
\frac{1}{4i}\Omega^{(0)\underline{\alpha\beta}}Y_{\underline\alpha}Y_{\underline\beta}
=
\frac{1}{4i}\left(
\omega^{(0)\a\b}y_\a y_\b
+\bar\omega^{(0)\ad\bd}\yb_\ad\yb_\bd 
+2e^{(0)\a\bd}y_\a\yb_\bd
\right)
\,,
\end{equation}
with the spinorial components defined as
\begin{align}
e^{(0)a}
&=
(\sigma^a)_{\a\ad}e^{(0)\a\ad}
\,,&
\omega^{(0)ab}
&=
-\frac12\left(
(\sigma^{ab})_{\a\a}\omega^{(0)\a\a}
+(\bar\sigma^{ab})_{\ad\ad}\bar\omega^{(0)\ad\ad}
\right)
\,,\\
e^{(0)\a\ad}
&=
-\frac12(\sigma_a)^{\a\ad}e^{(0)a}
\,,&
\omega^{(0)\a\a}
&=
-\frac14(\sigma_{ab})^{\a\a}\omega^{(0)ab}
\,,\quad
\bar\omega^{(0)\ad\ad}
=
-\frac14(\bar\sigma_{ab})^{\ad\ad}\omega^{(0)ab}
\,,
\end{align}
and satisfying
\begin{align}
d\omega^{(0)\a\a} + \omega^{(0)\a}_{\phantom{(0)\a}\b}\omega^{(0)\b\a} + e^{(0)\a}_{\phantom\a\bd} e^{(0)\a\bd}=0
\,,\quad 
d\bar\omega^{(0)\ad\ad} + \omega^{(0)\ad}_{\phantom{(0)\ad}\bd}\omega^{(0)\bd\ad} + e_{\phantom{(0)}\b}^{(0)\phantom\b\ad} e^{(0)\b\ad}=0
\,,\\
de^{(0)\a\ad} 
+\omega^{(0)\a}_{\phantom{\a}\b}e^{(0)\b\ad}
+\omega^{(0)\ad}_{\phantom{(0)\ad}\bd}e^{(0)\a\bd}
=0
\,.
\end{align}
$U^{(0)}$ is pure gauge, in the sense that it can be written
\begin{equation}
{U}^{(0)}= L^{-1}\star d L
\,,
\end{equation}
where $L$ is a $x$-dependent Gaussian in $Y$.
Its adjoint action on $Z$-independent symbols reads
\begin{align}
\label{eq:Lrot gen app}
L^{-1}\star f(x;Y)\star L
&=
f\left(x;Y^L_\au\right)
\,,&
Y^L_\au(x)
&=
L(x)_{\au}^{\phantom\au\bu}Y_\bu
\,,
\end{align}
where $L(x)_{\au}^{\phantom\au\bu}$ is a $x$-dependent $Sp(4)$ matrix corresponding to $L$.
\paragraph{Stereographic coordinates.}
Whenever we need to make the spacetime dependence explicit, we choose to do it with the help of stereographic coordinates,
that can be written in Lorentz covariant fashion
\begin{equation}
x^a\in\mathbb{R}^4
\,,\qquad 
x^2\neq1
\,,\qquad
ds^2=\frac{4}{(1-x^2)^2}\dx^2
\,.
\end{equation}
The radial coordinate of the global AdS${}_4$ spherical coordinate system, that appears in the analysis of the black hole states, can be related to the stereographic coordinates via
\begin{equation}
r=\frac{2\sqrt{(x_0)^2+x^2}}{(1-x^2)}
\,.
\end{equation}
Two useful definitions are
\begin{equation}
h:=\sqrt{1-x^2}
\,,\qquad 
\xi:=
\vert x^2\vert^{-\tfrac12}\tanh^{-1}\left(\sqrt{\frac{1-h}{1+h}}\right)
\,.
\end{equation}
The vacuum gauge function corresponding to stereographic coordinates can be given in terms of those quantities
\begin{equation}
L
=
\exp_\star\left(4i\xi x^aP_a\right)
=
\frac{2h}{1+h}
\exp\left(
-\frac{i}{1+h}yx\yb
\right)
\,.
\label{eq:L gf stereo}\end{equation}
The associated $Sp(4)$ matrix allows to rewrite Eq.\eqref{eq:Lrot gen app} as
\begin{align}
\label{eq:Lrot stereo}
L^{-1}\star f(x;y,\yb)\star L
&=
f\left(x;\frac{1}{h}(y+x\yb),\frac{1}{h}(\yb+\xb y)\right)
\,,&
y^L
&=
\frac1h(y+x\yb)
\,,&
\yb^L
&=
\frac1h(\yb+\xb y)
\,.
\end{align}

\section{Definition of the trace on ${\cal A}({\cal Y}_4)$}
\label{App:Tr}

Although the precise definition of the trace operation on ${\cal A}({\cal Y}_4)$ does not enter the derivation of the main results of this paper, the following remarks are in order.
A trace operation that is naturally associated with the Groenewold-Moyal product induced on ${\cal A}({\cal Y}_4)$ by \eqref{eq:NOprod} is
\begin{equation}
\label{eq:Tr*}
\Tr_\star f:= \int d^4Y f(Y)
\,.
\end{equation}
However, some of the functions of $Y$-oscillators that we work with in this paper are non-integrable.
For example, the twisted Fock-space endomorphisms $P_{\bf m\vert n}\star \k_y$ (see \cite{Iazeolla:2017vng} and Section \ref{sec:examples}) have infinite $\Tr_\star$.
To regularise their traces, one may introduce a projector $p$ to $L^1({\cal Y}_4)$, or equivalently a projector $\widetilde{p}$ to the space $\mathscr{F}_{Y=0}({\cal Y}_4)$ of functions that have a finite value at $Y=0$.
The associated linear map
\begin{align}
\Tr_p f
:=
\int d^4Y\,p f(Y)
=
\left.(2\pi)^2\,\widetilde{p}(f\star\k_y\kb_\yb)\right\vert_{Y=0}
\,,
\end{align}
is a trace operation provided that it is cyclic.
The existence of such a regularized trace depends on the details of the algebra ${\cal A}({\cal Y}_4)$.

An example is provided by the extension of the Weyl algebra \cite{Sundell:2016mxc} spanned by elements of the form 
\begin{equation}
f(Y)=f_{0,\bar0}+f_{1,\bar0}\star\kappa_y+f_{0,\bar1}\star\bar\kappa_{\yb}+f_{1,\bar1}\star\kappa_y\star\bar\kappa_{\yb}
\,,
\label{Wexp}\end{equation}
where $f_{\ell,\bar\ell}(Y)$ are analytic functions of $Y$ that form a star-product algebra, i. e. the algebra has basis elements $Y^k\star\k_y^p\kb_{\yb}^q$, that is, monomials in $Y$ of arbitrary degree together with monomials multiplied with delta functions in $Y$ and their derivatives. 
The extended Weyl algebra admits the non-degenerate trace operation
\begin{equation}
\label{eq:Tr reg}
\Tr_{\rho}(f)
:=
\left.(2\pi)^2f_{1,\bar{1}}\right\vert_{Y=0}
\,,
\end{equation}
where $\rho$ is the projection whose Fourier dual $\tilde\rho$ is defined by 
\be \tilde\rho(Y^k\star\k_y^p\kb_{\yb}^q)=\d_{p0}\d_{q0} Y^k 
\,.
\label{rhop}\ee
Thus, for instance, the regularized trace of a non-polynomial (anti-)Fock-space endomorphism $P_{m_1,m_2|n_1,n_2}$, where $m_i,n_i$ are all positive (negative) semi-integers (see \cite{Iazeolla:2011cb,Iazeolla:2017vng} and Section \ref{sec:examples}), is finite, and can be appropriately normalized as to give $\Tr_{\rho}(P_{m_1,m_2|n_1,n_2})\propto\langle n_1,n_2|m_1,m_2\rangle = \d_{n_1,m_1}\d_{n_2,m_2}$. 
From $P_{m_1,m_2|n_1,n_2}\star\k_y \sim P_{m_1,m_2|-n_2,-n_1}$, it follows that $\Tr_{\rho}(P_{m_1,m_2|n_1,n_2}\star\k_y)= 0$. 

The generalisation of the regularized trace \eqref{eq:Tr reg} to algebras that contain the Klein operators but are not a semi-direct product of ${1,\k,\kb,\k\kb}$ with an algebra of analytic function, needs more care.
Indeed, the expansion \eq{Wexp} is ambiguous for any $g$ such that $g\star \k_y^p\kb_{\yb}^q$ is itself an analytic function. One can resolve these ambiguities by prescribing that such functions enter the trace via a specific projection.
For instance, for the symmetry-enhanced projectors ${\cal P}_n(E)$ considered in \cite{Iazeolla:2011cb,Iazeolla:2017vng}, that are eigenstates of $\k_y\kb_{\yb}$, ${\cal P}_n(E)\star\k_y\kb_{\yb} = (-1)^n{\cal P}_n $, the ambiguity can be resolved by using $\Tr_{\rho\pi_n}$ instead, where $\pi_n:=\frac12(1+(-1)^n\k_y\kb_{\yb})$, hence $\Tr_{\rho\pi_n}({\cal P}_n(E))
:= 2\pi^2(-1)^n{\cal P}_n(E)\vert_{Y=0}$.
Elements of the group $SpH(8)$, that is discussed in section \ref{ssec:reg}, present the same difficulty, the study of which we defer to a future work.

\section{Perturbatively exact factorised solutions}
\label{App:fact}

In this appendix, we shall solve Eq. (\ref{eq:VE}) perturbatively around the anti-de Sitter background (\ref{eq:AdS})
using the Weyl-order resolution operator \eqref{eq:rhoF},
and more precisely taking advantage of its factorisation property \eqref{eq:rhoF fact}.
As we will see, the perturbation theory based on this homotopy integral can be performed to all orders.
Notice that, as this resolution operator is the only one we use in this appendix, we will drop the label $(E+i\partial^Y)$ on the various fields.

\subsection{Factorised perturbation theory}
\label{App:fact formal}
We will proof in this subsection that an\footnote{%
Having specified the resolution operator, the procedure should yield a unique solution.
However, $q^{(E)\ast}\k_z$ is not a uniquely defined symbol,
and we will see in the next subsection that for the family of representations of $\k_z$ that we will use,
the result depends on the chosen spin-frame.
} 
all-order result is
\begin{align}
\label{eq:Phi^m fact}
\Phi^{(m)}&=\delta_{m,1}\left(
\Psi^{(1)}\star\k_y+\bar\Psi^{(1)}\star\kb_y
\right)
\,,\\
\label{eq:V^m fact}
V^{(m)}&=
(\Psi^{(1)})^{\star m}\star v_m(z)+({\bar\Psi^{(1)}})^{\star m}\star\bar{v}_m(\zb)
\,,\\
\label{eq:U^m fact}
U^{(m)}&=\delta_{m,0}\,\Omega
\,,
\end{align}
where 
\begin{equation}
\Psi^{(1)}(Y):=C^{(1)}(Y)\star\k_y
\,,\qquad
\bar\Psi^{(1)}(Y):=C^{(1)}(Y)\star\kb_y
\,,
\end{equation}
where $v_m(z)$ and $\bar{v}_m(\zb)$ will be determined in the other subsections and where 
the integration constants have consistently been chosen to be zero, aside from $C^{(1)}(Y)$.
The solution (\ref{eq:Phi^m fact}--\ref{eq:U^m fact}) is factorised in accordance with Eq.\eqref{eq:fact sum}.
The bosonic projection \eqref{eq:BP} manifests itself on $\Psi^{(1)}$ and $\bar\Psi^{(1)}$ as
\begin{equation}
\label{eq:BP Psi}
\pi\bar\pi\left(\Psi^{(1)}\right)=\Psi^{(1)}
\,,\qquad
\pi\bar\pi\left(\bar\Psi^{(1)}\right)=\bar\Psi^{(1)}
\,,\qquad
\Psi^{(1)}\star\bar\Psi^{(1)}=\bar\Psi^{(1)}\star\Psi^{(1)}
\,.
\end{equation}

The generic solution to Eq.\eqref{eq:qPhi1} is $\Phi^{(1)}=C^{(1)}(Y)$ 
and it is straightforward to show that Eq.\eqref{eq:DPhi1} is equivalent to
\begin{equation}
\acd\Psi^{(1)} = \acd\bar\Psi^{(1)}=0
\,.
\end{equation}
Then all we want to use is Eq.\eqref{eq:rhoF}.
We rewrite Eq.\eqref{eq:qV1} as
\begin{equation}
q V^{(1)}=-\Psi^{(1)}\star j_z-\bar\Psi^{(1)}\star\jb_z
\,,\label{qV1}
\end{equation}
and solve it as
\begin{equation}
\label{eq:V1 fact}
V^{(1)}=
-\Psi^{(1)}\star q^{(E)\ast} j_z-\bar\Psi^{(1)}\star q^{(E)\ast} \jb_z
\,.
\end{equation}
Since $\acd$ annihilates $\Psi^{(1)}$ and $\bar\Psi^{(1)}$,
we can rewrite Eq.\eqref{eq:qU1+DV1} as
\begin{equation}
qU^{(1)}=\Psi^{(1)}\star\diff q^{(E)\ast} j_z+\bar\Psi^{(1)}\star\diff q^{(E)\ast} \jb_z\,.
\end{equation}
From now on, we assume that $q^{(E)\ast} j_z$ and $q^{(E)\ast} \jb_z$ are $x$-independent,
which is a non-trivial condition on the representation of $\k_z$ and $\kb_z$ that one uses to construct a solution.
Since $q^{(E)\ast}$ is linear, we have
\begin{equation}
\label{eq:U1=0}
U^{(1)}=W^{(1)}\,,
\end{equation}
which we decide to gauge fix to zero by virtue of Eq.\eqref{eq:DU1}, 
thereby proving Eqs.(\ref{eq:Phi^m fact}\,,\,\ref{eq:V^m fact}\,,\,\ref{eq:U^m fact}) for $m=1$.

Now we assume having proven it up to a certain order $n$ and will prove it for $m=n+1$. 
At order $n+1$, the equations (\ref{eq:qPhi1}--\ref{eq:DU1}) read
\begin{align}
\label{eq:qPhi(n+1)}
q\Phi^{(n+1)}
+\sum_{k=1}^{n}\picomm{V^{(k)}}{\Phi^{(n+1-k)}}
&=0
\,,\\\label{eq:DPhi(n+1)}
\tcd\Phi^{(n+1)}
+\sum_{k=1}^{n}\picomm{U^{(k)}}{\Phi^{(n+1-k)}}
&=0
\,,\\\label{eq:qV(n+1)}
qV^{(n+1)}
+\sum_{k=1}^{n}V^{(k)}\star V^{(n+1-k)}
+\Phi^{(n+1)}\star J
&=0
\,,\\\label{eq:qU(n+1)+DV(n+1)}
qU^{(n+1)}
+\acd V^{(n+1)}
+\sum_{k=1}^{n}\staracomm{V^{(k)}}{V^{(n+1-k)}}
&=0
\,,\\\label{eq:DU(n+1)}
\acd U^{(n+1)}
+\sum_{k=1}^{n} U^{(k)}\star U^{(n+1-k)}
&=0
\,.
\end{align}
We start solving the equations in that order using the previous results.
Eq.\eqref{eq:qPhi(n+1)} reads
\begin{align}
q\Phi^{(n+1)}
&=
-\picomm{V^{(n)}}{\Psi^{(1)}\star\k_y+\bar\Psi^{(1)}\star{\kb}_y}
\\\nonumber&=
-\starcomm{
(\Psi^{(1)})^{\star n}\star v_n(z)+({\bar\Psi^{(1)}})^{\star n}\star\bar{v}_n(\zb)}{\Psi^{(1)}}\star\k_y
\\\nonumber&\quad
-\starcomm{
(\Psi^{(1)})^{\star n}\star v_n(z)+({\bar\Psi^{(1)}})^{\star n}\star\bar{v}_n(\zb)}{\bar\Psi^{(1)}}\star{\kb}_y
=0
\,,
\end{align}
where Eq.\eqref{eq:BP} was used to get the second line and Eq.\eqref{eq:BP Psi} was used to get the conclusion.
We can chose $\Phi^{(n+1)}=0$ as a solution,
consistently with Eq.\eqref{eq:DPhi(n+1)} that has become
\begin{equation}
\acd{\Phi}^{(n+1)}=0
\,.
\end{equation}
This simplifies Eq.\eqref{eq:qV(n+1)} that, considering that $V_z(z)$ and ${\bar V}_z(\zb)$ anticommute,
takes the form
\begin{equation}
qV^{(n+1)}
=
-(\Psi^{(1)})^{\star n+1}\star \sum_{k=1}^{n}\left(
v_k(z)\star v_{n+1-k}(z)
\right)
-({\bar\Psi^{(1)}})^{\star n+1}\star\sum_{k=1}^{n}\left(
{\bar v}_k(\zb)\star {\bar v}_{n+1-k}(\zb)
\right)
\,.
\end{equation}
It admits as a solution Eq.\eqref{eq:V^m fact} with
\begin{align}
\label{eq:V(m+1)}
v_{n+1}(z)
:&=q^{(E)\ast}\left(
-\sum_{k=1}^{n}\left(
v_k(z)\star v_{n+1-k}(z)
\right)
\right)
\,,\\\label{eq:Vb(m+1)}
{\bar v}_{n+1}(\zb)
:&=q^{(E)\ast}\left(
-\sum_{k=1}^{n}\left(
{\bar v}_k(\zb)\star {\bar v}_{n+1-k}(\zb)
\right)
\right)
\,.
\end{align}
Assuming again that $v_m(z)$ are $x$-independent, the two remaining equations become 
\begin{equation}
qU^{(n+1)}=\acd U^{(n+1)}=0\,,
\end{equation}
that admit $U^{(n+1)}=0$ as a solution, thereby concluding the proof.

\subsection{Recursive solution using symbol calculus}
\label{App:fact param}
In this subsection we show that Eq.\eqref{eq:V1 fact} can give Eq.\eqref{eq:V1 unfact},
and that Eq.\eqref{eq:V(m+1)} can be solved recursively to provide this initial datum with an all order completion.
This is achieved by using a suitable representation\footnote{%
As detailed in Appendix \ref{App:param}, one can think of this representation as the $Z$-space Fourier transform of a similar representation of 1.
} of $\delta^2(z)$ and $\delta^2(\zb)$, 
that is described in more details in Appendix \ref{App:param}.

The all order solution to Eq.\eqref{eq:V(m+1)} is given by
\begin{align}
\label{eq:V_n sol}
v_n(z)
&=
\dz^\a\partial^\rho_\a\left.
\int_{-1}^1\frac{\diff s}{1+s}f_n(s)\exp\left(
\tfrac{i}{2}\tfrac{1-s}{1+s}\, z{\cal D}z
+\tfrac{i}{1+s}\,\rho z
\right)
\right\vert_{\rho=0}
\,,\\\label{eq:f_n sol}
f_n(s):&=
-\frac{b}{2}\frac{(2(n-1))!}{n!((n-1)!)^2}
\left(\frac{b}{8}\log\left(\tfrac{1}{s^2}\right)\right)^{n-1}
\,,
\end{align}
that can be plugged into Eq.\eqref{eq:V^m fact} an exact solution for the internal connection 
\begin{equation}
V=
-\frac{b}{2}\Psi^{(1)}\star 
\int_{-1}^1\frac{\diff s}{1+s}
\,{}^{\phantom\star}_1F_1^{\star}\left(
\frac{1}{2};2;\frac{b}{2}\log\left(\tfrac{1}{s^2}\right)\Psi^{(1)}
\right)\star\,
\dz^\a\partial^\rho_\a\left.
\exp\left(
\tfrac{i}{2}\tfrac{1-s}{1+s}\, z{\cal D}z
+\tfrac{i}{1+s}\,\rho z
\right)
\right\vert_{\rho=0}
-\hc
\,.
\end{equation}
This corresponds to the solution already studied in \cite{Iazeolla:2011cb,Iazeolla:2012nf,Sundell:2016mxc,Iazeolla:2017vng} in a particular gauge\footnote{%
This gauge, that is referred to as symmetric gauge,
corresponds to taking $f^+=f^-=f$ in the notation of Appendix C of \cite{Iazeolla:2017vng}}.
In particular, the holomorphic component of the first order part is
\begin{align}
V^{(1)}_\a
&=
-\Psi^{(1)}\star
\frac{b}{2}
\partial^\rho_{\alpha}
\left.\int_{-1}^1\frac{\diff s}{1+s}
\exp\left(
\tfrac{i}{2}\tfrac{1-s}{1+s}\, z{\cal D}z
+\tfrac{i}{1+s}\rho z
\right)
\right\vert_{\rho=0}
\nonumber\\&=
-\frac{b}{2}
\partial^\rho_{\a}
\int\frac{\diff^2u}{2\pi}
\Psi^{(1)}(y-u,\yb)e^{iuz}
\left.\int_{-1}^1\frac{\diff s}{1+s}
\exp\left(
\tfrac{i}{2}\tfrac{1-s}{1+s}\, u{\cal D}u
+\tfrac{i}{1+s}\rho{\cal D}u
\right)
\right\vert_{\rho=0}
\nonumber\\&=
-\frac{b}{2}
\partial^\rho_{\a}
\int\frac{\diff^2u}{2\pi}
\Phi^{(1)}(u-z,\yb)e^{iy(z-u)}
\left.\int_{-1}^1
\frac{\diff s}{1+s}
\exp\left(
\tfrac{i}{2}\tfrac{1-s}{1+s}\, u{\cal D}u
+\tfrac{i}{1+s}\rho u
\right)
\right\vert_{\rho=0}
\,,
\end{align}
which proves indeed Eq.\eqref{eq:V1 unfact}.

In the remaining part of this section we shall give the detailed proof of Eqs.(\ref{eq:V_n sol},\,\ref{eq:f_n sol}).

The right hand side of \eqref{eq:V1 fact} can be rewritten using \eqref{eq:def J} and \eqref{eq:kappaz param} as 
\begin{equation}
j_z = 
-\frac{ib}{2}\,\dz^{\alpha}\dz_{\alpha}
\int_{-1}^1\frac{\diff s}{1+s}\,\delta(1+s)\exp\left(
\tfrac{i}{2}\tfrac{1-s}{1+s}\, z{\cal D}z
\right)
\,,
\end{equation}
where one should think of this $\delta$ distribution as a limit $T\to-1$ rather than an evaluation at the singular point $T=-1$.
The homotopy integral is then performed using the lemma \eqref{eq:q*dz2}
\begin{align}
v_1(z)
= 
\label{eq:V1z param}
-\frac{b}{2}\dz^{\alpha}\partial^\rho_{\alpha}\int_{-1}^1\frac{\diff s}{1+s}
\exp\left(
\tfrac{i}{2}\tfrac{1-s}{1+s}\, z{\cal D}z
+\tfrac{i}{1+s}\rho z
\right)
\,,
\end{align}
which proves the result for $n=1$.
Notice that any spin frame allows to get this linear solution for $V$,
even one that depends on the spacetime coordinates $x$.
However, Eq. \eqref{eq:U1=0} is not guaranteed with a $x$-dependent spin-frame\footnote{%
In fact, performing the integral gives Eq. \eqref{eq:q*jz},
which shows that the result is $x$-dependent if and only if ${\cal D}$ is.
}.
For this reason, we assume from now on that the spin frame in Eq. \eqref{eq:D=uu+uu} is spacetime-independent.

The proof of the higher-order part of Eqs.(\ref{eq:V_n sol},\,\ref{eq:f_n sol})  begins with the $\star$-product in \eqref{eq:V(m+1)}, which is performed using the lemma \eqref{eq:circ 1form}
\begin{equation}
v_k(z)\star v_\ell(z)
=
-\frac{i}{4}\dz^\a\dz_\a
\int_{-1}^1\frac{\diff s}{(1+s)^2}
\,f_k\circ f_\ell(s)
\left(1
+\frac{i}{2}\frac{1-s}{1+s}
z{\cal D}z
\right)
\exp\left(\tfrac{i}{2}\tfrac{1-s}{1+s}\,z{\cal D}z\right)
\,,
\end{equation}
where $\circ$ is the commutative and associative product defined in \eqref{eq:def circ}.
Plugging that expression into \eqref{eq:V(m+1)} and using the lemma \eqref{eq:q*VV} gives
\begin{align}
v_{n+1}(z)
=
\left.
\dz^\a \partial^\rho_\a
\int_{-1}^1\frac{\diff s}{(1+s)^2}
\left(-\frac{1}{4}\sum_{k=1}^{n}
f_k\circ f_{n+1-k}(s)
\right)
\exp\left(\tfrac{i}{2}\tfrac{1-s}{1+s}\,z{\cal D}z
+\tfrac{i}{1+s}\rho z\right)
\right\vert_{\rho=0}
\,.
\end{align}
This translates \eqref{eq:V(m+1)} as a recursion for $f_n$
\begin{equation}
f_{n+1}(s)=
-\frac{1}{4}\sum_{k=1}^{n}
f_k\circ f_{n+1-k}(s)
\,,
\end{equation}
with the base case given by Eq.\eqref{eq:V1z param}:
\begin{equation}
f_1(s)=-\frac{b}{2}
\,.
\end{equation}
This is solved by
\begin{equation}
f_n(s)
=
-\frac{b}{2}C_{n-1}\left(\frac{b}{8}\right)^{n-1}1^{\circ n}
\,,
\end{equation}
where $C_n$ are the Catalan numbers, that are defined recursively as
\begin{equation}
C_0=1\,,\qquad 
C_{n+1}=\sum_{k=0}^n C_kC_{n-k}\,,
\end{equation}
and that can be written
\begin{equation}
C_n=\frac{(2n)!}{n!(n+1)!}\,.
\end{equation}
It can then be shown recursively that
\begin{equation}
1^{\circ k}=\frac{1}{(k-1)!}\left(\log\left(\tfrac{1}{s^2}\right)\right)^{k-1}
\,.
\end{equation}
Indeed, beyond the trivial k=1 case, one can use \eqref{eq:circ even} to get
\begin{align}
1^{\circ k+1}
&=
\frac{1}{(k-1)!}\left(\log\left(\tfrac{1}{s^2}\right)\right)^{k-1}
\circ 1
\nonumber\\&=
\frac{2}{(k-1)!}
\int_{\vert s\vert}^1\frac{\diff t}{t}
\left(\log\left(\tfrac{1}{t^2}\right)\right)^{k-1}
\nonumber\\&=
-\frac{1}{(k-1)!}
\int_{\log\left(\tfrac{1}{s^2}\right)}^0
\diff\left(\log\left(\tfrac{1}{t^2}\right)\right)
\left(\log\left(\tfrac{1}{t^2}\right)\right)^{k-1}
\nonumber\\&=
\frac{1}{k!}\left(\log\left(\tfrac{1}{s^2}\right)\right)^{k}
\,,
\end{align}
thereby concluding the proof of (\ref{eq:V_n sol},\,\ref{eq:f_n sol}).

\subsection{Alternative interpretations of $v_1(z)$}\label{App:altv1}

As it is intuitive from \eq{qV1}, given the distributional nature of $j_z$, the $z$-dependent coefficient $v_1(z)$ cannot be a regular function. Thus, according to the regular presentation scheme of Section \ref{ssec:reg}, a faithful representation, encoding the properties of $v_1(z)$ under star product, is given by \eq{eq:V1z param}, and only at the last step of the computation (that is, when all algebraic operations have been carried out and the resulting master fields are supposed to contain physical spacetime fields as coefficients of their power series expansion in oscillators) the auxiliary integrals are supposed to be evaluated. However, it can be interesting to compute the auxiliary integral in Eq. \eq{eq:V1z param} in order to understand what interpretation can be given to $v_1(z)$ as a distribution and how exactly the integral presentation takes care of it. 

\paragraph{As the inverse of a $z$ oscillator in Schwinger parametrization.} The integral defining $v_1(z)$ can be written as
\be v_{1\a} \ = \ -\frac{ib}{2}\, z_\a \int_{-1}^1 \frac{dt}{(t+1)^2}\,e^{-\ft{i}2\ft{t-1}{t+1}z{\cal D}z} \ = \ -\frac{ib}{4}\,z_\a\,\int_0^{+\infty} d\t \,e^{-i\t w_z} \,,\label{intpresv1}\ee
where we recall that $w_z = -\frac12 z{\cal D}z$, and converges provided that $\Im(z{\cal D}z)>0$. In other words, this integral can be regularized by means of a $-i\e$ prescription, and interpreted as a standard Schwinger parametrization, i.e., 
\be  \int_0^{+\infty} d\t \,e^{-i\t w_z}  =\lim_{\e\to 0}  \int_0^{+\infty} d\t \,e^{-i\t (w_z-i\e)} =\lim_{\e\to 0} \frac{-i}{w_z-i\e} \,.\ee
Indeed, this is how \eq{eq:int exp=1/z2} is obtained, as well as the last step of the computation leading to the COMST for particle states \eq{eq:COMST pt}. Hence, taking the limit $\e\to 0$,
\begin{equation}
\label{eq:q*jz}
v_1(z) = q^{(E)\ast}j_z = -\frac{b}{2}\frac{\dz^\a z_\a}{z{\cal D}z}
\,,
\end{equation}
or, projecting with the spin-frame,
\be v_{1}^\pm \ = \ -\frac{b}{4}\,\frac{1}{z^\mp}  \,.\label{v11z}\ee
Note however that the latter is not, per se, a faithful representation of $v_{1}^\pm $ as far as its star-product properties are concerned. For a start, \eq{v11z} does not satisfy $q v_1 = -j_z$ (see \eq{qV1} and \eq{eq:V^m fact}), whereas the integral presentation \eq{intpresv1} does. In particular we recall that, as shown in \cite{Iazeolla:2011cb}, the delta function comes from a boundary term: using that $w_z = -\frac12 z{\cal D}z$,
\bea 
&\displaystyle\frac{\partial}{\partial z^{[\a}}\, z_{\b]}  \int_0^{+\infty} d\t \,e^{-i\t w_z} \ = \ \e_{\a\b}  \int_0^{+\infty} d\t \,\left[1+\frac{i\t}2 z{\cal D}z\right]\,e^{-i\t w_z} &\nn\\
&\displaystyle  = \  \e_{\a\b}  \int_0^{+\infty} d\t\left[1+\t\frac{d}{d\t}\right]\,e^{-i\t w_z}  \ = \ \e_{\a\b} \left[\t e^{-i\t w_z}\right]^{+\infty}_0 &\nn\\
&\displaystyle  = \ \e_{\a\b}\lim_{\t\to +\infty}\t e^{-i\t w_z} \ = \ \e_{\a\b}\k_z \,, &
\eea
where the second equality on the second line is obtained via integration by parts. 

\paragraph{As a limit representation of a $\theta\delta$ distribution.} A different regularization of the integral defining $v_{1} $ indeed shows that, in getting \eq{v11z}, one is essentially neglecting a boundary term. One could in fact evaluate \eq{intpresv1} as
\be v_{1\a} \ = \ -\frac{ib}{2}\, z_\a \int_{-1}^1 \frac{dt}{(t+1)^2}\,e^{-\ft{i}2\ft{t-1}{t+1}z{\cal D}z} \ = \ \frac{b}{2}\, \frac{z_\a}{z{\cal D}z}\left(1-\lim_{\e\to 0}e^{\ft{i}{2\e}z{\cal D}z}   \right)  \,,\label{eq:V1znew} \ee
that is
\be v_{1}^\pm \ = \ -\frac{b}{4}\frac{1}{z^\mp}\left(1-\lim_{\e\to 0}e^{-\ft{i}{\e}z^+ z^-}   \right) \,.  \label{eq:V1bdry}\ee
The boundary term in \eq{eq:V1bdry} accounts for the delta function source term without having to use the integral presentation --- the proviso here being, naturally, that the limit $\e\to 0$ must be taken after $z$-derivatives. Indeed, 
\be  \partial_- v_{1}^- =-\frac{b}4\partial_-  \frac{1}{z^+}\left(1-\lim_{\e\to 0}e^{-\ft{i}{\e}z^+ z^-} \right)=-\frac{b}4 \lim_{\e\to 0} \frac{i}{\e}e^{-\ft{i}{\e}z^+ z^-} =-\frac{ib}4 \k_z \,.  \ee
Analogously for $\partial_+ v_{1}^+ $.

This result can also be obtained by considering the integral 
\be \lim_{\e\to 0}\int_0^{z^\pm} dz' \,\frac{1}{\e}  e^{-\ft{i}{\e}z' z^\mp} =  \frac{-i}{z^\mp}\left(1-\lim_{\e\to 0}e^{-\ft{i}{\e}z^+ z^-}\right) \,.\label{eq:V1intz}\ee
In fact, the evaluation of $v_{1} $ as in \eq{eq:V1bdry} is coherent with the interpretation of the integral in \eq{eq:V1intz} obtained by taking the limit on the integrand, 
\be \lim_{\e\to 0}\int_0^{z^\pm} dz' \,\frac{1}{\e}  e^{-\ft{i}{\e}z' z^\mp} =\int_0^{z^\pm}dz' \d(z')\d(z^\mp) =\theta(z^\pm )\d(z^\mp) \,, \label{thetadelta}\ee
which obviously lead to a two-dimensional delta function after taking a $z$-curl. By following this route, one can operate with $z$-derivative on the result of the integral \eq{eq:V1znew} without invoking any integral presentation; hence, one obtains immediately the correct identity \eq{eq:correctderivative}.   

Hence, the result we derived via the Schwinger parametrization \eq{v11z} makes sense as long as it is possible to consider
\begin{equation}
\lim_{\epsilon\to 0}e^{-\frac{i}{\epsilon}z^+z^-}=0 \,,\label{epsilondelta}
\end{equation}
which actually follows from the delta sequence $\k_z = \lim_{\e\to 0} \frac1{\e}e^{-\frac{i}{\e}w_z}$. The discussion above  suggests that the limit $\e\to 0$, enforcing \eq{epsilondelta}, should be taken only after all oscillator derivatives (or star products) have been performed, lest the $z$-dependence of $v_1$ be altered. This is in accordance, for example, with the way the result in \eq{eq:COMST pt} was obtained. Whether a limit representation like \eq{eq:V1bdry} can be considered truly faithful from the star product point-of-view, thus providing an alternative to the integral presentation \eq{intpresv1}, is yet to be fully investigated.

\paragraph{As a $z$-space integral.} It is also possible to write $v_1(z)$ as an integral of a delta sequence in $z$ (as shown before in \eq{eq:V1intz} and \eq{epsilondelta} as intermediate steps in order to relate the limit representation \eq{eq:V1bdry} to \eq{thetadelta}). One can thus identify 
\begin{equation}
\label{eq:1/z as int}
v^\pm_1(z)
=
-\frac{b}4 \lim_{\epsilon\to0}\frac{i}{\epsilon}
\int_0^{z^\pm}dz' e^{-\frac{i}{\epsilon}z'z^\mp}
\,.
\end{equation}
The nature of $v^\pm_1(z)$ as potential for $j_z$ is then clear from referring to their integral definition \eq{eq:1/z as int}, since for instance
\begin{equation}
\frac{\partial}{\partial z^-}v^-_1(z)
=
-\frac{b}4\lim_{\epsilon\to0}\frac{i}{\epsilon}
e^{-\frac{i}{\epsilon}z^-z^+}
=
-\frac{ib}2 \pi\d^2(z)
\,. \label{dz1z}
\end{equation}
Moreover, the representation \eq{eq:1/z as int} is also consistent with differentiating with respect to the other $z$-component. In particular, note that, in all cases in which one is allowed to use \eq{epsilondelta}, one has
\begin{align}
\frac{\partial}{\partial z^+}v^-_1
& \propto \frac{\partial}{\partial z^+}\frac{1}{z^+} =
\lim_{\epsilon\to0}\frac{i}{\epsilon}
\int_0^{z^-}dz' \left(-\frac{i}{\epsilon}z'\right)
e^{-\frac{i}{\epsilon}z'z^+}
\nonumber\\&=
\lim_{\epsilon\to0}\frac{i}{\epsilon z^+}
\int_0^{z^-}dz' z'\frac{\partial}{\partial z'}
e^{-\frac{i}{\epsilon}z'z^+}
\nonumber\\&=
\lim_{\epsilon\to0}\frac{i}{\epsilon z^+}\left(
z^-e^{-\frac{i}{\epsilon}z'z^+}
-\frac{\epsilon}{iz^+}
+\frac{\epsilon}{iz^+}e^{-\frac{i}{\epsilon}z'z^+}
\right)
\nonumber\\&=
2i\pi\frac{z^-}{z^+}\delta^2(z)-\frac{1}{(z^+)^2}
\,.
\end{align}
The first term is a bit unexpected, but it can be regularised to zero by using 
\begin{equation}
\frac{1}{z^+}\delta(z^+)=-\delta'(z^+)\,.
\end{equation}
The latter identity can be derived by letting both sides act on an analytic test function:
\begin{align*}
\int_{-\infty}^{+\infty}dz\frac{1}{z}\delta(z)f(z)
=
\int_{-\infty}^{+\infty}dz\frac{1}{z}\delta(z)\left(f_0+zf_1+O(z^2)\right)
=
f_1
=
-\int_{-\infty}^{+\infty}dz\delta'(z)f(z)
\,,
\end{align*}
where the vanishing of the first term comes from Cauchy's principal value prescription,
using that $\delta(z^+)=\delta(-z^+)$.
Hence, we have
\begin{equation}
\frac{\partial}{\partial z^+}\frac{1}{z^+}
=
-\frac{1}{(z^+)^2}
\,,
\label{eq:correctderivative}\end{equation}
hence ensuring consistency between this integral representation \eqref{eq:1/z as int} and the previous ones proposed within the regular scheme.

\paragraph{As the degree-one cohomology of $S^1$.} Finally, note that \eq{dz1z} is suggestive of yet another interpretation of $v_1(z)$ as a distributional potential of $j_z$, provided 
the coordinates of the $z$ complex sphere can be rewritten as
\begin{equation}
z^{\pm}=re^{\mp i\varphi}
\,,
\end{equation}
where $\varphi$ is real and $r$ is possibly complex\footnote{%
The compatibility with the convergence of integrals such as Eq.\eqref{intpresv1} would require $\Im(r^2)<0$, in turn compatible with the regularization $r=\Re(r)-i\e$.
}.
Then, rewriting \eqref{eq:q*jz} as
\begin{equation}
\label{eq:q*jz=dz/z}
q^{(E)\ast}j_z = 
\frac{b}{4}
\left(
\frac{\dz^+}{z^+}-\frac{\dz^-}{z^-}
\right)
=
-
\frac{ib}{2}
\diff\varphi 
\end{equation}
one can observe that, while the action of $q$ on this expression vanishes away from the origin of $z$-space,
the analysis at $z=0$ requires the additional use of either the residue formula or Stokes' theorem.
One has
\begin{align}
\int q\,q^{(E)\ast}j_z
&=
\frac{b}{4}
\left(
\oint\frac{\dz^-}{z^-}
-\oint\frac{\dz^+}{z^+}
\right)
=
-\frac{ib}{2}
\int_0^{2\pi}\diff\varphi
=
-i\pi b
\,,
\end{align}
from which one deduces
\begin{equation}
j_z
=
-i\pi b\,d^2z\,\delta^2(z)
=
-\frac{ib}{4}\k_zdz^\a dz_\a
\,,
\end{equation}
which is consistent with Eqs.(\ref{eq:def J},\,\ref{eq:d2z}).

\section{Parametric integrals}
\label{App:param}
The exact solution that we find using $q^{(E)\ast}$ uses a representation of $\k_z$ that we shall detail below.
Every time such integrals appear, the prescription is to perform all star products and derivatives between integrands
and perform the integral as the very last step.

\subsection{$\circ$-product algebra}
This formalism relies on the introduction of a spin-frame $(u^{+\a},u^{-\a})$ satisfying $u^{+\a}u^-_{\a}=1$\,.
It has the property
\begin{equation}
\epsilon^{\a\b}=u^{-\a}u^{+\b}-u^{+\a}u^{-\b}.
\end{equation}
Hence for a spinor $v^{\a}$, we have the following relations
\begin{equation}
\label{eq:spin frame}
v^\a=u^{-\a}v^+-u^{+\a}v^-
\,,\qquad
v^{\pm}:=u^{\pm\a}v_\a 
\,.
\end{equation}
It can then be used to defined the following metric in twistor space:
\begin{equation}
\label{eq:D=uu+uu}
{\cal D}_{\a\b}=
u^+_\a u^-_\b+u^-_\a u^+_\b
\,.
\end{equation}

The prescription is to expand functions of $z$ over the following basis:
\begin{equation}
G_{\a_1,...,\a_n}(s):=
\partial^\rho_{\a_1}...\partial^\rho_{\a_n}
\left.
\exp\left(
\tfrac{i}{2}\tfrac{1-s}{1+s}\, z{\cal D}z
+\tfrac{i}{1+s}\rho z
\right)
\right\vert_{\rho=0}
\,.
\end{equation}
In particular
\begin{equation}
G_{s}:=
\exp\left(
\tfrac{i}{2}\tfrac{1-s}{1+s}\, z{\cal D}z
\right)
=
\exp\left(
-i\tfrac{1-s}{1+s}\,z^+z^-
\right)
\,.
\end{equation}
Most of the functions the appear in the construction of the factorised exact solution are of the form:
\begin{align}
\label{eq:param expand}
F
:&=
\int_{-1}^{1}\frac{\diff s}{1+s}\,f(s)
\exp\left(
\tfrac{i}{2}\tfrac{1-s}{1+s}\, z{\cal D}z
\right)
\,,\\
F_{(a,b)}
:&=
\dz^\a(a\epsilon_\a^{\phantom\a\b}+b{\cal D}_\a^{\phantom\a\b})\partial^\rho_\b
\left.
\int_{-1}^{1}\frac{\diff s}{1+s}\,f(s)
\exp\left(
\tfrac{i}{2}\tfrac{1-s}{1+s}\, z{\cal D}z
+\tfrac{i}{1+s}\rho z
\right)
\right\vert_{\rho=0}
\,,
\end{align}
where $f(t)$ is a function or distribution,
referred to as symbol.
What makes this representation interesting is the following self-replication property:
\begin{align}
&
\frac{1}{1+s}\exp\left(
\tfrac{i}{2}\tfrac{1-s}{1+s}\, z{\cal D}z
+\tfrac{i}{1+s}\,\rho z
\right)
\star
\frac{1}{1+s'}\exp\left(
\tfrac{i}{2}\tfrac{1-s'}{1+s'}\, z{\cal D}z
+\tfrac{i}{1+s'}\,\rho' z
\right)
\nonumber\\&=
\frac{1}{2(1+ss')}\exp\left(
\tfrac{i}{2}\tfrac{1-ss'}{1+ss'}\, z{\cal D}z
+\tfrac{i}{1+ss'}\,\left(
\rho\left(\tfrac{1+s'}{2}+\tfrac{1-s'}{2}{\cal D}\right)
+\rho'\left(\tfrac{1+s}{2}-\tfrac{1-s}{2}{\cal D}\right)
\right)z
\right)\times\nonumber\\&\quad\times\exp\left(
\tfrac{i}{4}\tfrac{1-s'}{(1+s)(1+ss')}\rho {\cal D}\rho
+\tfrac{i}{4}\tfrac{1-s}{(1+s')(1+ss')}\rho' {\cal D}\rho'
-\tfrac{i}{2(1+ss')}\rho\rho'
\right)
\,.
\label{eq:self repl}
\end{align}
In the context of this work we are particularly interested in one of its consequences:
\begin{align}
\label{eq:circ 1form}
F_{(1,0)}\star F'_{(1,0)}
&=
-\frac{i}{4}\dz^\a\dz_\a
\int_{-1}^1\frac{\diff S}{(1+S)^2}
\,f\circ f'(S)
\left(1
+\frac{i}{2}\frac{1-S}{1+S}
z{\cal D}z
\right)
\exp\left(\tfrac{i}{2}\tfrac{1-S}{1+S}\,z{\cal D}z\right)
\,,
\end{align}
where the $\circ$-product was defined as in \cite{Prokushkin:1998bq}
\begin{equation}
\label{eq:def circ}
f\circ f'(S)
:=
\int_{-1}^{1}\diff s
\int_{-1}^{1}\diff s'\,
f(s)\,f'(s')\,\delta(S-ss')
\,.
\end{equation}
It is commutative and associative.
Moreover, in the case of two even functions of $s$, it can be reexpressed as
\begin{align}
f^{(+)}\circ g^{(+)}(S)
&=
\int_{-1}^1\diff s\int_{-1}^1\diff s'\,
f^{(+)}(s)\,g^{(+)}(s')\,\delta(S-ss')
\nonumber\\&=
\int_{-1}^1\frac{\diff s}{\vert s\vert}
f^{(+)}(s)\,g^{(+)}\left(\frac{S}{s}\right)
\theta\left(1-\left\vert\frac{S}{s}\right\vert\right)
\nonumber\\&=
2\int_{\vert S\vert}^1\frac{\diff s}{s}\,
f^{(+)}(s)\,g^{(+)}\left(\frac{S}{s}\right)
\,.
\label{eq:circ even}
\end{align}

\subsection{Fourier transform and $\delta$ distribution}
As for any function of $z$, the Fourier transform is given by the star-multiplication by $\k_z$, as can be seen using equations (\ref{eq:NOprod},\,\ref{eq:k fact})
\begin{equation}
f(z)\star\k_z
=
\int\frac{\diff^2 u}{2\pi}f(u)\,e^{-iuz}
\,.
\end{equation}
This property translates on symbols as
\begin{align}
F\star\k_z
&=
\int_{-1}^{1}\frac{\diff s}{1+s}\,f(-s)
\exp\left(
\tfrac{i}{2}\tfrac{1-s}{1+s}\, z{\cal D}z
\right)
\,,\\
F_{(a,b)}\star\k_z
&=
\dz^\a(a{\cal D}_\a^{\phantom\a\b}+b\epsilon_\a^{\phantom\a\b})\partial^\rho_\b
\left.
\int_{-1}^{1}\frac{\diff s}{1+s}\,f(-s)
\exp\left(
\tfrac{i}{2}\tfrac{1-s}{1+s}\, z{\cal D}z
+\tfrac{i}{1+s}\rho z
\right)
\right\vert_{\rho=0}
\,,
\end{align}
In particular, from
\begin{equation}
1 = 2
\int_{-1}^{1}\frac{\diff s}{1+s}\,\delta(1-s)
\exp\left(
\tfrac{i}{2}\tfrac{1-s}{1+s}\, z{\cal D}z
\right)
\,,
\end{equation}
one deduces
\begin{equation}
\label{eq:kappaz param}
\delta^2(z) = 
\frac{1}{\pi}
\int_{-1}^{1}\frac{\diff s}{1+s}\,\delta(1+s)
\exp\left(
\tfrac{i}{2}\tfrac{1-s}{1+s}\, z{\cal D}z
\right)
\,.
\end{equation}
The latter expression is nothing but a rewriting of the delta sequence
\begin{equation}
\delta^2(z)
=
\lim_{\epsilon\to0^+}\frac{1}{2\pi\epsilon}
\exp\left(-\tfrac{i}{\epsilon}z^+z^-\right)\,.
\end{equation}
Its derivative can can expressed in the same basis as
\begin{equation}
\label{eq:delta'}
\partial^z_\a
\delta^2(z) = 
\frac{2i}{\pi}
({\cal D}z)_\a
\int_{-1}^{1}\frac{\diff s}{(1+s)^2}\,\delta(1+s)
\exp\left(
\tfrac{i}{2}\tfrac{1-s}{1+s}\, z{\cal D}z
\right)
\,.
\end{equation}
Other useful identities (which, according to the regular scheme, are meant to be used after all star-products have been evaluated, see Appendix \ref{App:altv1}) can be recovered using standard integration tools:
\begin{align}
\label{eq:Tr exp param}
\int d^2z\exp\left(
\tfrac{i}{2}\tfrac{1-s}{1+s}z{\cal D}z
+\tfrac{i}{1+s}\rho z
\right)
=2\pi\frac{1+s}{1-s}
\exp\left(
\tfrac{i}{2}\tfrac{1-s}{1+s}\rho{\cal D}\rho
\right)
\,,\\\label{eq:int exp=1/z2}
\int_{-1}^{1}\frac{\diff s}{(1+s)^2}
\exp\left(
\tfrac{i}{2}\tfrac{1-s}{1+s}\, z{\cal D}z
\right)
=
\frac{i}{z{\cal D}z}
\,,\\\label{eq:lem int(1+...)exp}
\int_{-1}^{1}\frac{\diff s}{(1+s)^2}
\left(1+\frac{i}{2}\frac{1-s}{1+s}\right)
\exp\left(
\tfrac{i}{2}\tfrac{1-s}{1+s}\, z{\cal D}z
\right)
=
\pi\delta^2(z)
\,.
\end{align}
In particular, the change of variables $\zeta=\tfrac{1-s}{1+s}$ gives Eq.\eqref{eq:int exp=1/z2} as a Schwinger integral.
It is important to stress that Eq.\eqref{eq:lem int(1+...)exp} holds only when it has no singular prefactor,
as it makes use of the fact that $z_\a\delta^2(Z)=0$.

\subsection{Homotopy contractions}
We will now show how the resolution operation $q^{(E)\ast}$ acts on two different kind of sources expanded in this basis.
First, on 
\begin{equation}
J(z;\dz)
:=
\dz^\a\dz_\a
\int_{-1}^{1}\frac{\diff s}{1+s}\,
j(s)
\exp\left(
\tfrac{i}{2}\tfrac{1-s}{1+s}\, z{\cal D}z
\right)
\,,
\end{equation}
one gets
\begin{align}
q^{(E)\ast} J
:&=
\int_0^1\frac{\diff t}{t}
2t^2z^\a\dz_\a
\int_{-1}^{1}\frac{\diff s}{1+s}\,
j(s)
\exp\left(
\tfrac{i}{2}t^2\tfrac{1-s}{1+s}\, z{\cal D}z
\right)
\nonumber\\&=
-\dz^\a z_\a
\int_{-1}^{1}\frac{\diff s}{1+s}\,
j(s)
\int_{1}^{s}\diff S\,
\frac{(-2)}{(1+S)^2}
\frac{1+s}{1-s}
\exp\left(
\tfrac{i}{2}\tfrac{1-S}{1+S}\, z{\cal D}z
\right)
\nonumber\\&=
\label{eq:q*dz2}
\dz^\a \partial^\rho_\a
\int_{-1}^{1}
\frac{\diff S}{1+S}\,
\left(2i
\int_{-1}^{S}\frac{\diff s}{1-s}\,
j(s)\right)
\exp\left(
\tfrac{i}{2}\tfrac{1-S}{1+S}\, z{\cal D}z
+\tfrac{i}{1+S}\rho z
\right)
\,.
\end{align}
Then, on a source that might result from \eqref{eq:circ 1form}
\begin{equation}
J(z;\dz)
:=
\dz^\a\dz_\a
\int_{-1}^1\frac{\diff s}{(1+s)^2}
\,j(s)
\left(1
+\frac{i}{2}\frac{1-s}{1+s}
z{\cal D}z
\right)
\exp\left(\tfrac{i}{2}\tfrac{1-s}{1+s}\,z{\cal D}z\right)
\,,
\end{equation}
one finds
\begin{align}
q^{(E)\ast} J
&=
\int_0^1\frac{\diff t}{t}
2t^2z^\a\dz_\a
\int_{-1}^1\frac{\diff s}{(1+s)^2}
j(s)
\left(1
+\frac{i}{2}t^2\frac{1-s}{1+s}
z{\cal D}z
\right)
\exp\left(\tfrac{i}{2}\tfrac{1-s}{1+s}t^2\,z{\cal D}z\right)
\nonumber\\&=
-\dz^\a z_\a
\int_{-1}^1\frac{\diff s}{(1+s)^2}
j(s)
\int_0^1\diff\tau
\left(1
+\tau\partial_\tau
\right)
\exp\left(\tfrac{i}{2}\tfrac{1-s}{1+s}\tau\,z{\cal D}z\right)
\nonumber\\&=
\label{eq:q*VV}
\dz^\a \partial^\rho_\a\left.
\int_{-1}^1\frac{\diff s}{1+s}
(i\,j(s))
\exp\left(\tfrac{i}{2}\tfrac{1-s}{1+s}\,z{\cal D}z
+\tfrac{i}{1+s}\rho z
\right)
\right\vert_{\rho=0}
\,.
\end{align}

\section{Black-hole-like solutions}
\label{App:BH}
In this appendix we generalize some results first presented in \cite{Iazeolla:2011cb} about the spherically symmetric black-hole-like solutions
to the black-hole state generating function \eqref{eq:Phi' bh}, endowed with polarization spinors,
and we provide some additional useful lemmas.

It is convenient to start from the particle initial datum \eqref{eq:Phi' pt} and compute
\begin{equation}
\Psi_{\rm bh}'^{(1)}
=
\Psi_{\rm pt}'^{(1)}\star\kappa_y
=
\Phi_{\rm pt}'^{(1)}
=
\exp\left(\eta y\sigma_0\yb+\chi y+\bar\chi\yb\right)
\,.
\end{equation}
Focusing on stereographic coordinates, the $L$-rotation is performed using Eq.\eqref{eq:Lrot stereo}
and reads
\begin{align}
\Psi_{\rm bh}^{(1)}(Y;X)
&=
\exp\left(
\tfrac{\eta}{2}y\vark^Ly
+\eta yv^L\yb
+\tfrac{\eta}{2}\yb\varkb^L\yb
+\chi^Ly+\bar\chi^L\yb
\right)
\,,\\
\chi^L:&=\frac{1}{h}\left(\chi-x\bar\chi\right)
\,,\qquad
\bar\chi^L:=\frac{1}{h}\left(\bar\chi-\xb\chi\right)
\,,
\end{align}
in terms of the matrices 
\begin{equation}
\vark^L:=
\frac{1}{h^2}\left(\sigma_0\xb-x\bar\sigma_0\right)
\,,\qquad
\varkb^L:=
\frac{1}{h^2}\left(\bar\sigma_0x-\xb\sigma_0\right)
\,,\qquad
v^L:=
\frac{1}{h^2}\left(\sigma_0-x\bar\sigma_0x\right)
\,.
\end{equation}
They have the following properties:
\begin{alignat}{2}
(\vark^L)^2=(\varkb^L)^2=r^2
\,,&\qquad
&v^L\vb^L=\vb^Lv^L=-(1+r^2)&
\,,\\
\vark^Lv^L=-v^L\varkb^L
\,,&\qquad
&\vb^L\vark^L=-\varkb^L\vb^L&
\,.
\end{alignat}
where the products and squares are meant in terms of matrix notation, with the NW-SE contraction \eqref{eq:NWSE}.
These relations are reflected in the possibility the write the matrices as
\be
\label{eq:uE}
(\vark^L)^{\a\b}
=
r(
u_E^{+\a}u_E^{-\b}+u_E^{-\a}u_E^{+\b})
\,,\qquad
(\varkb^L)^{\ad\bd}
=
r(
\ub_E^{+\ad}\ub_E^{-\bd}+\ub_E^{-\ad}\ub_E^{+\bd})
\,,
\ee\be
(v^L)^{\a\bd}
=
\sqrt{1+r^2}(
u_E^{+\a}\ub_E^{+\bd}+u_E^{-\a}\ub_E^{-\bd})
=
(\bar{v}^L)^{\bd\a}
\,,
\ee
in terms of the \textit{E-adapted spin-frame}, whose expression can be found in App.E of \cite{Iazeolla:2011cb}.

The corresponding Weyl zero-form then reads
\begin{align}
\Phi_{\rm bh}^{(1)}(x;Y;X)
=&
-\frac{i}{\sqrt{\eta^2} r}
\exp\left(
-\tfrac{1}{2\eta}y(\vark^L)^{-1}y
+iy(\vark^L)^{-1}v^L\yb
-\tfrac{i}{\eta}y(\vark^L)^{-1}\chi^L
\right)\times\\\nonumber&\times\exp\left(
\tfrac{\eta}{2}\yb(\varkb^L-\bar{v}^L(\vark^L)^{-1}v^L)\yb
+\tfrac{1}{2\eta}\chi^L(\vark^L)^{-1}\chi^L
+\yb\bar{v}^L(\vark^L)^{-1}\chi^L+\bar\chi^L\yb
\right)
\\=&
\exp\left(
-\tfrac{1}{2\eta}y(\vark^L)^{-1}y
+iy(\vark^L)^{-1}v^L\yb
-\tfrac{i}{\eta}y(\vark^L)^{-1}\chi^L
\right)
\Phi_{\rm bh}^{(1)}(x;0,\yb;X)
\,.
\end{align}
The internal connection can be computed using Eq.\eqref{eq:V1 unfact}
\begin{align}
V_{\rm bh}^{(1,E+i\partial_Y)}
=&
\left.
-\frac{b}{2}
\left.\Phi_{\rm bh}^{(1)}\right\vert_{y=0}
\dz\partial^\rho
\int_{-1}^1\frac{\diff s}{(1+s)\sqrt{\det G}}
\exp\left(
-\tfrac{1}{2}\varsigma\,G^{-1}\varsigma
+\tfrac{i}{2}\tfrac{1-s}{1+s}z{\cal D}z
+\tfrac{i}{1+s}\rho z
\right)
\right\vert_{\rho=0}
\nonumber\\&
-\hc
\,,\\
\varsigma:=&iy+i(\vark^L)^{-1}v^L\yb-\frac{1}{\eta}(\vark^L)^{-1}\chi^L-i\tfrac{1-s}{1+s}z{\cal D}-\tfrac{i}{1+s}\rho
\,,\\
G:=&\frac{1}{\eta r^2}\vark^L-i\frac{1-s}{1+s}{\cal D}
\,.
\end{align}
Up to the polarisation term, this result agrees with the one presented in \cite{Iazeolla:2017vng}.
If we further define
\begin{equation}
\ytb:=
y+(\vark^L)^{-1}v^L\yb-\frac{1}{\eta}(\vark^L)^{-1}\chi^L
\,,
\end{equation}
we can rewrite the result 
\begin{align}
V_{\rm bh}^{(1,E+i\partial_Y)}
&=
-\frac{b}{2}
\left.\Phi^{(1)}_{\rm bh}\right\vert_{y=0}
\partial^\rho_\a
\int_{-1}^1\frac{\diff s}{(1+s)\sqrt{\det G}}
\exp\left(
\tfrac{1}{2}\ytb\,G^{-1}\ytb 
+\tfrac{i}{1+s}\rho\left(
1+i\tfrac{1-s}{1+s}G^{-1}{\cal D}
\right)z
\right)\times\nonumber\\&\qquad\left.\times\exp\left(
-\tfrac{1}{1+s}\rho\,G^{-1}\ytb 
-\tfrac{1-s}{1+s}z{\cal D}G^{-1}\ytb
+\tfrac{i}{2}\tfrac{1-s}{1+s}
z\left(
{\cal D}+i\tfrac{1-s}{1+s}{\cal D}G^{-1}{\cal D}
\right)z
\right)
\right\vert_{\rho=0}
\label{eq:V1 BH app}
\,,
\end{align}
The square and determinant of the matrix $G$ are given by
\begin{equation}
\label{eq:detG}
G^2=\frac{(1+s)^2-i\eta(1+s)(1-s)\Tr(\vark^L{\cal D})-\eta^2r^2(1-s)^2}{\eta^2r^2(1+s)^2}
=-\det G
\,.
\end{equation}
Let us give a series of additional lemmas, relevant for the derivation of the COMST for black-hole states.
\begin{small}
\begin{align}
\frac{\diff}{\diff s}\left(\frac{1}{\sqrt{\det G}}\right)
&=
-\frac{1}{(1+s)\sqrt{\det G}}
\frac{i\eta(1+s)\Tr(\vark^L{\cal D})+2(1-s)\eta^2r^2}{(1+s)^2-i\eta(1-s)(1+s)\Tr(\vark^L{\cal D})-\eta^2r^2(1-s)^2}
\,,\\
\frac{\diff}{\diff s}\left(\frac{1}{\sqrt{\det G}}\frac{1-s}{1+s}\right)
&=
-\frac{1}{(1+s)\sqrt{\det G}}
\frac{2(1+s)-i\eta(1+s)\Tr(\vark^L{\cal D})}{(1+s)^2-i\eta(1-s)(1+s)\Tr(\vark^L{\cal D})-\eta^2r^2(1-s)^2}
\,,\\
\Tr(G^{-1}{\cal D})
&=
G^{-2}\left(\frac{1}{\eta r^2}\Tr(\vark^L{\cal D})
-2i\frac{1-s}{1+s}\right)
\,,\\
\Tr\left(1+i\frac{1-s}{1+s}G^{-1}{\cal D}\right)
&=
-\sqrt{\det G}(1+s)^2
\frac{\diff}{\diff s}\left(\frac{1}{\sqrt{\det G}}\frac{1-s}{1+s}\right)
\,,\\
(G^{-1}{\cal D}G^{-1})_{\a\b} 
&=
\frac{i}{2}(1+s)^2\frac{\diff}{\diff s}(G^{-1}_{\a\b})
\,.
\end{align}
\end{small}

\bibliography{biblio.bib}
\bibliographystyle{utphys}

\end{document}